\documentclass[a4paper,11pt]{article}
\pdfoutput=1 

\usepackage{jheppub} 

\usepackage[T1]{fontenc} 
\title{\boldmath Top quark mass studies with $t\bar{t}j$  at
  the LHC}

\author[a]{G. Bevilacqua,}
\author[b,\,c]{H. B. Hartanto,}
\author[d]{M. Kraus,}
\author[d]{M. Schulze}
\author[c]{and M. Worek}

\affiliation[a]{MTA-DE Particle Physics Research Group, University of
Debrecen, H-4010 Debrecen, PBox 105, Hungary} 
\affiliation[b]{Institute for
Particle Physics Phenomenology, Department of Physics, 
Durham University, Durham, DH1 3LE, UK} 
\affiliation[c]{ Institute for Theoretical Particle Physics
and Cosmology, RWTH Aachen University, D-52056 Aachen, Germany}
\affiliation[d]{Humboldt-Universit\"at zu
Berlin, Institut f\"ur Physik, Newtonstra\ss{}e 15, D-12489 Berlin,
Germany}

\emailAdd{giuseppe.bevilacqua@science.unideb.hu}
\emailAdd{heribertus.b.hartanto@durham.ac.uk}
\emailAdd{manfred.kraus@physik.hu-berlin.de}
\emailAdd{markus.schulze@physik.hu-berlin.de}
\emailAdd{worek@physik.rwth-aachen.de}

\abstract{
A precise measurement of the top quark mass, a fundamental parameter
of the Standard Model, is among the most important goals of top quark
studies at the Large Hadron Collider. Apart from the standard methods,
numerous new observables and reconstruction techniques are employed to
improve the overall precision and to provide different sensitivities
to various systematic uncertainties.  Recently, the normalised inverse
invariant mass distribution of the $t\bar{t}$ system and the leading
extra jet not coming from the top quark decays has been proposed for
the $pp \to t\bar{t}j$ production process, denoted as ${\cal
R}(m_t^{pole},\rho_s)$. In this paper, a thorough study of different
theoretical predictions for this observable, however, with top quark
decays included, is carried out. We focus on fixed order NLO QCD
calculations for the di-lepton top quark decay channel at the LHC with
$\sqrt{s}=13$ TeV. First, the impact on the extraction of $m_t$ is
investigated and afterwards the associated uncertainties are
quantified. In one approach we include all interferences, off-shell
effects and non-resonant backgrounds. This is contrasted with a
different approach with top quark decays in the narrow width
approximation. In the latter case, two cases are employed: NLO QCD
corrections to the $pp\to t\bar{t}j$ production process with leading
order decays and the more sophisticated case with QCD corrections and
jet radiation present also in top quark decays. The top quark mass
sensitivity of ${\cal R}(m_t^{pole},\rho_s)$ is investigated and
compared to other observables: the invariant mass of the top anti-top
pair, the minimal invariant mass of the $b$-jet and a charged lepton
as well as the total transverse momentum of the $t\bar{t}j$
system. Once top quark decays are included the invariant mass of the
$t\bar{t}$ system shows better sensitivity to the top quark mass
extraction and smaller dependence on the off-shell effects and
non-resonant contributions of the top quark and the $W$ gauge boson
than the ${\cal R}(m_t^{pole},\rho_s)$ observable.}

\dedicated{\rm TTK-17-28, HU-EP-17/19, IPPP/17/62}

\begin{document} 
\maketitle
\flushbottom

\tableofcontents
%
\section{Introduction}
%

The top quark is the most massive of all observed elementary
particles. As a result it has a very short lifetime and decays before
hadronic bound states can be formed. With a mass of the order of the
electroweak scale, the top quark decays through the weak interaction
into a $W$ boson and a down-type quark, most frequently into the
bottom quark. This gives us an opportunity to study the unstable top
quark via its decay products, i.e charged leptons, bottom- and
light-jets as well as missing transverse momentum. The top quark
Yukawa coupling to the Higgs boson, expressed as $\lambda_t =
\sqrt{2}\left( m_t/v\right)$, is of the order of unity. This alone
makes the top quark unique among the fermions and its closer relation
to physics Beyond the Standard Model (BSM) is anticipated. In the
Standard Model (SM) of particle physics the top quark coupling to the
Higgs boson is controlled by the Yukawa sector of the model. Moreover,
its couplings to the other particles are fixed through the gauge
structure of the SM.  On the other hand, the top quark mass is a
fundamental parameter of QCD, which furthermore influences electroweak
processes through virtual effects.  Thus, the numerical value of the
top quark mass affects theory predictions of cross sections and
various differential distributions that are indispensable for example
in studying the Higgs boson properties or in searching for BSM
effects. Additionally, the stability of the electroweak vacuum depends
crucially on the precise numerical value of $m_t$, see
e.g. Ref. \cite{Degrassi:2012ry,Alekhin:2012py}.  Therefore, $m_t$ is
a crucial input for the self-consistency of the SM.

The mass of the top quark can be measured in a variety of ways.
However, the most precise experimental determinations are based on the
direct kinematic reconstruction of the measured top quark decay
products. Various differential distributions sensitive to $m_t$ are
employed to perform multi-observable analyses. To this end,
differential observables inferred from data are typically normalised
to the inclusive $t\bar{t}$ cross section and compared to theoretical
predictions generated with different top quark masses. 
The current standard for the latter comprises next-to-leading-order
(NLO) QCD corrections to the $pp \to t\bar{t}$ production process
matched to parton shower (PS) Monte Carlo (MC) programs, see
e.g. Ref.~\cite{Frixione:2003ei, Frixione:2007nw}.  Only recently
NLO+PS matching techniques that deal with radiation from top quark
decays and which allow for a consistent treatment of top quark
resonances have been introduced in Ref.~\cite{Campbell:2014kua,
Jezo:2015aia,Jezo:2016ujg}.   Apart from parton shower effects,
non-perturbative physics must also be incorporated into $m_t$
measurements. Here, choices must be made for example on the proton
parton distribution functions (PDFs), the hadronisation model, the
underlying event, the modelling of colour re-connection and the
description of additional interactions accompanying the hard
scattering process, the so-called pile-up.  Even though
the definition and implementation of the top quark mass in NLO+PS MC
tools is based on the on-shell renormalisation scheme of $m_t$ at one
loop and it is identical to what is used in parton-level calculations,
above mentioned effects play an important role as they enter in the
relation between $m_t$ and physical observables. The top quark mass
can also be extracted indirectly from the inclusive total cross
section for the top quark pair production process. However, even the
total cross section, $\sigma_{t\bar{t}}$, is not free from
uncertainties due to the above mentioned non perturbative effects. Due
to the extrapolation of the fiducial cross section to the full phase
space the measured $\sigma_{t\bar{t}}$ depends on hadronisation
effects as it relies on the MC modelling of these phenomena. The
dependence on non-perturbative effects is smaller than for exclusive
observables, but, unfortunately, top quark mass determinations based
on the mass dependence of the inclusive $t\bar{t}$ production cross
section are less precise.

Since the discovery of the top quark, direct measurements of
$t\bar{t}$ production have already been made at five different
center-of-mass system energies, two at the Tevatron and three at the
LHC.  The top-quark mass has been measured in various decay channels,
i.e. the $\ell$+jets, the di-lepton, and the all-jets channel by all
four experiments: CDF, D0, ALTAS and CMS.  A combination of Tevatron
and LHC measurements has been performed in 2014 and resulted in
\begin{equation} 
m_t=173.34 \pm 0.27 ~{\rm (stat)} \pm 0.71 
~{\rm (syst) } ~{\rm GeV} ~~~~~~~~~~{\rm Tevatron ~+ ~LHC}
~\text{\cite{ATLAS:2014wva}}\,,
\end{equation}
with a total uncertainty of $0.76$ GeV.  The latest and most precise
combinations of various measurements presented separately by ATLAS and
CMS collaborations, from June 2016 and September 2015 respectively,
can be summarised as follows
\begin{equation}
\begin{split}
m_t & =172.84 \pm 0.34 ~{\rm (stat)} \pm 0.61 ~{\rm (syst)} 
~ {\rm GeV} ~~~~~~~~~~{\rm  ATLAS}
~\text{\cite{Aaboud:2016igd}}\,,\\[0.2cm]
m_t &= 172.44 \pm 0.13 ~{\rm (stat)}  \pm 0.47 ~{\rm (syst)} 
 ~{\rm GeV}  ~~~~~~~~~~{\rm CMS}
 ~\text{\cite{Khachatryan:2015hba}}\,,\\
\end{split}
\end{equation}
with a total uncertainty of $0.70$ GeV (ATLAS) and $0.49$ GeV
(CMS). The world's best measurements by the ATLAS and CMS
collaborations are in good agreement with the 2014 world average.
These results can be further compared to $m_t$ extracted from the
inclusive top quark pair production cross-section $\sigma(t\bar{t})$
at $\sqrt{s}=7,8$ and $13$ TeV. Using the expected
dependence of the cross section on the top quark mass and comparing it
to theoretical predictions at the next-to-next-to-leading order level
including the resummation of next-to-next-to-leading logarithmic soft
gluon effects (NNLO+NNLL) \cite{Czakon:2013goa} the following values
of $m_t$ have been determined 
\begin{equation}
\begin{split}
m_t & = 172.90^{+ 2.50}_{- 2.60} ~{\rm GeV}   ~~~~~~~~~~{\rm  ATLAS} ~
7+8 ~{\rm TeV} ~\text{\cite{Aad:2014kva}}\,,\\[0.2cm]
m_t &= 173.80^{+1.70}_{-1.80} ~{\rm GeV}  ~~~~~~~~~~{\rm  CMS} ~
7+8 ~{\rm TeV}  ~\text{\cite{Khachatryan:2016mqs}}\,,\\[0.2cm]
m_t &= 170.60^{+2.70}_{-2.70} ~{\rm GeV} ~~~~~~~~~~{\rm  CMS} ~
13 ~{\rm TeV}  ~\text{\cite{Sirunyan:2017uhy}}\,.\\
\end{split}
\end{equation}
Predictions for $t\bar{t}$ production at NNLO+NNLL also
employ the on-shell scheme for mass renormalisation since the scheme
is commonly used for calculations of perturbative higher order
predictions in top quark physics.  However, the top quark pole mass,
$m^{pole}_t$, has an uncertainty of its own, which is of the order of
${\cal O}(\Lambda_{QCD})$.  For example, the intrinsic uncertainty on
the $m^{pole}_t$ definition due to renormalons has been recently
estimated to be of the order of ${\cal O}(100)$ MeV
\cite{Beneke:2016cbu, Hoang:2017btd}.  On the experimental side, the
main systematic uncertainties contributing to the top quark mass
measurements typically originate from the understanding of the jet
energy scale for light-quark and $b$-quark originated jets and from
modelling of the performance of the $b$-tagging algorithms. Thus,
various alternative methods to extract $m_t$ have been proposed to
give a further insight by providing different sensitivities to various
systematic uncertainties. Such methods, which can also help to improve
the overall precision, comprise either new observables or new
reconstruction techniques, see e.g. Ref.
\cite{Biswas:2010sa,Heinrich:2013qaa,
Frixione:2014ala,Agashe:2016bok,Heinrich:2017bqp,Corcella:2017rpt,
Ravasio:2018lzi} for $pp\to t\bar{t}$ production.  Among others, a
novel method to determine $m_t$ in the $pp \to t\bar{t}j$ production
process has been proposed in Ref. \cite{Alioli:2013mxa,
Fuster:2017rev} for on-shell top quarks. It uses the normalised
differential cross section as a function of the invariant mass of the
$t\bar{t}$ system and the leading extra jet not coming from the top
quark decays. To be more precise it is defined according to
\begin{equation} \rho_s = \frac{2m_0}{M_{t\bar{t}j}}\,,~~~
~~~~~~~~~~~~~~~~~~{\rm with}
~~~~~{m_0=170} ~{\rm GeV}\,,
\end{equation}
where $m_0$ is a parameter that is of the order of the top quark mass
and $M_{t\bar{t}j}$ is the invariant mass of the $t\bar{t}j$
system. In Ref.  \cite{Alioli:2013mxa, Fuster:2017rev} 
 NLO QCD corrections to on-shell $t\bar{t}j$ production
are matched with parton shower programs that are responsible for top
quark decays, shower effects and non-perturbative physics. Since
additional radiation depends on the mass of the top quark, the
$\rho_s$ distribution should impact the $m_t$ extraction differently
than for example the invariant mass of the top anti-top pair alone.
As a consequence it should be studied in the context of a precise
determination of the top quark mass.  Indeed, the method, has already
been applied by ATLAS and CMS experimental groups \cite{Aad:2015waa,
CMS:2016khu}. The measured differential cross sections have been
compared to the predicted cross sections for each bin of the $\rho_s$
observable for the full phase-space using different top quark
masses. In the end the most probable top quark mass has been extracted
yielding
\begin{equation}
\begin{split} m_t & = 173.70^{+2.28}_{-2.11} ~{\rm GeV}
~~~~~~~~~~~~~{\rm ATLAS ~ at ~7 ~ TeV},\\[0.2cm] m_t
& =169.90^{+4.52}_{-3.66} ~{\rm GeV} ~~~~~~~~~~~~~{\rm CMS ~at ~8 ~
  TeV}.\\
\end{split}
\end{equation}
 In this paper we investigate the sensitivity of the
$\rho_s$ observable even further by including NLO QCD corrections also
in top quark decays.  The main goal is to study the impact of top
quark decay modelling on the extraction of the top quark mass in the
di-lepton top quark decay channel. To this end we concentrate on fixed
order NLO QCD predictions, which rigorously allows us to define the
top quark pole mass as the input parameter. We compare three distinct
theoretical predictions for the $pp\to t\bar{t}j$ process at the NLO
level in QCD. First a complete description of the $e^+\nu_e
\mu^-\bar{\nu}_\mu b\bar{b} j$ final state as explained in
Ref. \cite{Bevilacqua:2015qha, Bevilacqua:2016jfk} is employed, which
takes into account all possible contributions, i.e. double (top
quark), single (top quark) and non (top quark) resonant contributions
together with their interferences and off-shell effects. Off-shell
effects and non-resonant contributions due to the $W$ gauge boson are
also properly taken into account. From the quantum field theory point
of view this is the most comprehensive description of the $t\bar{t}j$
production process at NLO QCD because all effects that are
perturbatively calculable at ${\cal O}(\alpha^4 \alpha_s^4)$ are
accounted for. We dub this approach the {\it Full} approach. As a
second case, we consider the narrow width approximation (NWA)
description for top quarks and $W$ gauge bosons
\cite{Melnikov:2011qx}, with the following decay chains $pp \to
t\bar{t}j \to W^+W^- b\bar{b} j \to e^+\nu_e \mu^-\bar{\nu}_\mu
b\bar{b} j$ and $pp \to t\bar{t} \to W^+W^- b\bar{b} j \to e^+\nu_e
\mu^-\bar{\nu}_\mu b\bar{b} j$. Thus, NLO QCD corrections to top quark
pair production with a hard jet are incorporated together with QCD
radiative corrections to top quark decays including also the
possibility that this hard jet is emitted in the decay stage. Even
though $t$ and $W$ decays are treated in the NWA, NLO spin
correlations are retained throughout the entire decay chain. This
approach is dubbed {\it NWA}. Finally, mostly for comparisons, we
employ calculations from Ref. \cite{Melnikov:2010iu}, where the NLO
QCD corrections to on-shell $t\bar{t}j$ production are provided,
however, top quark decays are included only at the leading order (LO)
in perturbative QCD. Thus, the following decay chain is investigated
$pp \to t\bar{t}j ~\stackrel{\rm LO}{\longrightarrow} ~ W^+W^-
b\bar{b} j \to e^+\nu_e \mu^-\bar{\nu}_\mu b\bar{b} j$, hence spin
correlations are only contained at the LO level.  This third approach
is dubbed {\it {\it NWA}${}_{Prod.}$}. At the end of the paper we are
going to compare the normalised $\rho_s$ differential distribution to
other observables that are commonly used in the top quark mass
measurements, namely the invariant mass of the $t\bar{t}$ system,
$M_{t\bar{t}}$, and the (minimal) invariant mass of the charged lepton
and the $b$-jet, $M_{b\ell}$. We shall also present results for the
total transverse momentum of the $t\bar{t}j$ system, $H_T$, owing to
its similar sensitivity to $m_t$ as observed in the case of $\rho_s$.

The paper is organised as follows.  The general setup of our analysis
is described in Section 2. In Section 3 we depict the main observable
and discuss the details of methods used in the top quark mass
extraction.  In Section 4 we present our results on the $m_t$
extraction and assess theoretical uncertainties stemming from the
scale dependence and various assumptions that enter into the
parameterisation of the PDFs. For the latter case we follow PDF4LHC
recommendations for LHC Run II \cite{Butterworth:2015oua} by employing
CT14, MMHT14 and NNPDF3 PDF sets. Results for a slightly modified
version of $\rho_s$ are discussed in Section 5. In Section 6 a
comparison between $\rho_s$ and other observables, that are also
sensitive to $m_t$, is performed.  Following our conclusions that are
given in Section 7, we include an appendix that presents the
comparison between {\it Full}, {\it NWA} and {\it NWA}${}_{Prod.}$
obtained using a fixed scale choice, for several observables.

%
\section{Setup of the Analysis}
%

Numerical results with complete top quark and $W$ gauge boson
off-shell effects and non-resonant backgrounds included, which are the
basis for our top quark mass extraction studies, are obtained with the
help of the \textsc{Helac-Nlo} Monte Carlo program
\cite{Bevilacqua:2011xh}, that comprises \textsc{Helac-Dipoles}
\cite{Czakon:2009ss, Bevilacqua:2013iha} and \textsc{Helac-1Loop}
\cite{vanHameren:2009dr}. Theoretical aspects related to the complex
mass scheme introduced in our calculations are explained in details in
Ref. \cite{Bevilacqua:2010qb}. On the other hand, a comprehensive
description of NLO calculations in the NWA for the top quarks is given
in Ref. \cite{Melnikov:2009dn}.  We, therefore, do not repeat these details
here, but rather refer interested readers to our earlier
publications. We consider the $pp \to e^+\nu_e\mu^- \bar{\nu}_\mu b
\bar{b}j+X$ process at ${\cal O}(\alpha_s^4 \alpha^4)$ for the LHC Run
II energy of $\sqrt{s}$ = 13 TeV. Throughout, for the masses and
widths of the $W$ and $Z$ gauge bosons we use the following values
\begin{equation}
\begin{split}
  m_{W} &=80.399 ~{\rm GeV}\,,  \quad \quad \quad \quad \quad  
\Gamma_{W} = 2.09875 ~{\rm GeV}\,,
\vspace{0.2cm}\\
  m_{Z}&=91.1876  ~{\rm GeV}\,,  \quad \quad \quad \quad \quad  
\Gamma_{Z} = 2.50848 ~{\rm GeV}\,,
\end{split}
\end{equation}
where, in the total decay rates for the $W$ and $Z$ gauge bosons, the
NLO QCD corrections to $W\to f_1 \bar{f}_2$ and $Z\to f \bar{f}$ have
been included.  Further electroweak parameters such as the electroweak
coupling and the weak mixing angle are computed in the so called
$G_\mu$ scheme with the Fermi constant $G_\mu=1.16637 \cdot 10^{-5}
~{\rm GeV}^{-2}$ through the following formulae
\begin{equation}
\alpha = \frac{\sqrt{2}}{\pi} \, G_\mu\, m^2_W  \, \sin^2\theta_W\,,
\quad \quad \quad \quad 
\sin^2\theta_W= 1-\frac{m^2_W}{m^2_Z}\,.
\end{equation}
 The mass and the width of the top
quark are set to
\begin{equation}
m_{t}=173.2 ~{\rm GeV}\,,  \quad \quad \quad \quad\Gamma^{\rm
  NLO}_t=1.35146 ~{\rm  GeV}\,, \quad \quad \quad \quad \Gamma^{\rm
  NLO}_{tW} =1.37276 ~{\rm GeV}\,, 
\end{equation}
where $\Gamma^{\rm NLO}_t$ refers to the top quark width with $W$
gauge boson off-shell effects included and $\Gamma^{\rm NLO}_{tW}$ to
the top quark width with an on-shell $W$ gauge boson as used in the
NWA \cite{Jezabek:1988iv, Denner:2012yc}. Both values are derived for
massless $b$ quarks since all leptons and $u,d,c,s,b$ partons are
considered to be massless.  The normalised $\rho_s$ differential
distribution and other observables are also evaluated  with different
top quark masses to be used for the fitting procedure. Generally we
shall use the following five values of $m_t$: $168.2$ GeV, $170.7$
GeV, $173.2$ GeV (the default value), $175.7$ GeV and $178.2$
GeV. This corresponds to the following spread $m_t = 173.2 \pm 5$ GeV
in steps of $2.5$ GeV.  For completeness, corresponding top quark decay
widths are shown in Table \ref{topwidths}.
%
\begin{table}[t!]
\begin{center}
\begin{tabular}{ccc}
 \hline\hline 
  $m_t~[{\rm GeV}]$ & $\Gamma_{t}^{\rm
NLO}~[{\rm GeV}]$ & $\Gamma_{tW}^{\rm NLO}~[{\rm GeV}]$ \\
  \hline\hline 
  $168.2$ & $1.21823$ & $1.23792$ \\[0.1cm]
  $170.7$ & $1.28389$ & $1.30438$ \\[0.1cm]
  $173.2$ & $1.35146$ & $1.37276$ \\[0.1cm]
  $175.7$ & $1.42097$ & $1.44309$ \\[0.1cm]
  $178.2$ & $1.49243$ & $1.51538$ \\
 \hline\hline 
\end{tabular}
\end{center}
\caption{\it Top quark decay widths for five values of
$m_t$. $\Gamma^{\rm NLO}_t$ refers to the top quark width with $W$
gauge boson off-shell effects included and $\Gamma^{\rm NLO}_{tW}$ to
the top quark width with the on-shell $W$ gauge boson as used in the
NWA. Massless bottom quarks are assumed.}
\label{topwidths}
\end{table}
%
The value of $\alpha_s(m_{t})$ needed for $\Gamma^{\rm NLO}_{t}$ and
$\Gamma^{\rm NLO}_{tW}$ is obtained from $\alpha_s (m_{Z})=0.118$ via
LHAPDF \cite{Buckley:2014ana}.  In general, the running of the strong
coupling constant $\alpha_s$ with two-loop accuracy is provided by the
LHAPDF library and the number of active flavours is set to $N_F =
5$. We employ the CT14nlo \cite{Dulat:2015mca}, NNPDF30-nlo-as-0118
\cite{Ball:2014uwa} and MMHT2014nlo68clas118
\cite{Harland-Lang:2014zoa} PDF sets that we dubbed as CT14, NNPDF3
and MMHT14.  Suppressed contributions from bottom quarks in PDFs are
not included.  All final-state partons with pseudo-rapidity $|\eta|
<5$ are recombined into jets with a separation parameter $R$ in the
rapidity-azimuthal-angle plane via the IR-safe anti$-k_T$ jet
algorithm \cite{Cacciari:2008gp}. The value of the jet radius $R$ is
set to $R=0.5$. When merging particles during the clustering
procedure one must specify how to combine the momenta. We use the
simplest procedure, currently used by the LHC experiments,
and add the four-vectors of combined partons (the so called
$E$-scheme). Finally, we require exactly two $b$-jets, at least one
light jet, two charged leptons and missing transverse momentum,
$p^{miss}_T$.  These final states have to fulfil  the following
criteria
\begin{equation}
\begin{array}{lcl}
 p_{T}( \ell)>30 ~{\rm GeV}\,,    & & p_{T}(j)>40 ~{\rm GeV}\,,
\vspace{0.2cm}\\
p^{miss}_{T} >40 ~{\rm GeV} \,,   & \quad \quad \quad \quad \quad
                                    \quad
& \Delta R_{jj}>0.5\,,
\vspace{0.2cm}\\
\Delta R_{\ell\ell}>0.4 \,,  &&
 \Delta R_{\ell j}>0.4 \,,
\vspace{0.2cm}\\
 |y_\ell|<2.5\,,&&
|y_j|<2.5 \,,
\end{array}
\end{equation}
where $\ell$ stands for $\mu^-$ and $e^+$ whereas $j$ corresponds to
light- and $b$-jets. For renormalisation and factorisation scales,
$\mu_R$ and $\mu_F$, three cases are considered. Specifically, we use
a fixed scale $\mu_R = \mu_F =\mu_0= m_{t}$ and two dynamical ones
$\mu_R = \mu_F =\mu_0=E_T/2$ and $\mu_R = \mu_F =\mu_0=H_T/2$, where
the transverse energy of the $t\bar{t}$ system and the total
transverse momentum of the $t\bar{t}j$ system are defined according to
\begin{equation}
\begin{split}
E_T
&=\sqrt{m^2_{t}+p_T^2(t)}+\sqrt{m^2_{t}+p_T^2(\,\bar{t}\,)}\,,\\[0.2cm]
H_T&=p_T(e^+) + p_T(\mu^-) + p_T(j_{b_1}) + p_T(j_{b_2}) + p_T(j_1) +
p_T^{miss}\,.\\
\end{split}
\end{equation}
The dynamical scales are evaluated using the momenta after the
application of the jet-algorithm.  Thus, $j_{b_1}$ and $j_{b_2}$ are
the $b$-jets and $j_1$ is the light (hard) jet. In the case of two
resolved jets the jet with the highest transverse momentum is chosen.
Additionally, momenta of $t$ and $\bar{t}$ are reconstructed from
their decay products, i.e. $p(t) =p( j_{b_1}) + p(e^+) + p(\nu_e)$ and
$p(\,{\bar t}\,) = p( j_{b_{2}}) + p(\mu^-) +p(\bar{\nu}_\mu)$ where
$j_{b_1}$ comes from the $b$-quark and $j_{b_2}$ from anti-$b$
quark. 

%
\section{Description of the Observable and the 
Methods Used}
%

\begin{figure}[t!]
\begin{center}
\includegraphics[width=1.0\textwidth]{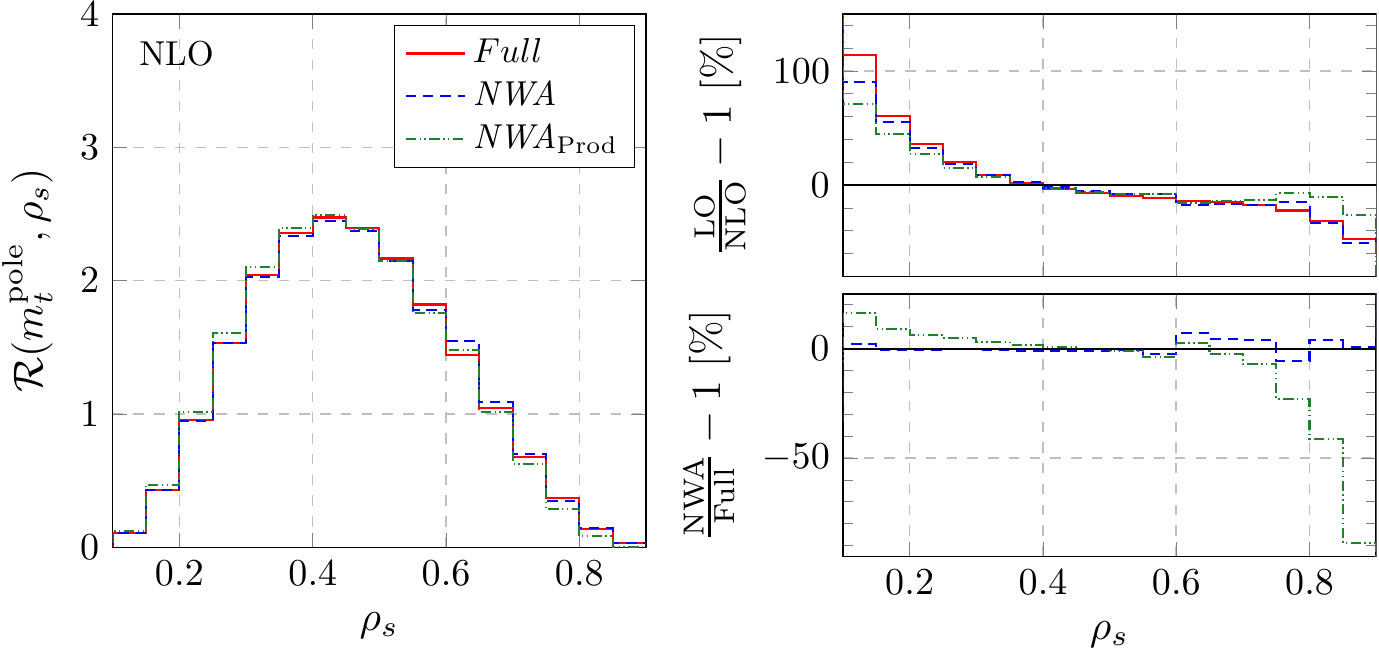}
\end{center}
\caption{\it Normalised $\rho_s$ differential distribution at NLO QCD
for the $pp \to e^+\nu_e \mu^- \bar{\nu}_\mu b\bar{b} j+ X$ production
process at the LHC with $\sqrt{s} =$ 13 TeV. Three different
theoretical descriptions with $m_t=$ 173.2 GeV are shown.  Also given
are the relative size of NLO QCD corrections and the combined relative
size of finite-top-width and finite-W-width effects for the normalised
$\rho_s$ observable. Renormalisation and factorisation scales are set
to the common value $\mu_R =\mu_F =\mu_0$ where $\mu_0 =m_t$.  
The CT14 PDF set is employed. }
\label{fig:1observable}
\end{figure}

We start with the ${\cal R}(m_t^{pole},\rho_s)$ observable defined as
normalised differential distribution of the $t\bar{t}+1$ jet cross
section with respect to the inverse invariant mass of the final
state, $M_{t\bar{t}j}$, that can be written in the following form
\begin{equation}
{\cal R}(m_t^{pole},\rho_s) =\frac{1}{\sigma_{t\bar{t}j}} 
\frac{d\sigma_{t\bar{t}j}}{d\rho_s} (m_t^{pole},\rho_s)\,, 
~~~~~~~~~~~~~~~~~{\rm with} ~~~~~~
\rho_s = \frac{2m_0}{M_{t\bar{t}j}}\,,
\end{equation}
where $m_0 = 170$ GeV is a scale of the order of $m_t$. 
We note here that top quarks are reconstructed from their decay
products assuming exact $W$ gauge boson reconstruction and taking $j$
as the leading light jet irrespectively of its origin (production or
decay). This corresponds to the invariant mass of the $WWb\bar{b}j$
system, $M_{WWb\bar{b}j}$, which for brevity we dub $M_{t\bar{t}j}$.
In Figure \ref{fig:1observable}, we present the NLO predictions for
${\cal R}(m_t^{pole},\rho_s)$ considering the following three cases,
namely {\it Full} (red solid line), {\it NWA} (blue dashed line) and
{\it NWA}${}_{Prod.}$ (green dotted-dashed line) for
$\mu_R=\mu_F=\mu_0=m_t$ that is a scale choice commonly used for the
$pp\to t\bar{t}j$ production process at the LHC and with the CT14 PDF
set.  Also shown are the relative NLO QCD corrections $\sigma^{\rm
LO}/\sigma^{\rm NLO} -1$ (upper right panel) and the relative
deviation of the NWA results from the full calculation (lower right
panel). Both are given in percent. To be more precise, in the former
and the latter case shape differences are shown, since we consider
normalised differential cross sections. For completeness in Table
\ref{tab:NLO_CS} integrated NLO cross sections are provided. Combined
finite top quark and $W$ gauge boson width effects change the NLO
cross section by $2\%$, which is consistent with the expected
uncertainty of the NWA, that is of the order of ${\cal
O}(\Gamma_t/m_t)$.
%
\begin{figure}[t!]
\begin{center}
\includegraphics[width=0.7\textwidth]
{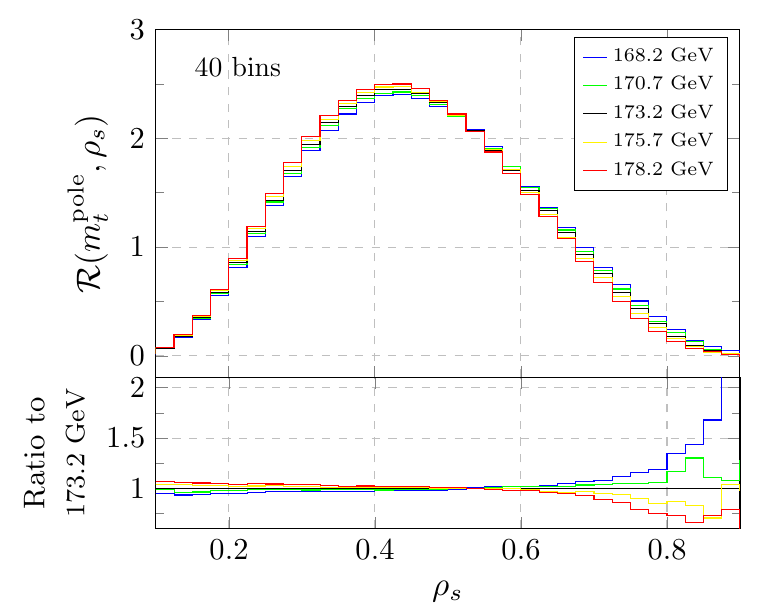}
\end{center}
\caption{\it 
 Normalised $\rho_s$ differential distribution at NLO QCD for the $pp
\to e^+\nu_e \mu^- \bar{\nu}_\mu b\bar{b} j+ X$ production process at
the LHC with $\sqrt{s} =$ 13 TeV. The Full case for five different top
quark masses is presented. We also plot the ratio to the
result with the default value of $m_t$, i.e. $m_t=$ 173.2 GeV.
Renormalisation and factorisation scales are set to the common value
$\mu_R =\mu_F =\mu_0$ where $\mu_0 =H_T/2$. The CT14 PDF set is
employed. }
\label{fig:1masses}
\end{figure}
%
In Figure \ref{fig:1masses}, we show ${\cal R}(m_t^{pole},\rho_s)$ as
given by the best theoretical predictions ({\it Full}\,) with $\mu_R
=\mu_F =\mu_0=H_T/2$ for five different top quark masses, that is $m_t
\in \left\{168.2, 170.7, 173.2, 175.7, 178.2\right\}$
GeV. We also plot the ratio to the result with the
default value of the top quark mass, i.e. $m_t=173.2$ GeV. To obtain
these results the CT14 PDF set has been used, however, any of the PDF
sets recommended by the PDF4LHC group can be employed here. 
A significant mass dependence can be observed for the
regions $\rho_s < 0.4$ and $\rho_s> 0.6$. The regions that are the
most sensitive to the top quark mass extraction are above $\rho_s >
0.7$. The latter finding is a consequence of the fact
that, the tail of the $\rho_s$ distribution is very sensitive to the
position of the $t\bar{t}j$ threshold, which in turn is sensitive to
$m_t$.   On the other hand, the crossing of various curves that
happens around $\rho_s \approx 0.55$ marks a point where the
normalised $\rho_s$ distribution is essentially insensitive to the top
quark mass. We can observe from Figure \ref{fig:1observable} that in
the most sensitive region deviations of {\it NWA} from the {\it Full}
case are below $15\%$. On the contrary, substantial differences,
$55\%-85\%$ up to even $100\%$, are visible for {\it NWA}${}_{Prod.}$
in that region.  These differences should have a considerable impact
on the extraction of $m_t$. The comparison of {\it NWA}
and {\it NWA}${}_{Prod}$ shows that in the region $\rho_s \approx 0.8$
(close to production threshold) more than $50\%$ of the events
originate from the radiative treatment of top quark
decays. Conversely, in the region around $\rho_s \approx 0.2$ this
treatment leads to $10\%$ negative corrections with respect to the
approximation where top quarks do not radiate hard jets in the decay
stage. The correct perturbative description of this observable,
therefore, requires hard jet emission in production and decay (and
their mixed contributions). Note here that in
\cite{Alioli:2013mxa, Fuster:2017rev} jet radiation by top quark
decays is not included. Additionally, in the most sensitive region,
sizeable NLO QCD corrections (shape differences), of the order of
$50\%$, are obtained for {\it Full} and {\it NWA} theoretical
predictions. In the case of {\it NWA}${}_{Prod.}$ they are around
$20\%$. The dominant source of the large ${\cal K}$
factor is final state radiation.  Nevertheless, in each case
differential ${\cal K}$ factors are indeed far from constant.  Thus,
LO calculations together with a suitably chosen global ${\cal
K}$-factor can not be applied to obtain results that well approximate
the full NLO QCD calculation. As a consequence, great caution has to
be taken for merging LO samples with parton shower programs to obtain
realistic hadronic events, directly comparable with the experimental
data. Instead, predictions with NLO QCD corrections included should be
used in $m_t$ studies where the shape of the $\rho_s$ observable is
important.
%
\begin{figure}[t!]
\begin{center}
\includegraphics[width=1.0\textwidth]{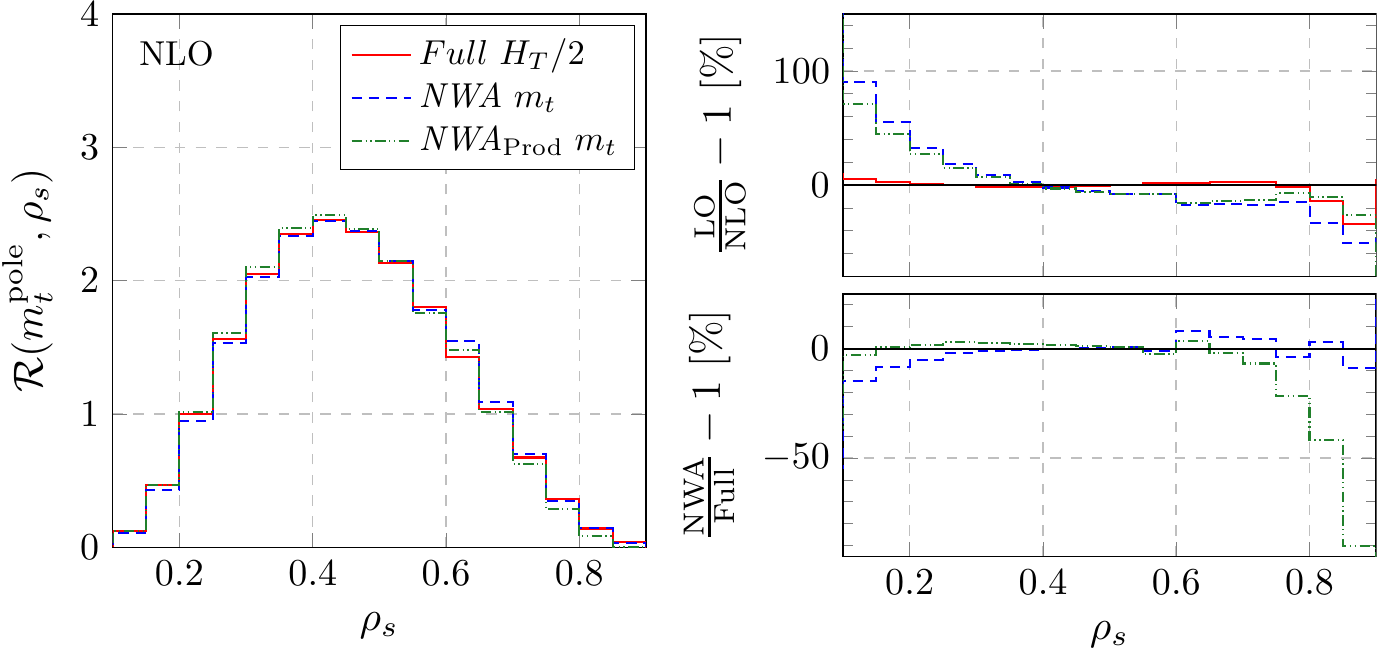}
\end{center}
\caption{\it Normalised $\rho_s$ differential distribution at NLO QCD
for the $pp \to e^+\nu_e \mu^- \bar{\nu}_\mu b\bar{b} j+ X$ production
process at the LHC with $\sqrt{s} =$ 13 TeV. Three different
theoretical descriptions with $m_t=$ 173.2 GeV are shown.  Also given
are the relative size of NLO QCD corrections and the combined relative
size of finite-top-width and finite-W-width effects for the
normalised $\rho_s$ observable. Renormalisation and factorisation
scales are set to the common value $\mu_R =\mu_F =\mu_0$ where $\mu_0
=m_t$ for both NWA cases and $\mu_0=H_T/2$ for the Full case.  The
CT14 PDF set is employed. }
\label{fig:1observable_ht}
\end{figure}
\begin{figure}[t!]
\begin{center}
\includegraphics[width=0.8\textwidth]{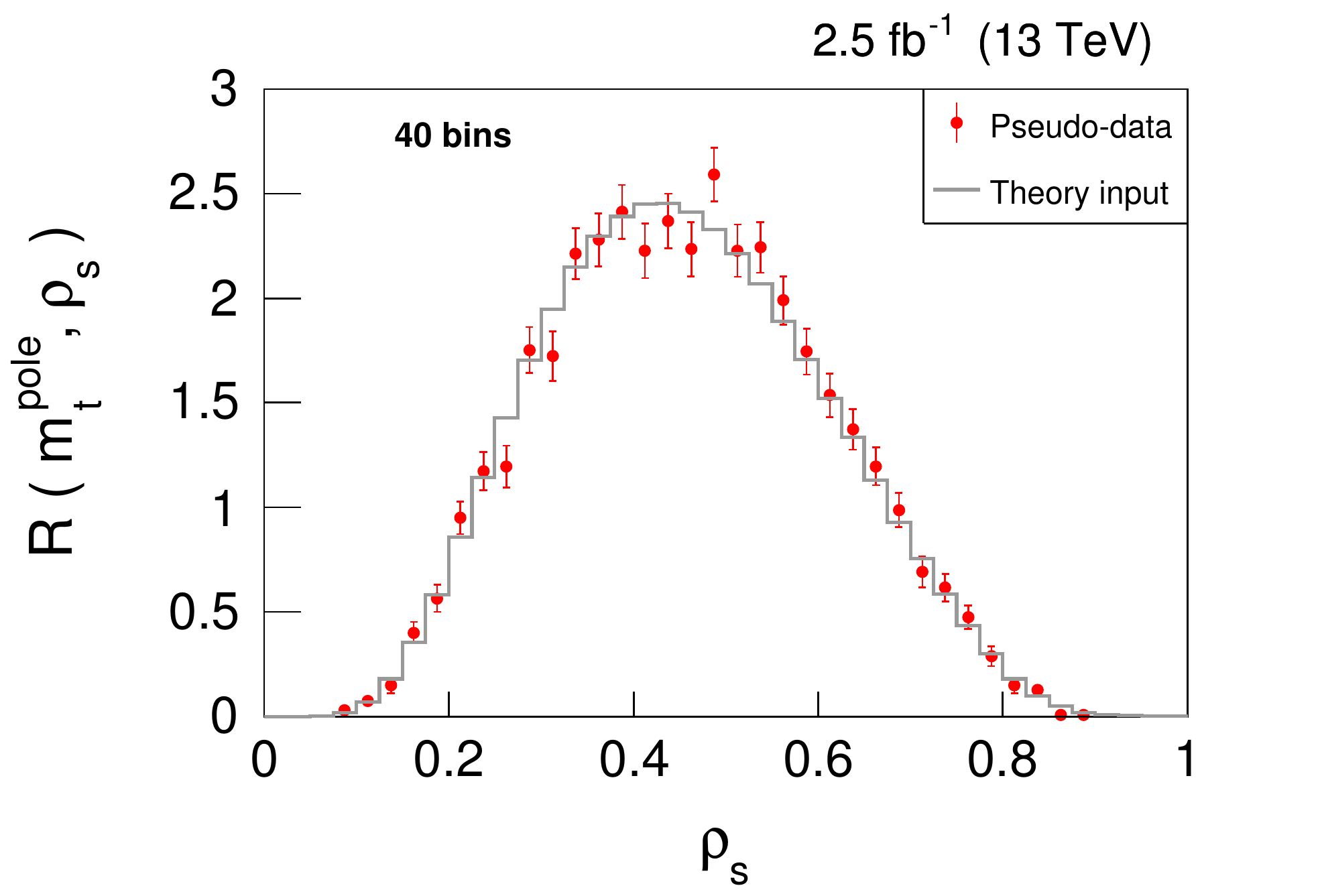}
\includegraphics[width=0.8\textwidth]{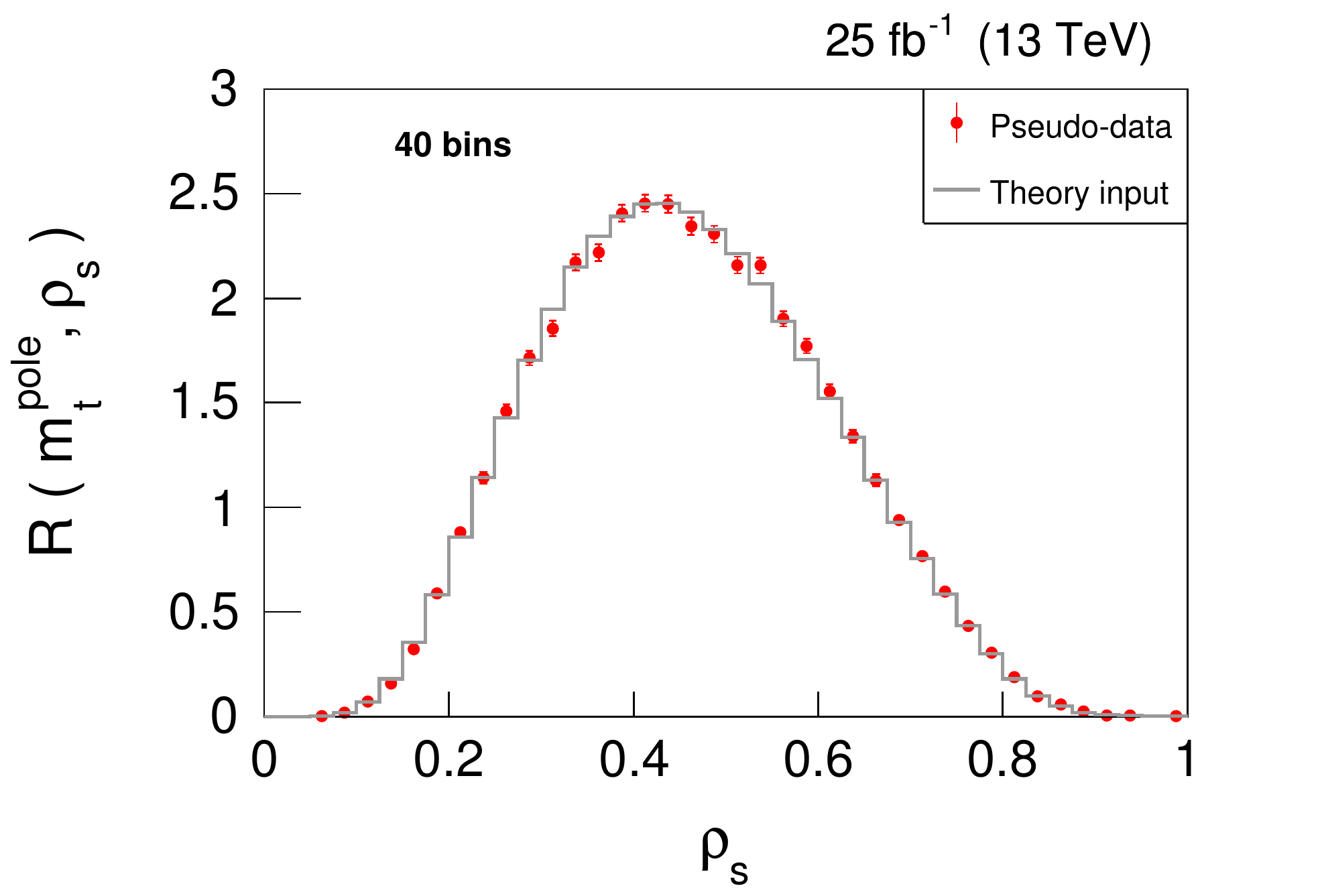}
\end{center}
\caption{\it One pseudo-data set (red points) as generated from the
NLO QCD calculations for the $pp \to e^+\nu_e \mu^- \bar{\nu}_\mu
b\bar{b} j+ X$ production process at the LHC with $\sqrt{s} =$ 13
TeV. The Full approach with $m_t=$ 173.2 GeV is used. The underlying
NLO template (grey histogram) is also shown. Two cases of the
integrated luminosity are presented, namely 2.5 fb${}^{-1}$ and 25
fb${}^{-1}$.}
\label{fig:1pseudodata}
\end{figure}
\begin{table}[ht!]
\begin{center}
\begin{tabular}{c||c|c|c|c||c|c}
& {\it Full} &{\it Full}  & {\it NWA} 
& {\it NWA}${}_{Prod.}$ 
& $\frac{NWA}{Full}-1 $
&$\frac{NWA}{Full}-1 $\\[0.2cm]
&$[m_t]$  &$[\frac{1}{2}H_T]$&$[m_t]$&$[m_t]$
& $[m_t/m_t]$ & $[m_t/\frac{1}{2}H_T]$\\[0.2cm]
\hline\hline
&&&&&&\\
$\sigma^{\rm NLO}_{t\bar{t}j}$ [fb]& 537.2 &538.6 & 527.1 & 656.1
& $-1.9 \%$& $-2.1 \%$\\
 \end{tabular}
\end{center}
\caption{\it Integrated NLO cross sections for the $pp \to e^+\nu_e
\mu^- \bar{\nu}_\mu b\bar{b} j+ X$ production process at the LHC with
$\sqrt{s} =$ 13 TeV. Results are evaluated using $\mu_R=\mu_F=\mu_0$,
where $\mu_0=m_t$ or $\mu_0=H_T/2$. The CT14 PDF set and
$m_t =$ 173.2 GeV are used. In the last two columns the combined relative
size of off-shell effects of $t$ and $W$ is also given.}
\label{tab:NLO_CS}
\end{table}
\begin{table}[t!]
\begin{center}
\begin{tabular}{ccc}
 \hline\hline 
5 equal size bins  & ATLAS binning   &
CMS binning  \\
\hline \hline
$0.00-0.20$  &$0.000 - 0.250$ & $0.00-0.20$ \\[0.1cm]
$0.20-0.40$ &$0.250 - 0.325$ & $0.20-0.30$\\[0.1cm]
$0.40-0.60$ &$0.325 - 0.425$ & $0.30-0.45$\\[0.1cm]
$0.60-0.80$ &$0.425 - 0.525$ & $0.45 -0.60$ \\[0.1cm]
$0.80-1.00$ &$0.525 - 0.675$ & $0.60-1.00$ \\[0.1cm]
$-$ &$0.675 - 1.000$ & $-$ \\
 \hline\hline 
\end{tabular}
\end{center}
\caption{\it Various binnings used in the $m_t$ extraction from the
normalised $\rho_s$ distribution. The three cases are shown: 5 equal
size intervals as well as ATLAS \cite{Aad:2015waa} and CMS
\cite{CMS:2016khu} intervals. The latter two have been optimised for
the $\ell+$jets and di-lepton channels at $\sqrt{s}=$ 7 TeV and
$\sqrt{s}=$ 8 TeV respectively.}
\label{binning}
\end{table}
\begin{figure}[t!]
\begin{center}
\includegraphics[width=0.55\textwidth]
{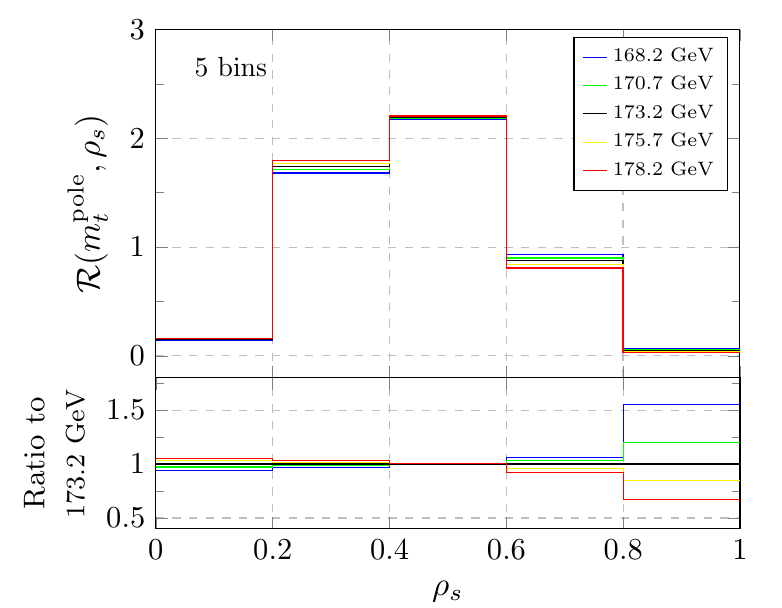}
\includegraphics[width=0.55\textwidth]
{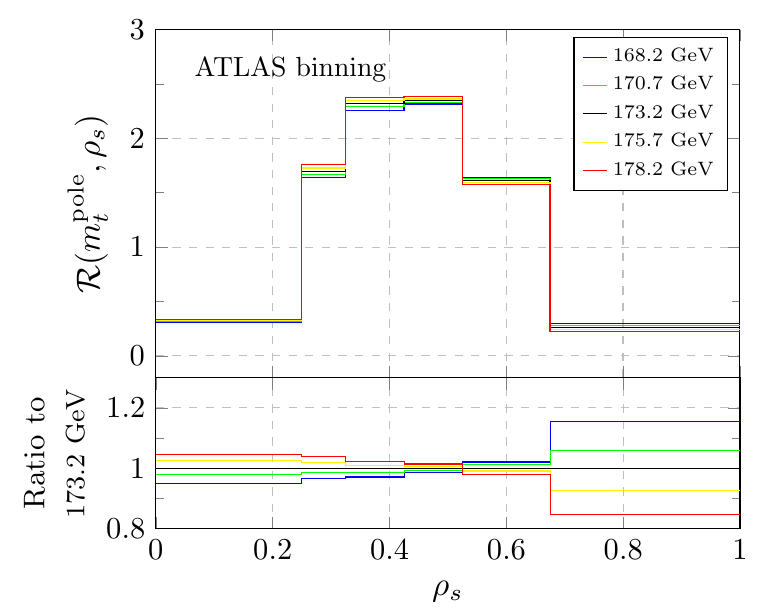}
\includegraphics[width=0.55\textwidth]
{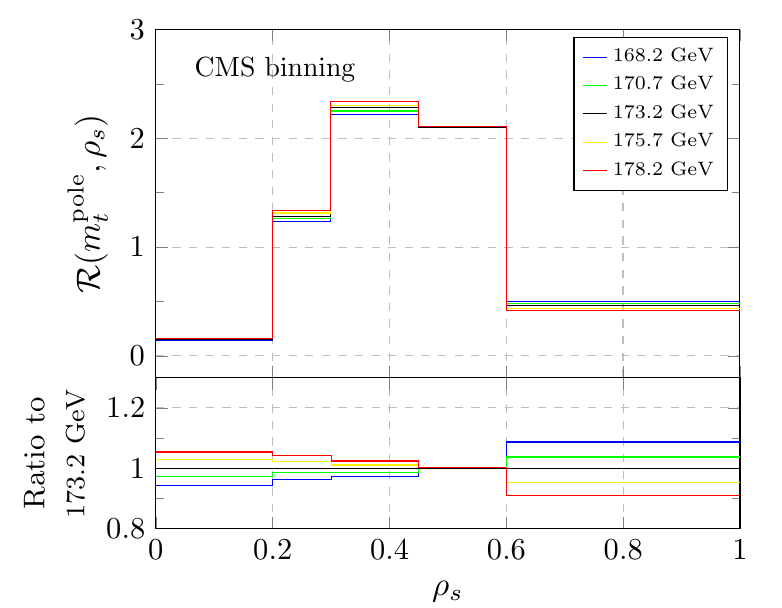}
\end{center}
\caption{\it 
 Normalised $\rho_s$ distribution at NLO QCD for the $pp \to e^+\nu_e
\mu^- \bar{\nu}_\mu b\bar{b} j+ X$ production process at the LHC with
$\sqrt{s} =$ 13 TeV. The Full case for five different top quark mass
values with various bin sizes is presented. We also plot
the ratio to the result with the default value of $m_t$, i.e. $m_t =$
173.2 GeV. Renormalisation and factorisation scales are set to the
common value $\mu_R =\mu_F =\mu_0$ where $\mu_0 =H_T/2$ and the CT14
PDF set is employed. }
\label{5-ATLAS-CMS}
\end{figure}
\begin{figure}[t!]
\begin{center}
\includegraphics[width=0.49\textwidth]{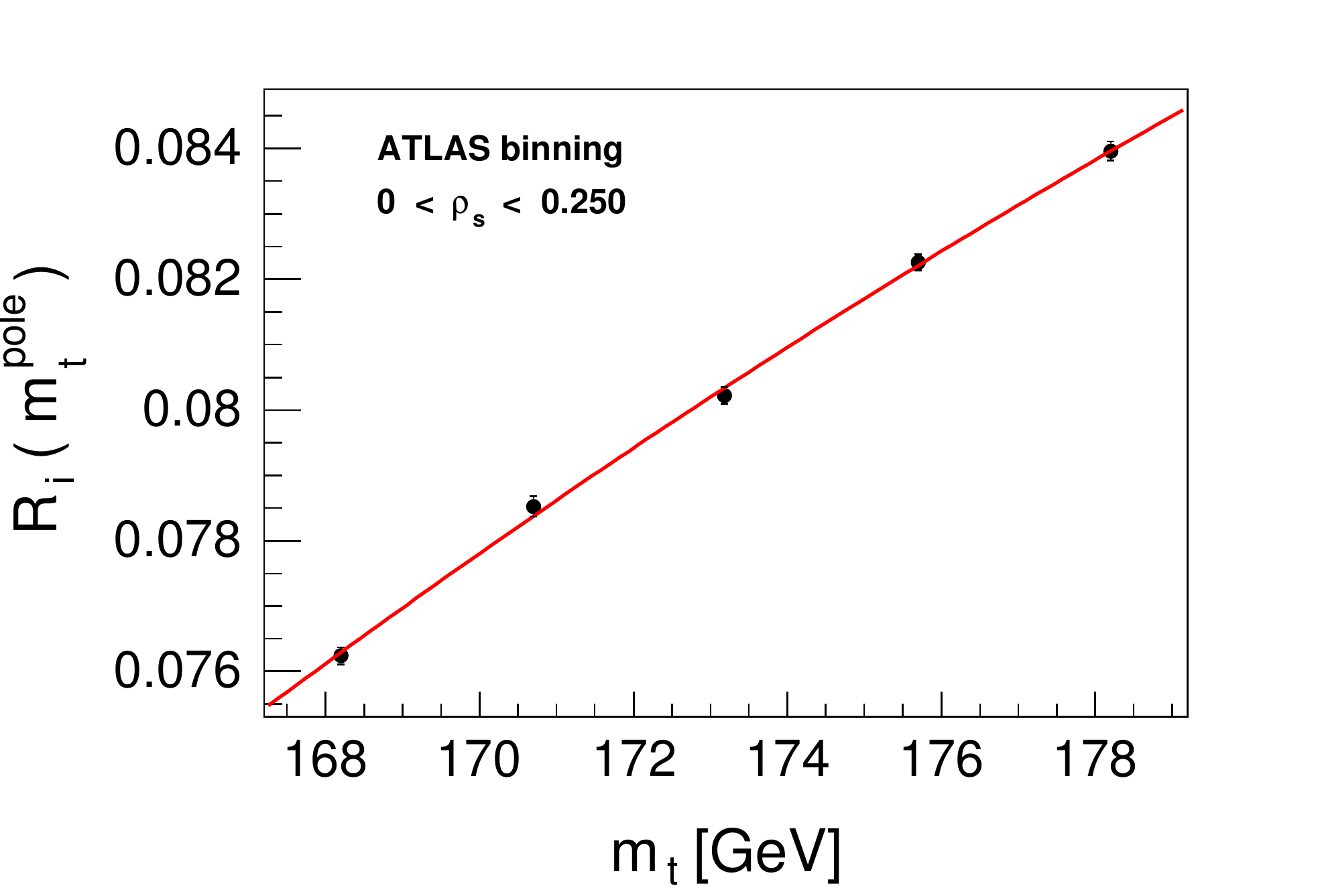}
\includegraphics[width=0.49\textwidth]{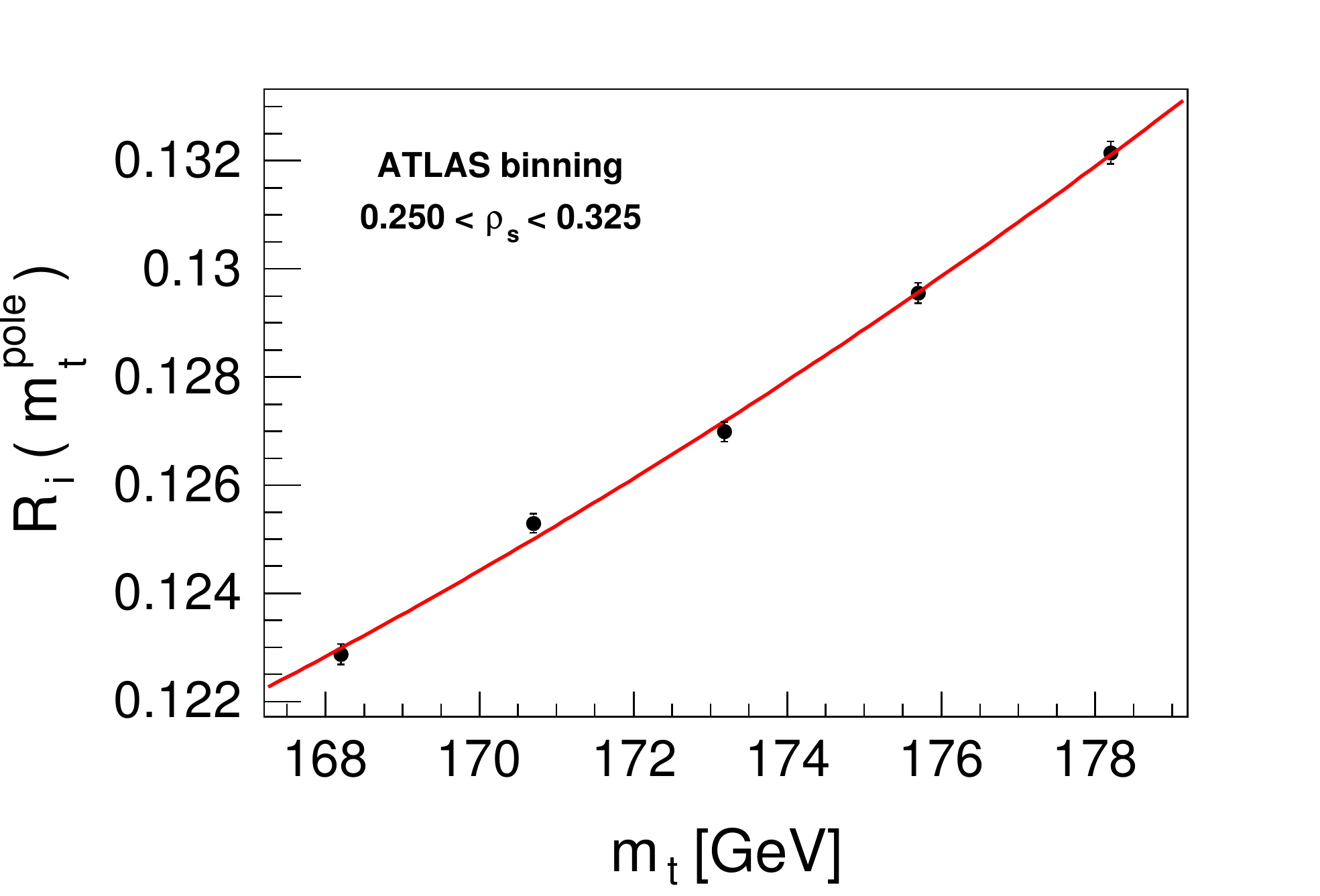}
\includegraphics[width=0.49\textwidth]{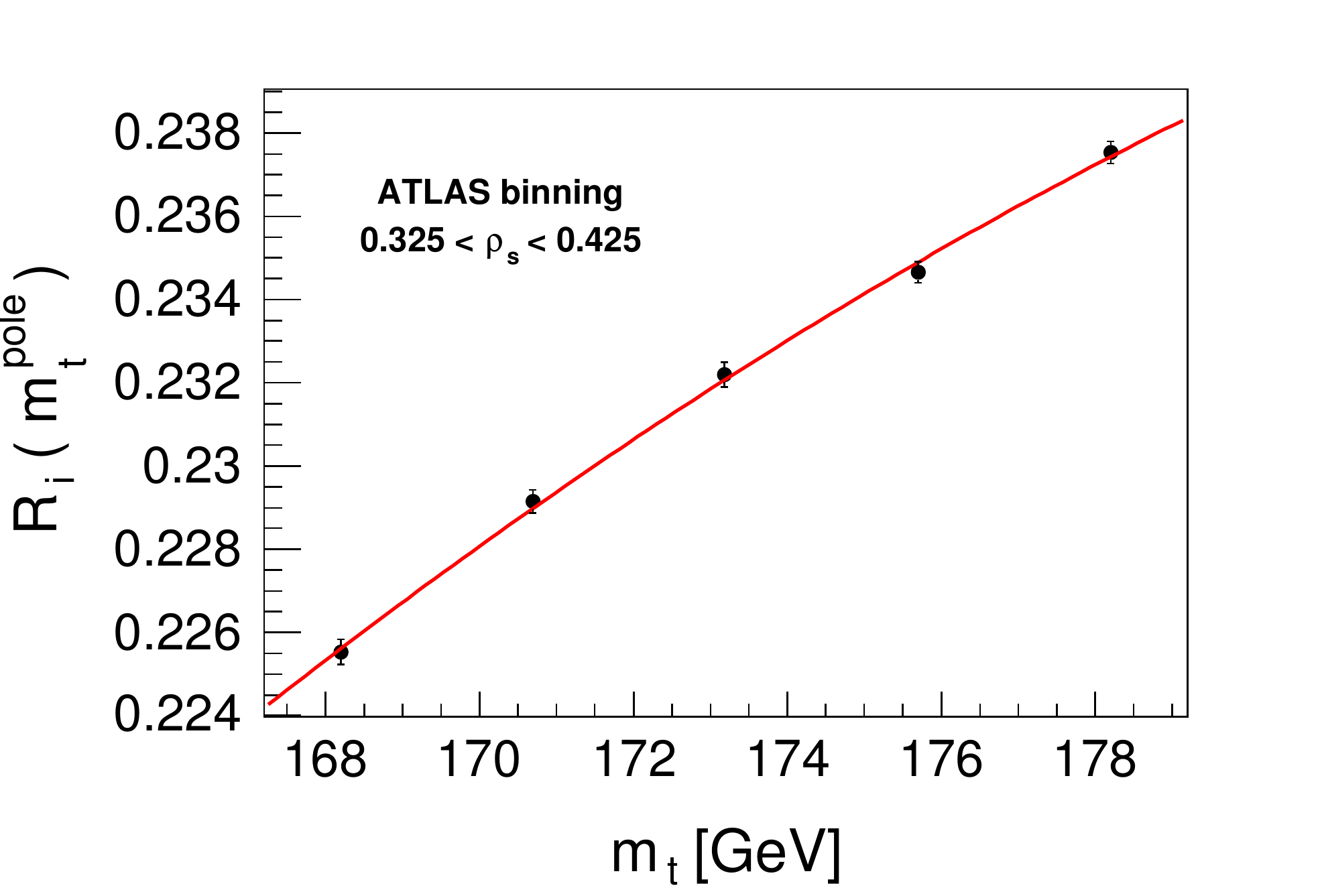}
\includegraphics[width=0.49\textwidth]{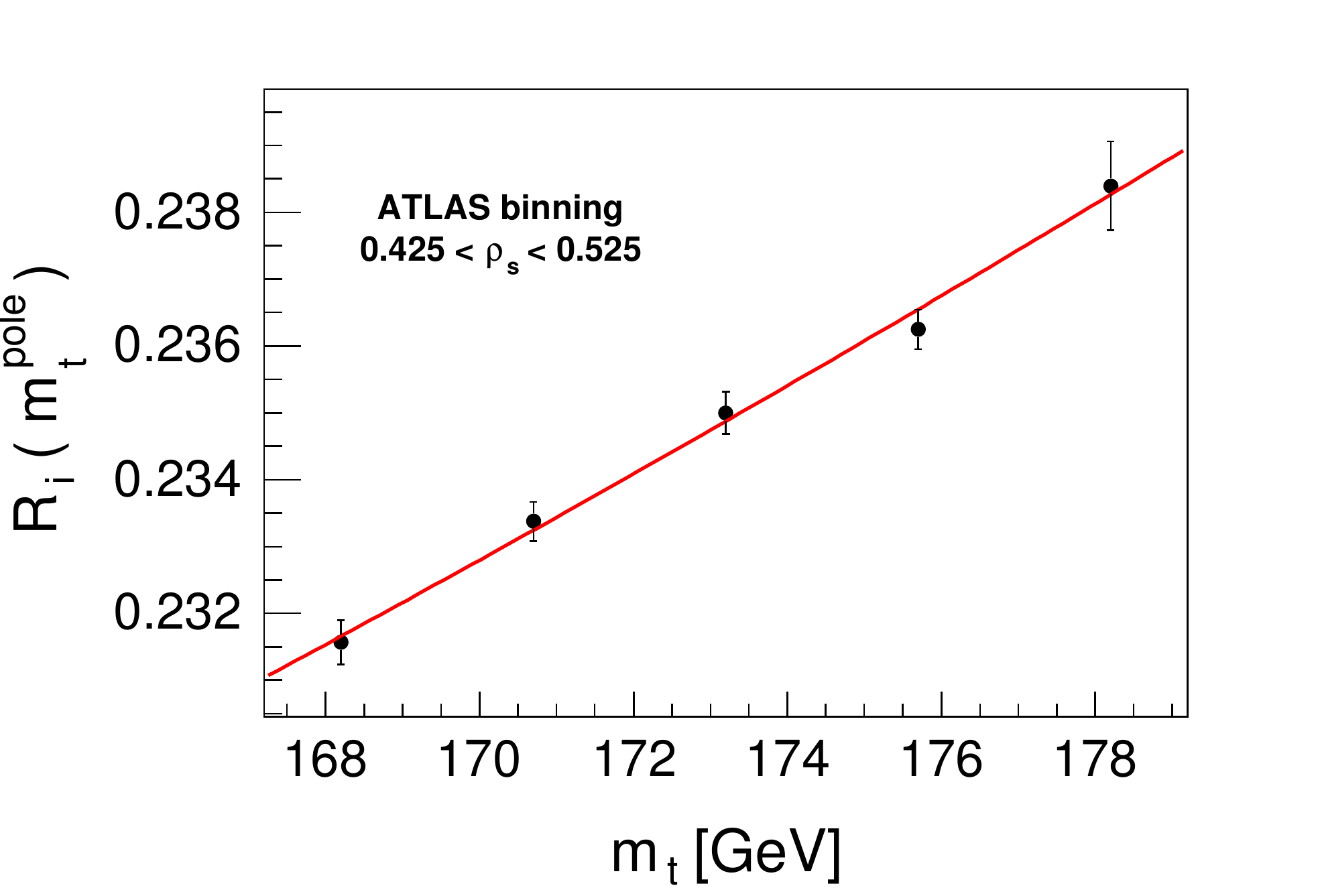}
\includegraphics[width=0.49\textwidth]{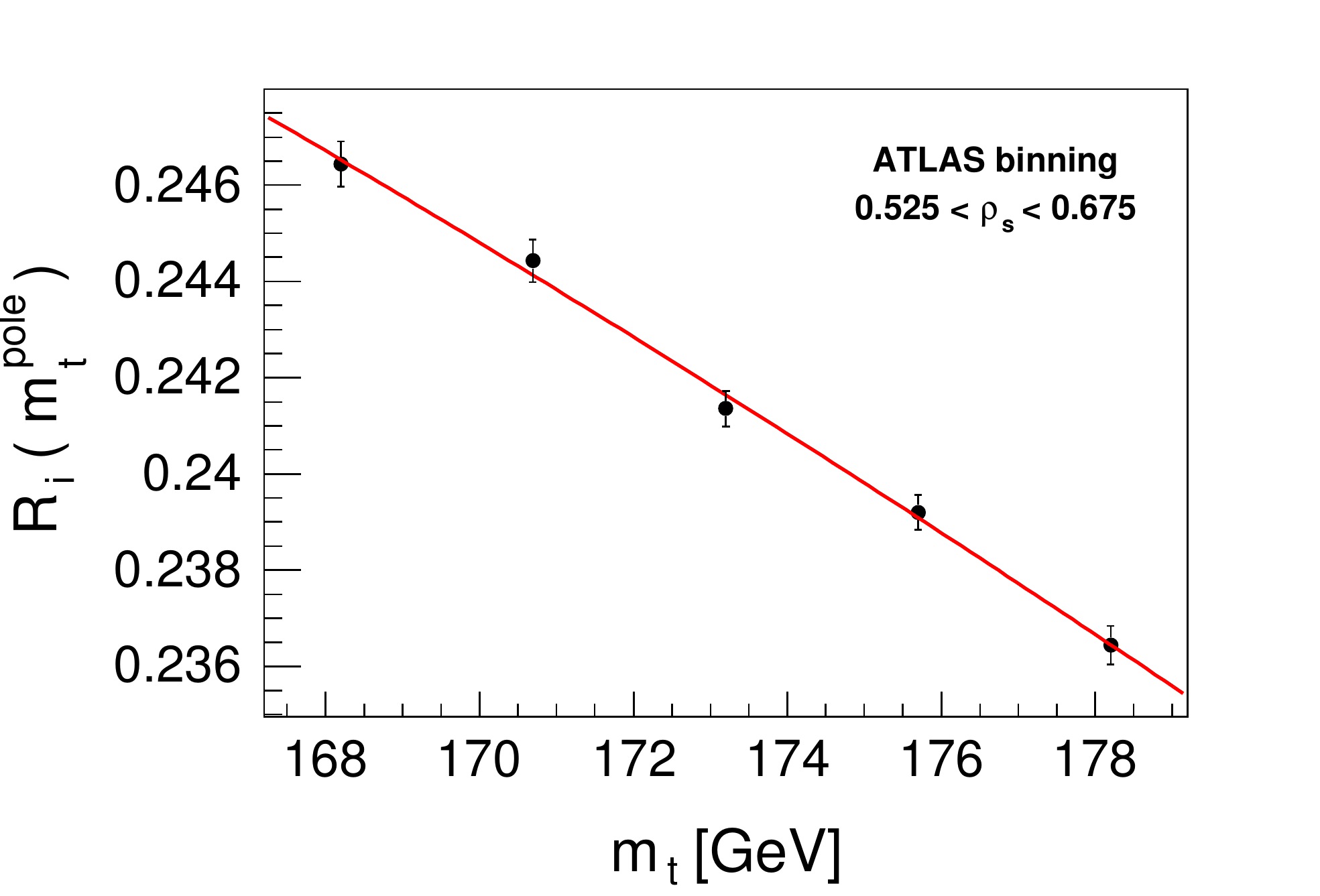}
\includegraphics[width=0.49\textwidth]{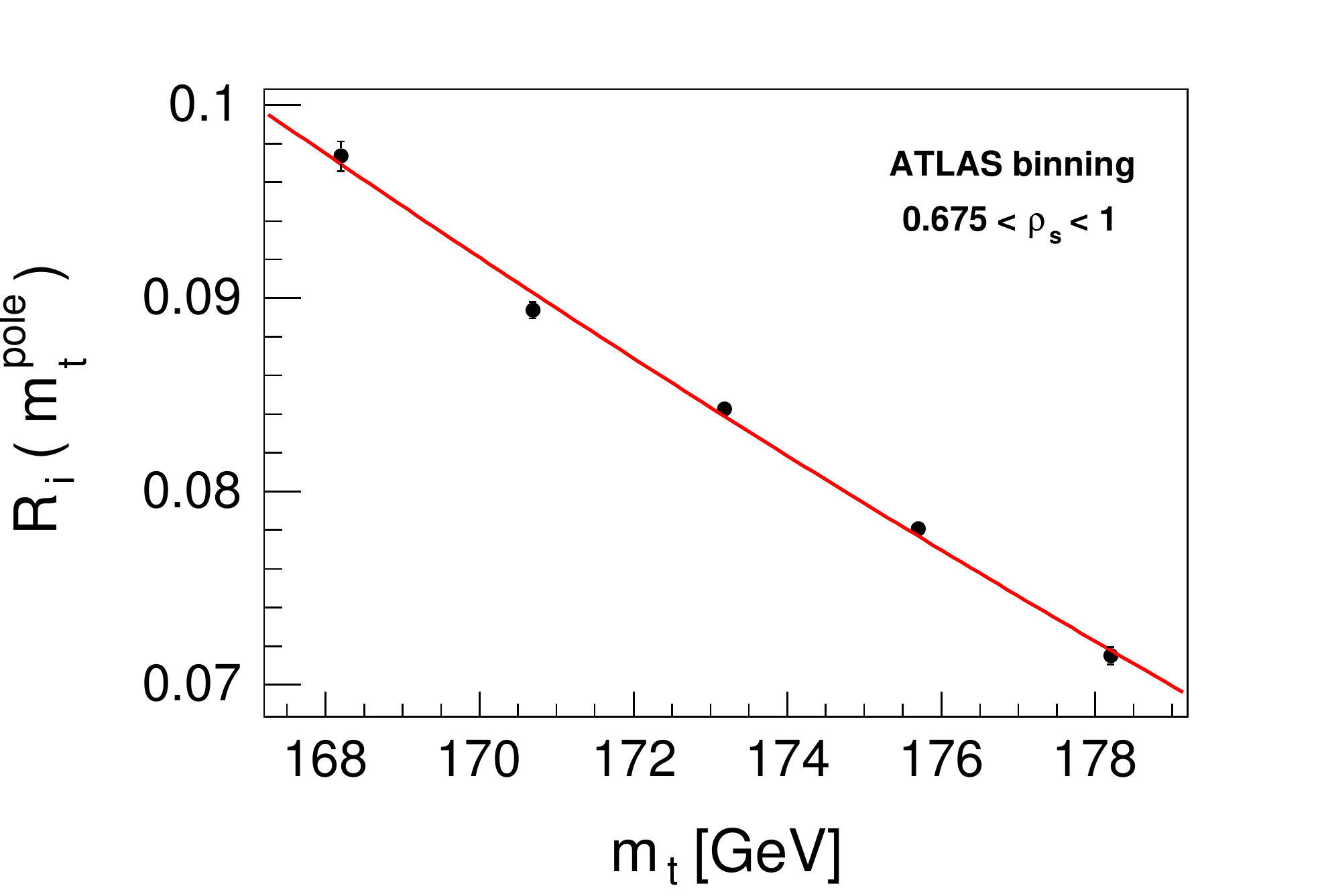}
\caption{\it 
Bin-by-bin fit of template distributions for the
normalised $\rho_s$ observable as given by full theory at NLO in QCD
for the $pp \to e^+\nu_e \mu^- \bar{\nu}_\mu b\bar{b} j+ X$ production
process at the LHC with $\sqrt{s} =$ 13 TeV. The CT14 PDF set and
$\mu_R=\mu_F=\mu_0 = H_T/2$ are used. The ATLAS binning is
assumed. The error bars denote statistical uncertainties.}
\label{ATLASbinning}
\end{center}
\end{figure}
\begin{figure}[t!]
\begin{center}
\includegraphics[width=0.49\textwidth]{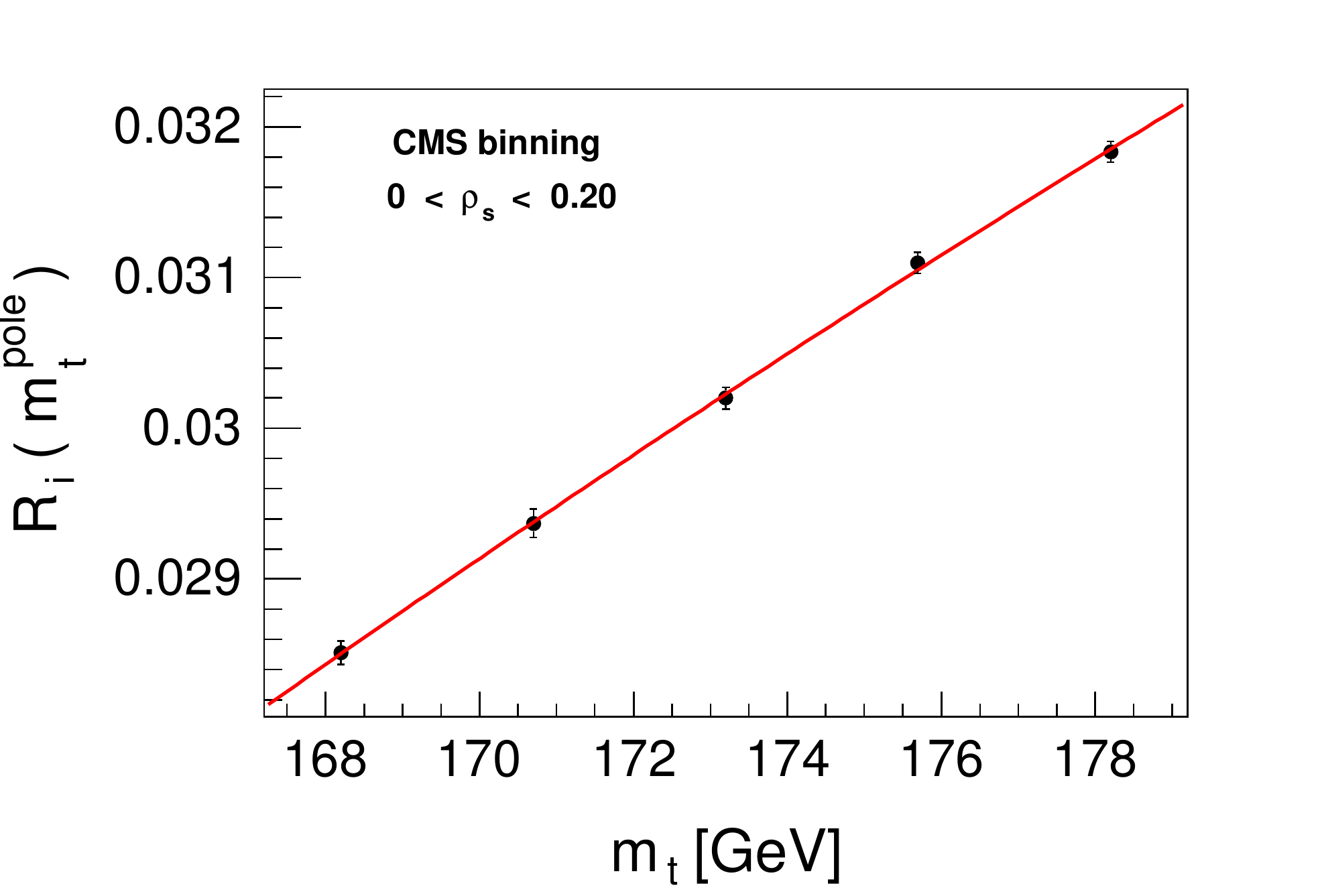}
\includegraphics[width=0.49\textwidth]{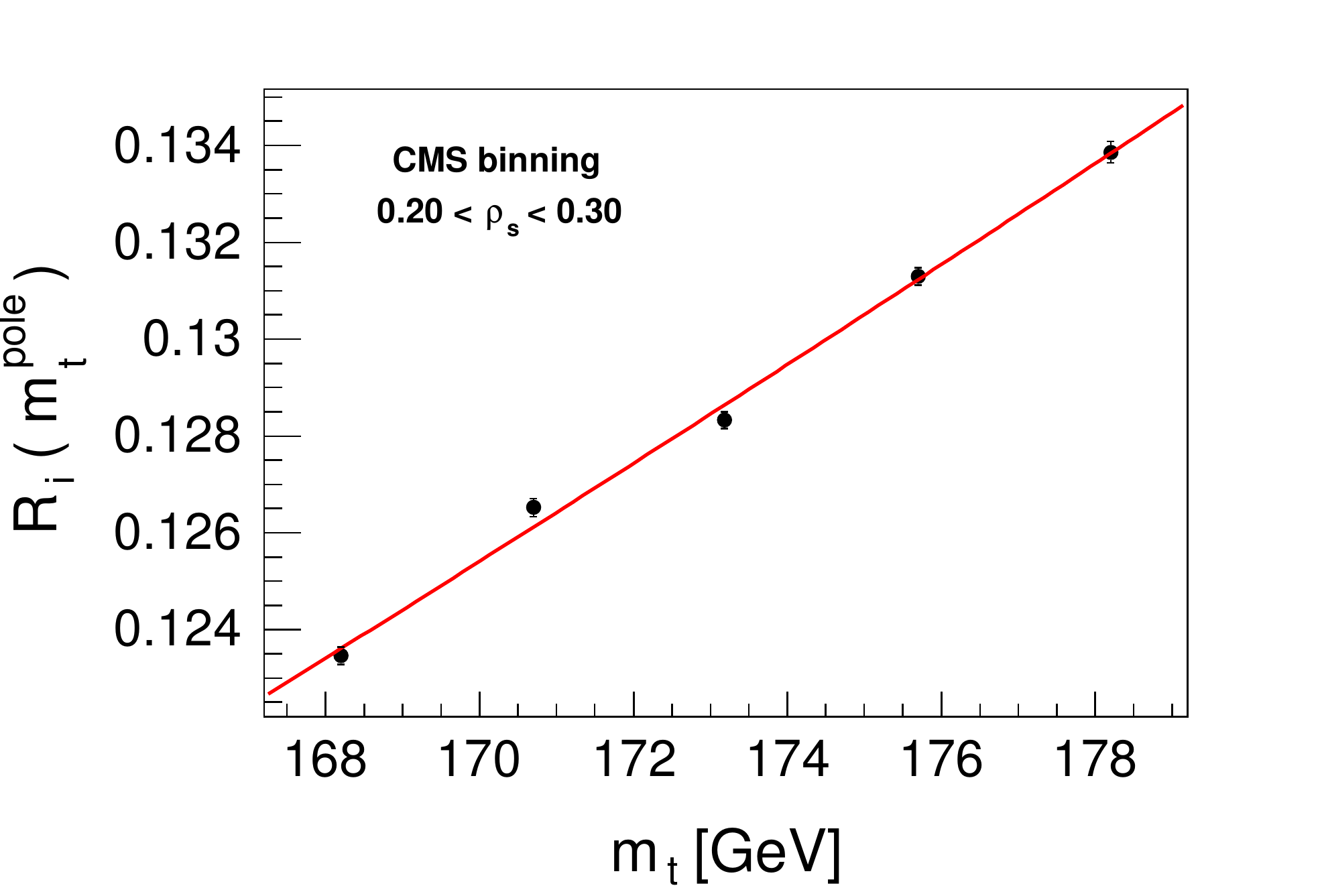}
\includegraphics[width=0.49\textwidth]{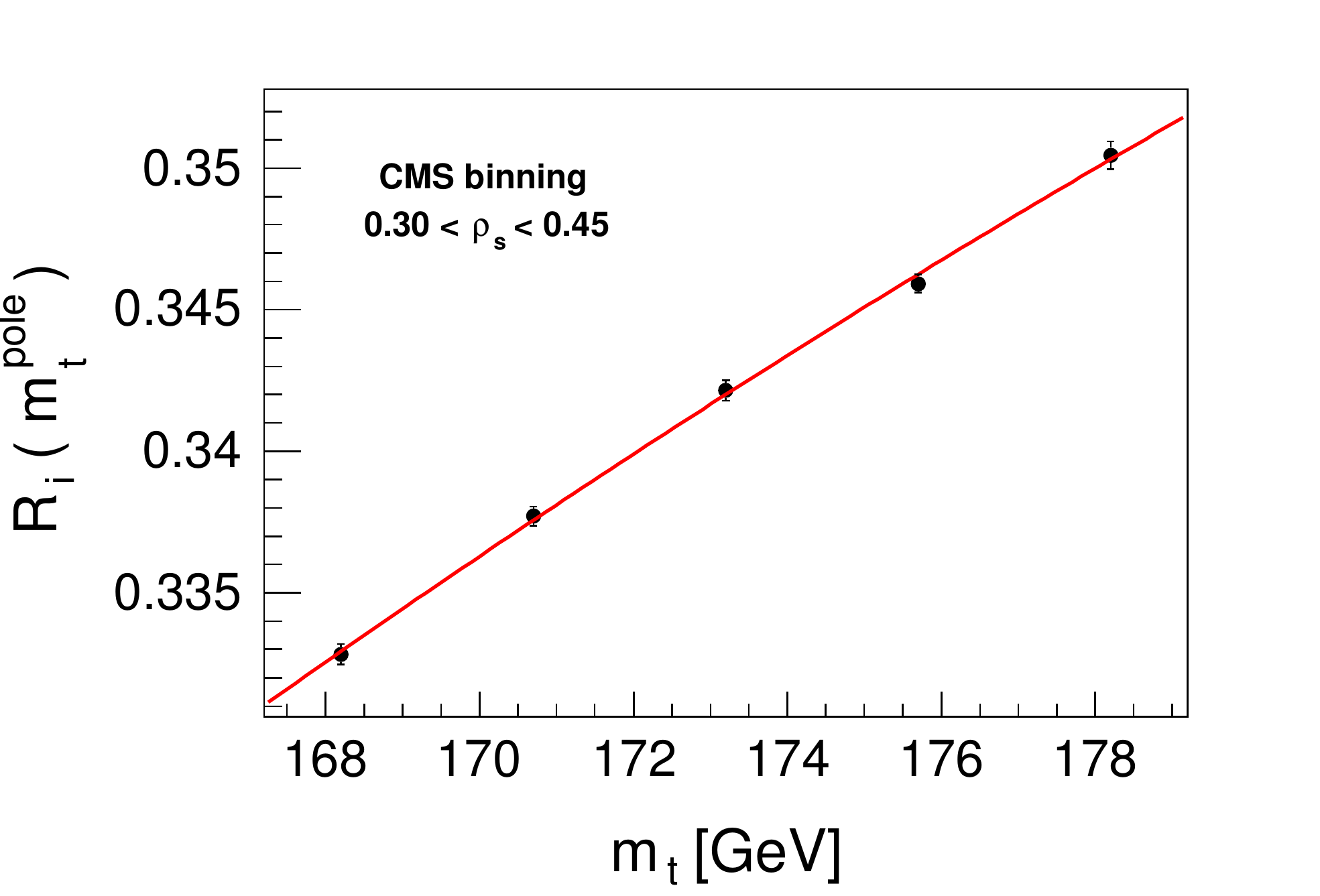}
\includegraphics[width=0.49\textwidth]{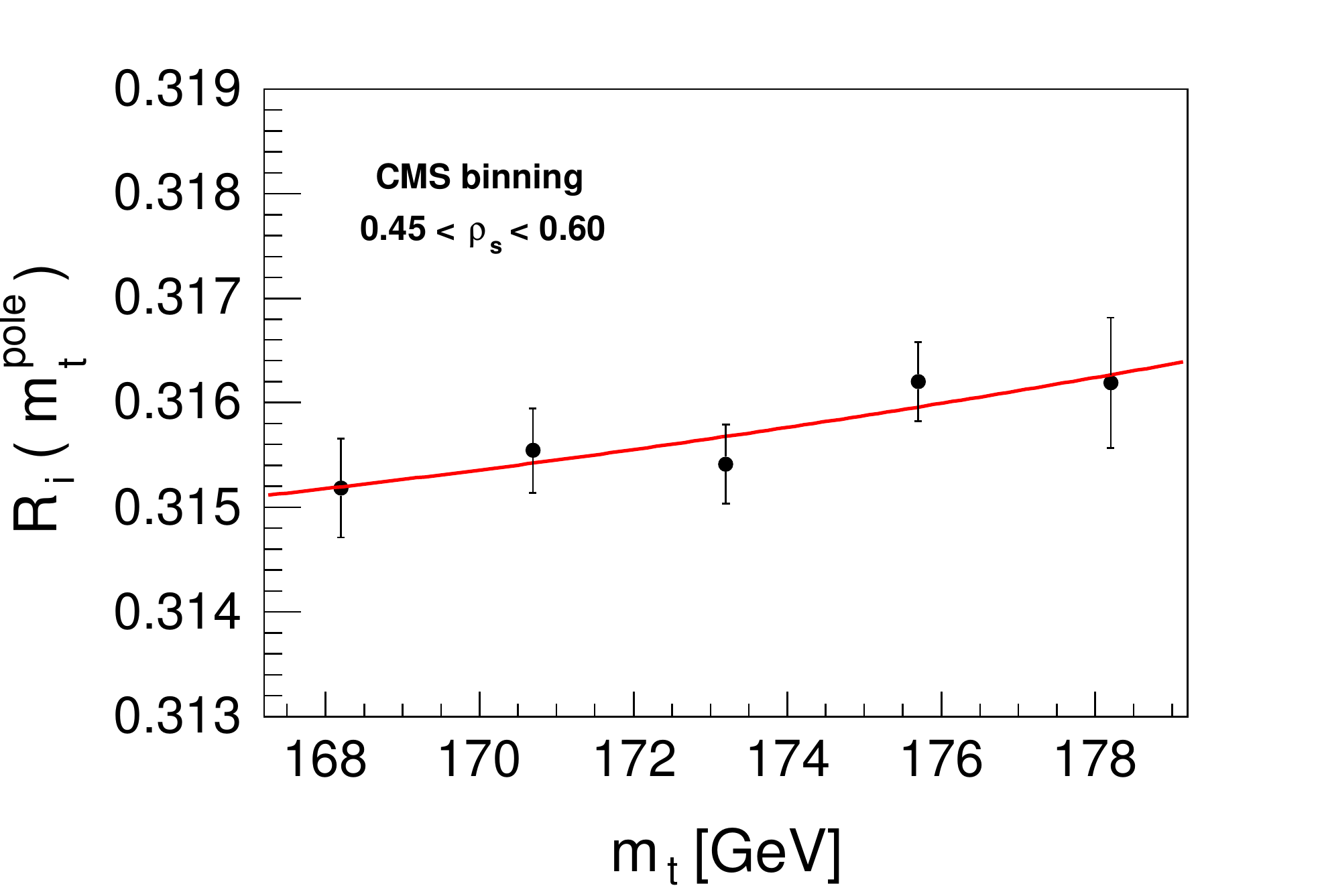}
\includegraphics[width=0.49\textwidth]{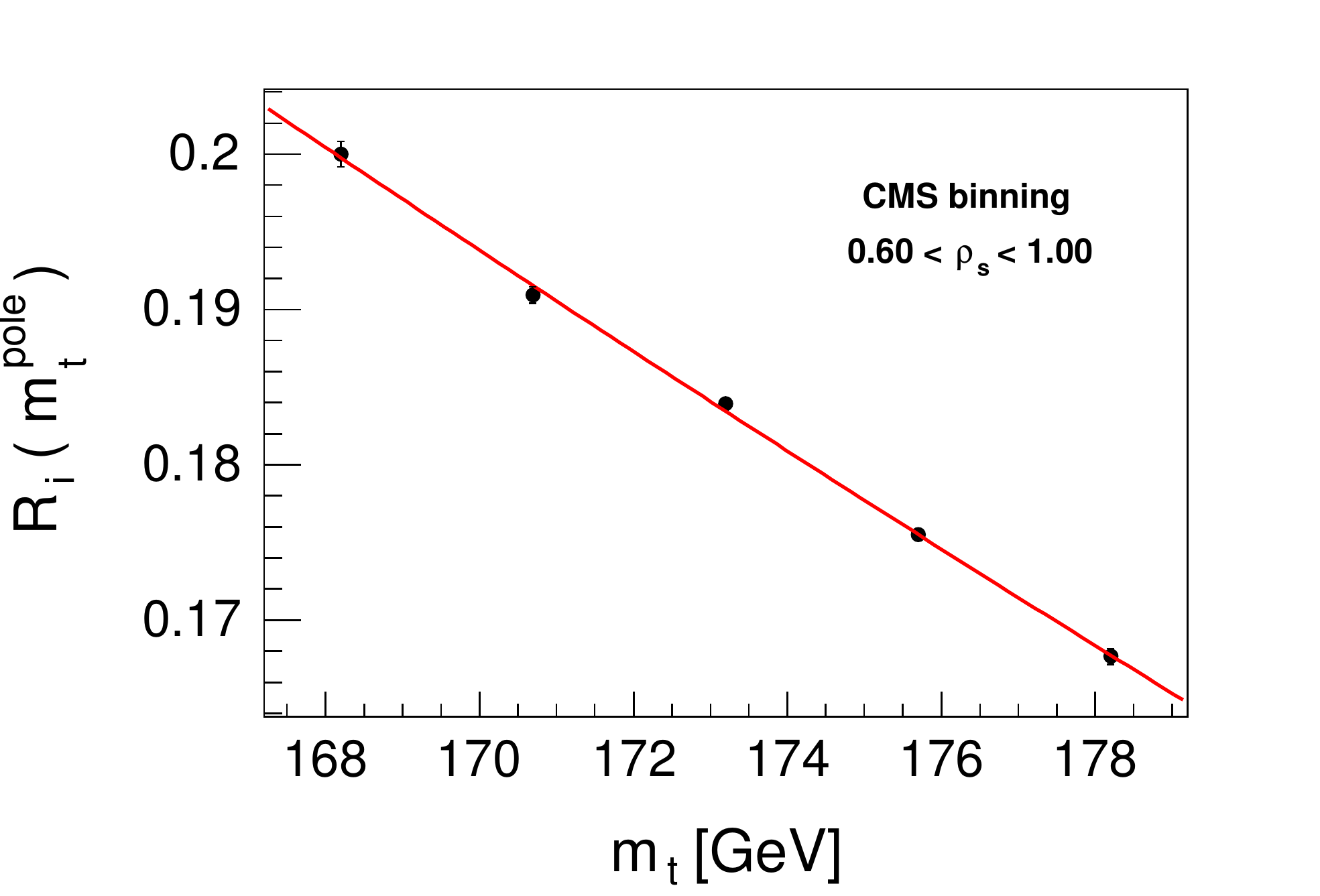}
\caption{\it 
Bin-by-bin fit of template distributions for the
normalised $\rho_s$ observable as given by full theory at NLO in QCD
for the $pp \to e^+\nu_e \mu^- \bar{\nu}_\mu b\bar{b} j+ X$ production
process at the LHC with $\sqrt{s} =$ 13 TeV. The CT14 PDF set and
$\mu_R=\mu_F=\mu_0 = H_T/2$ are used. The CMS binning is
assumed. The error bars denote statistical uncertainties.}
\label{CMSbinning}
\end{center}
\end{figure}
\begin{figure}[t!]
\begin{center}
\includegraphics[width=0.49\textwidth]
{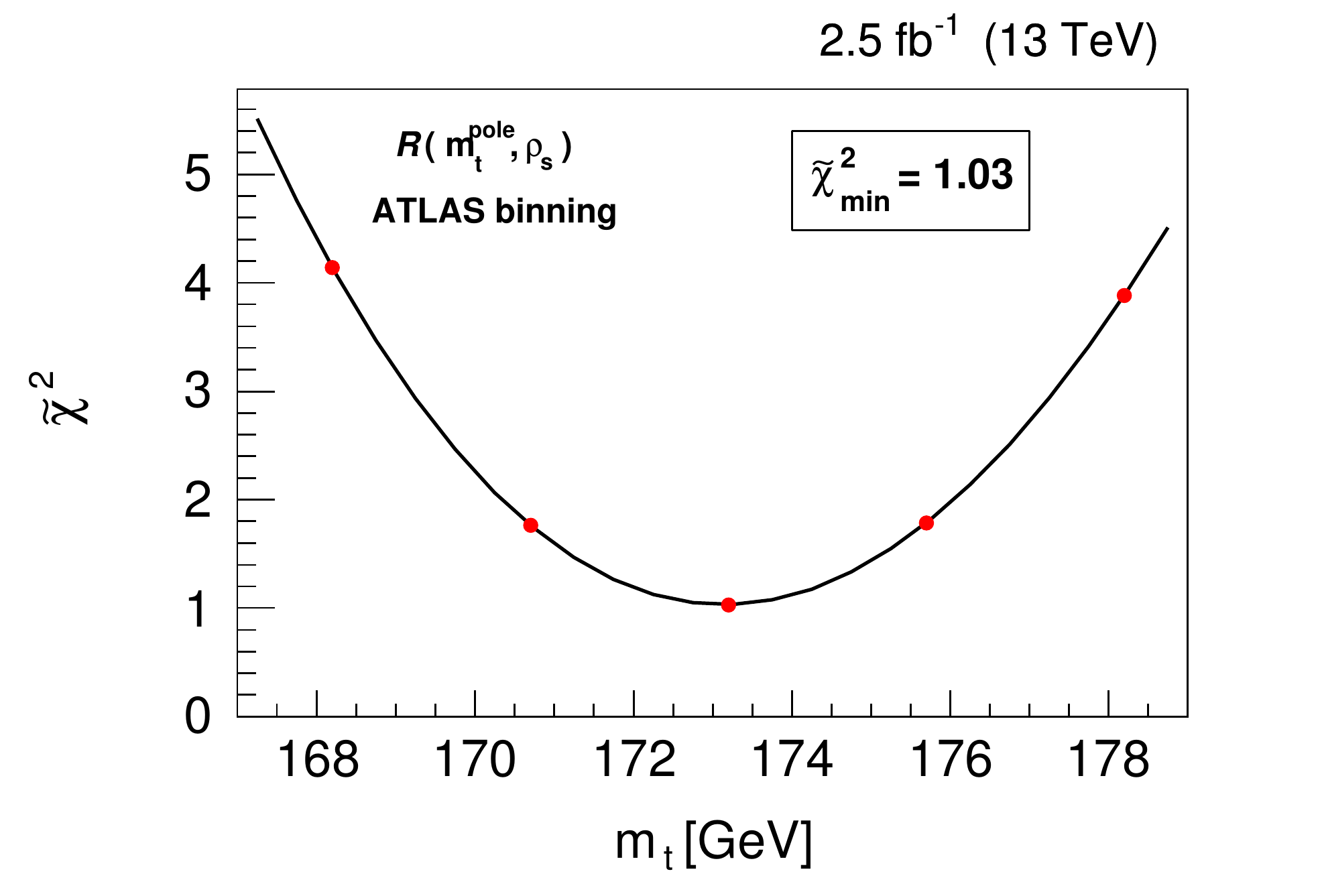}
\includegraphics[width=0.49\textwidth]
{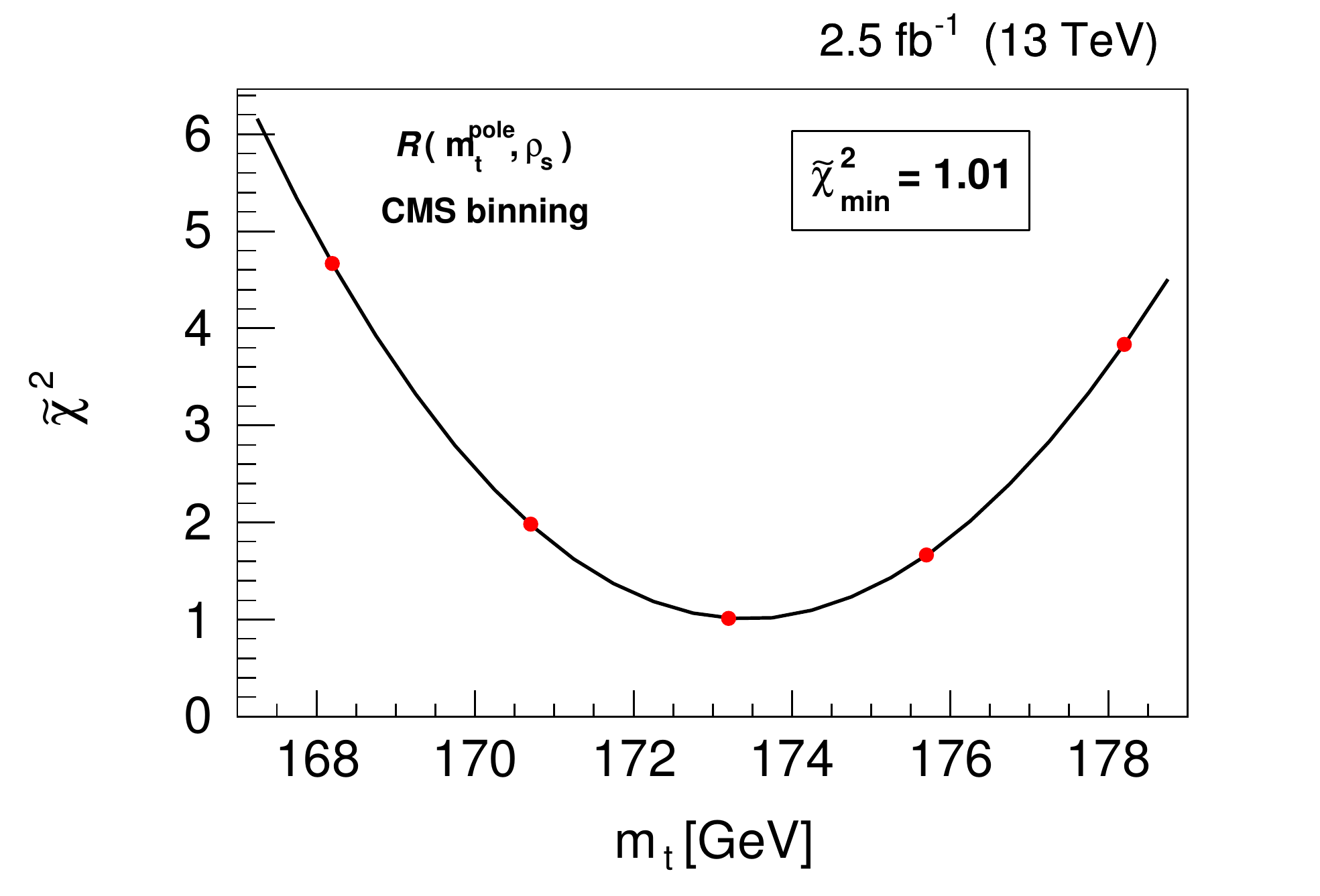}
\includegraphics[width=0.49\textwidth]
{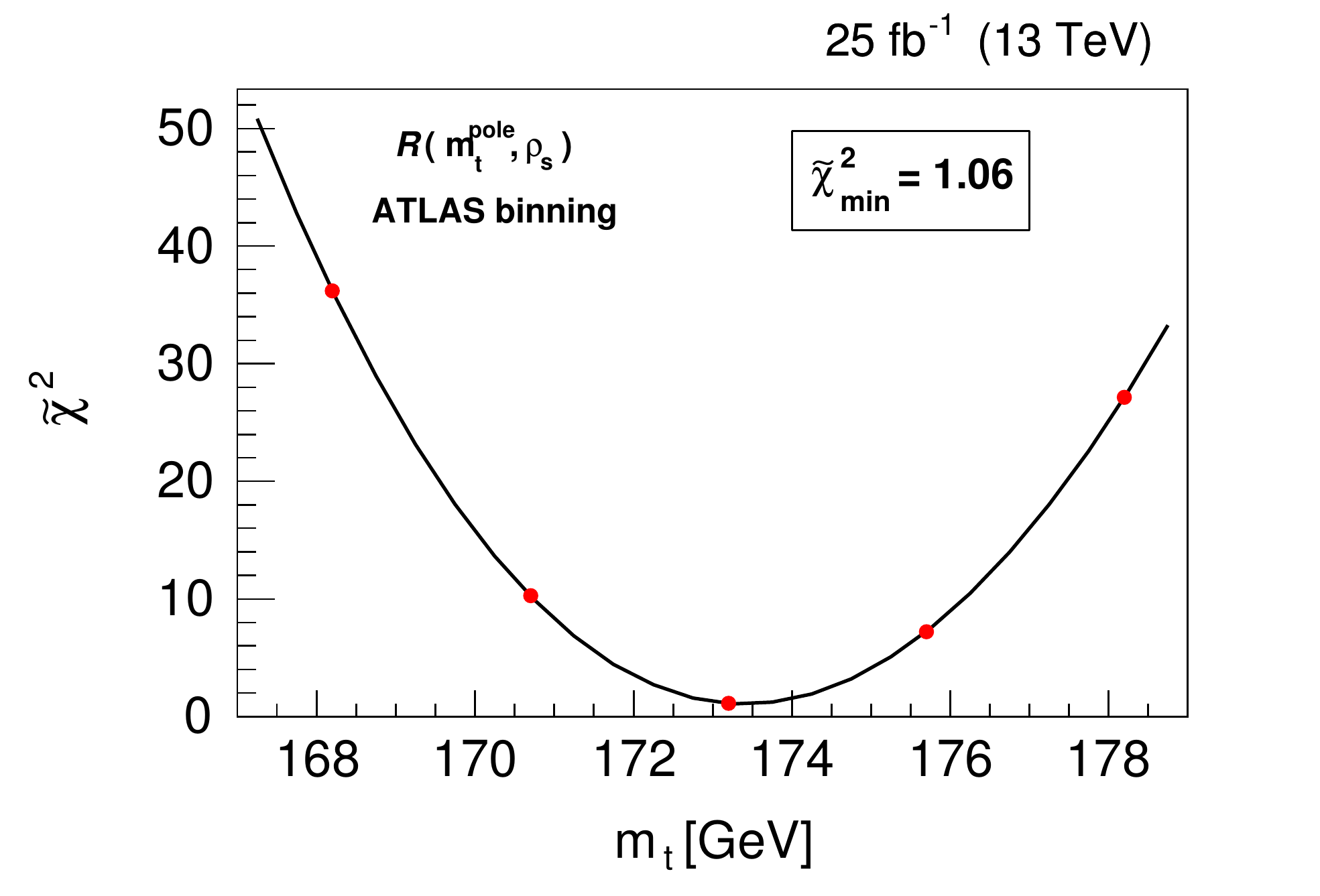}
\includegraphics[width=0.49\textwidth]
{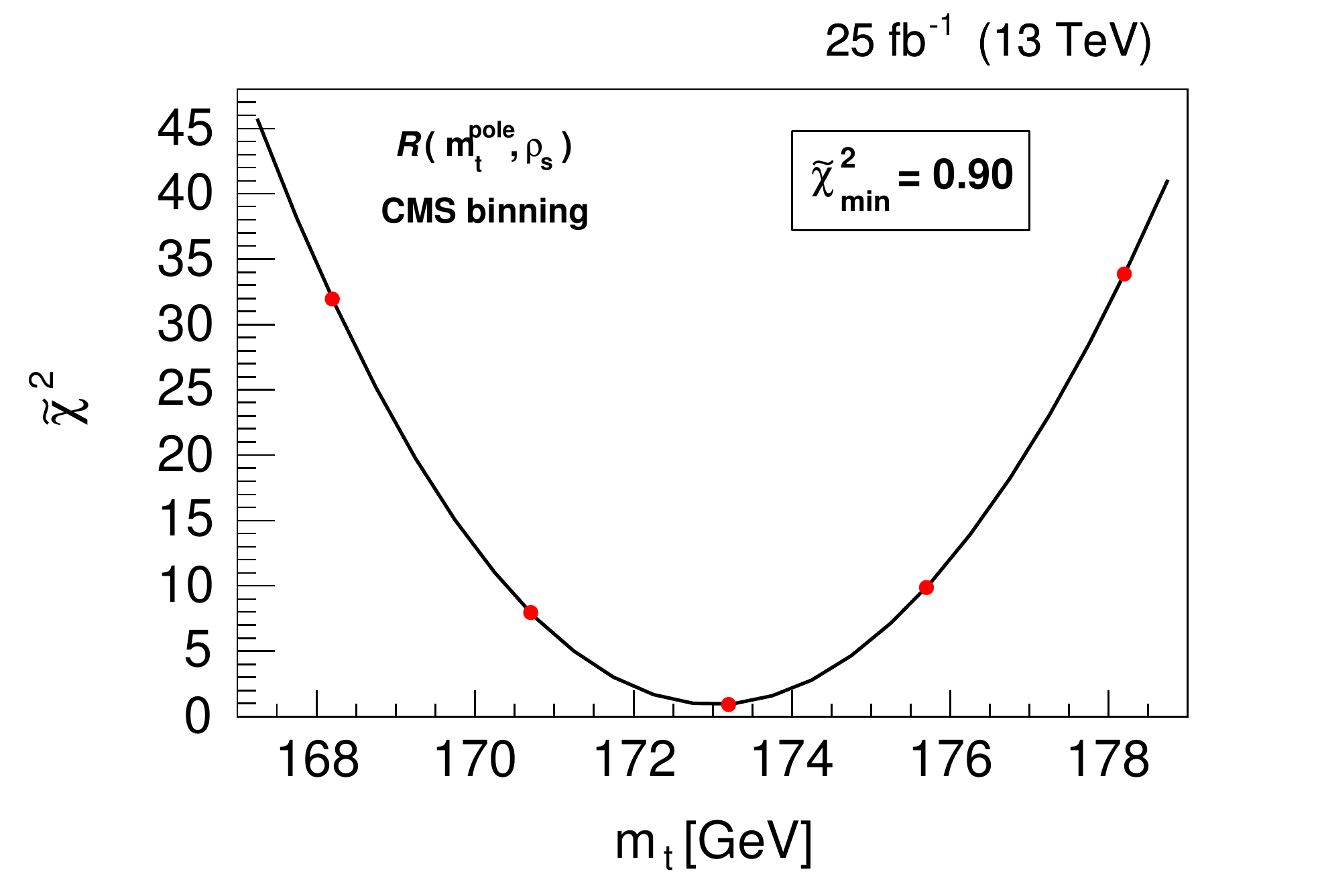}
\caption{\it 
Representative $\tilde{\chi}^2 \: ( \: \equiv
\chi^2/d.o.f.)$ for the normalised $\rho_s$ observable for the $pp \to
e^+\nu_e \mu^- \bar{\nu}_\mu b\bar{b} j+ X$ production process at the
LHC with $\sqrt{s} =$ 13 TeV. The binning of ATLAS and CMS is used.
Luminosity of $\mathcal{L} = 2.5 \mbox{ fb}^{-1}$ and $\mathcal{L} =
25 \mbox{ fb}^{-1}$ is assumed and $f_i(m_t)$ are obtained from the
full theory at NLO in QCD with $\mu_R=\mu_F=\mu_0 = H_T/2$ and with
the CT14 PDF set.}
\label{chi2}
\end{center}
\end{figure}
\begin{figure}[t!]
\begin{center}
\includegraphics[width=0.47\textwidth]{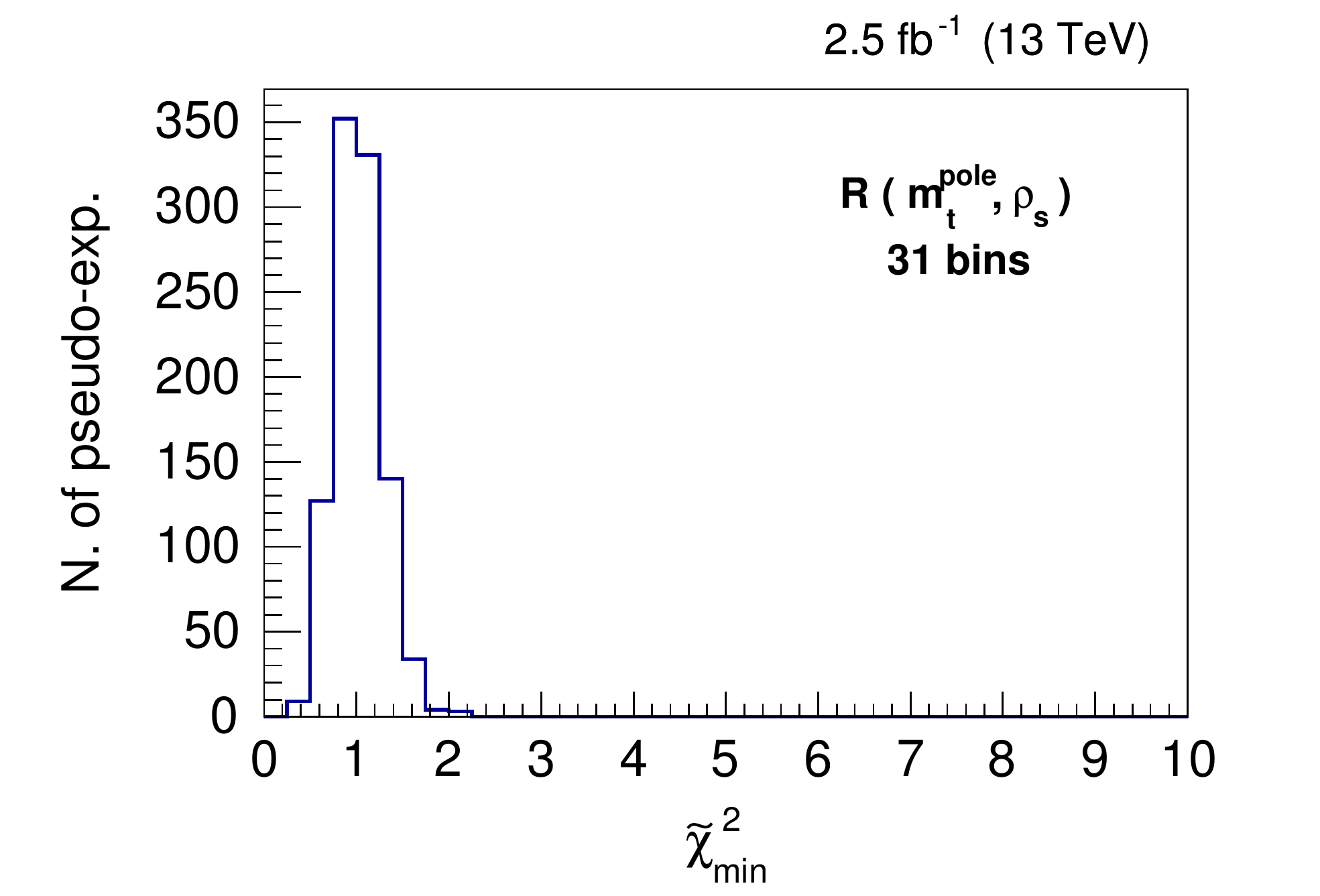}
\includegraphics[width=0.47\textwidth]{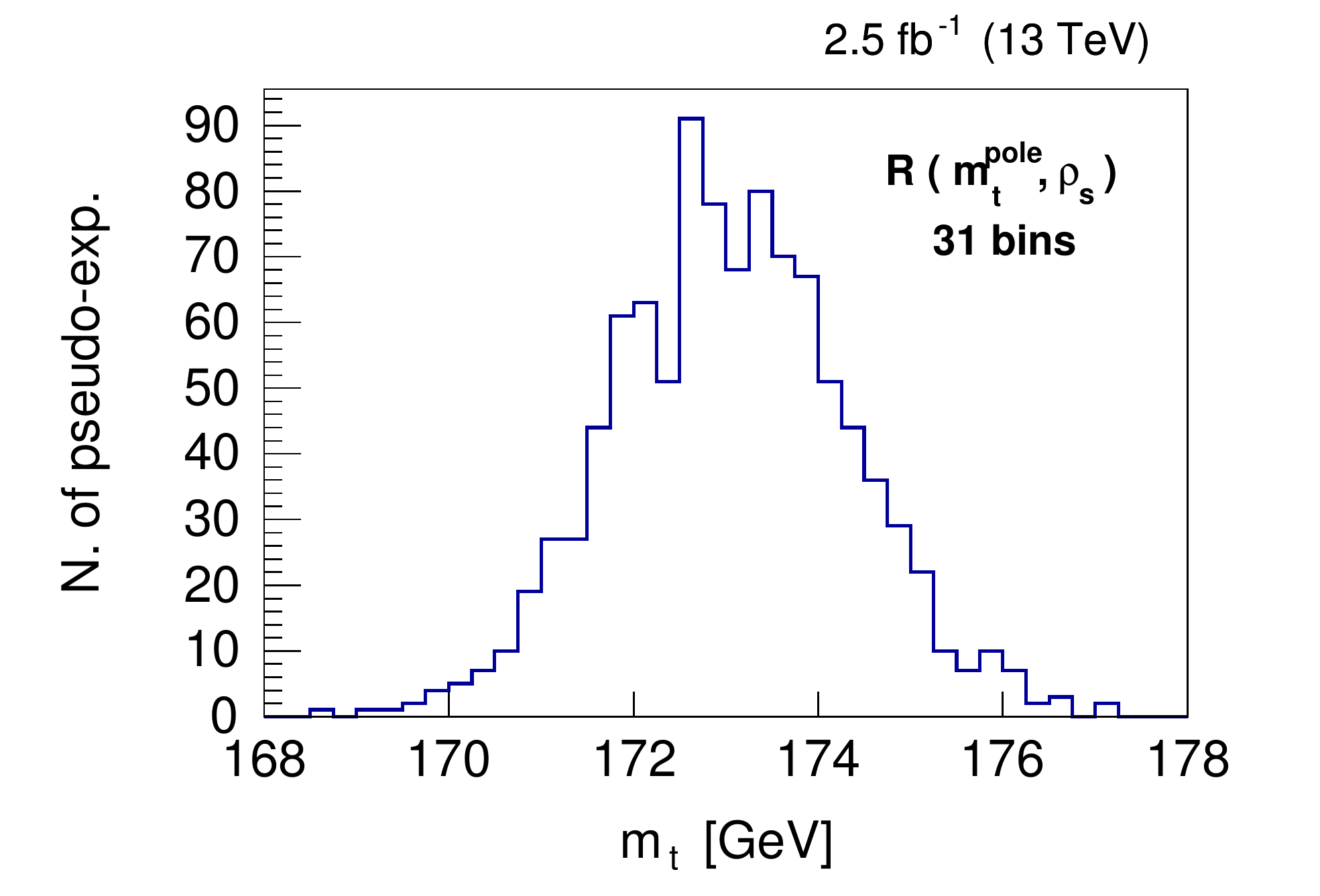}
\includegraphics[width=0.47\textwidth]{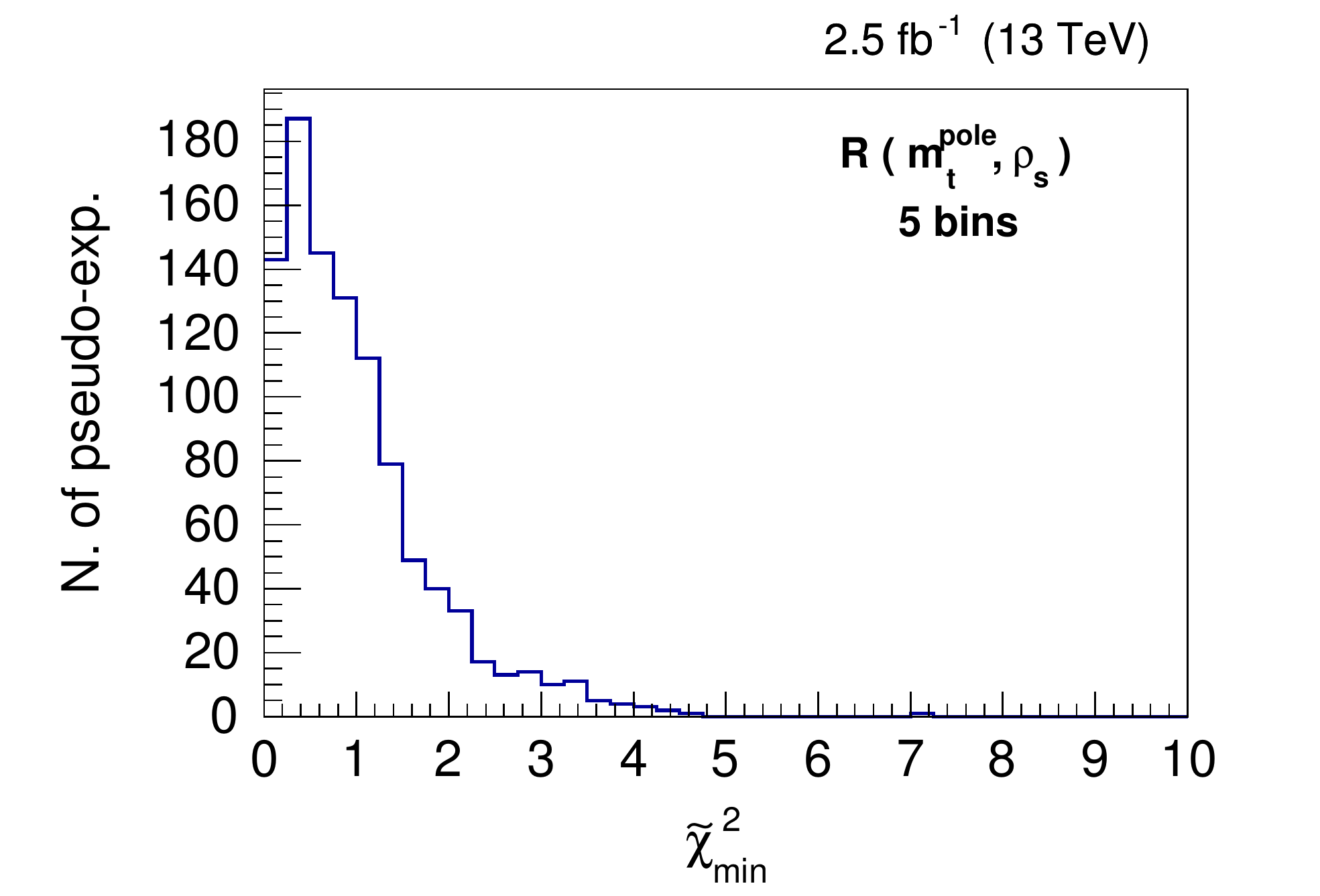}
\includegraphics[width=0.47\textwidth]{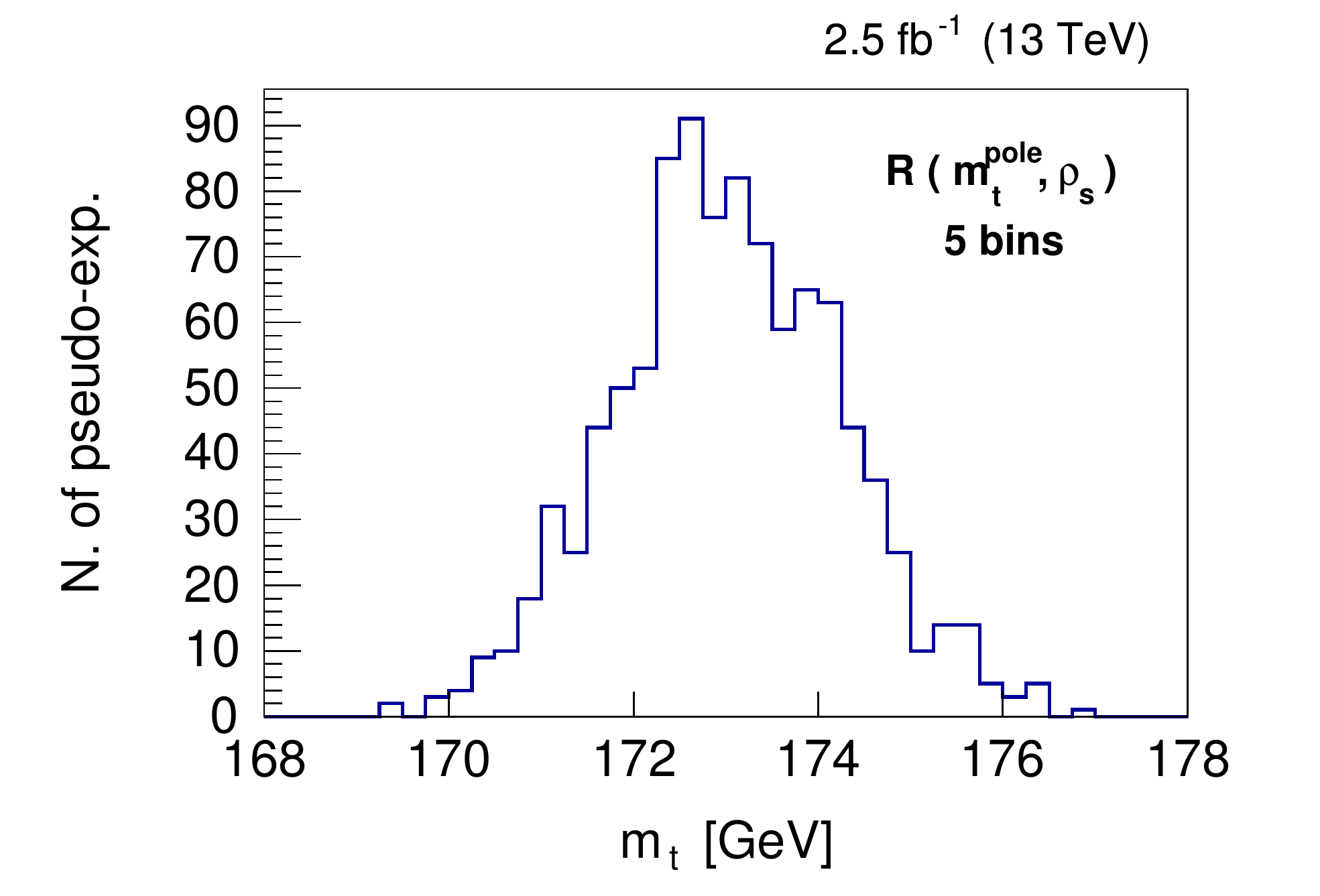}
\includegraphics[width=0.47\textwidth]{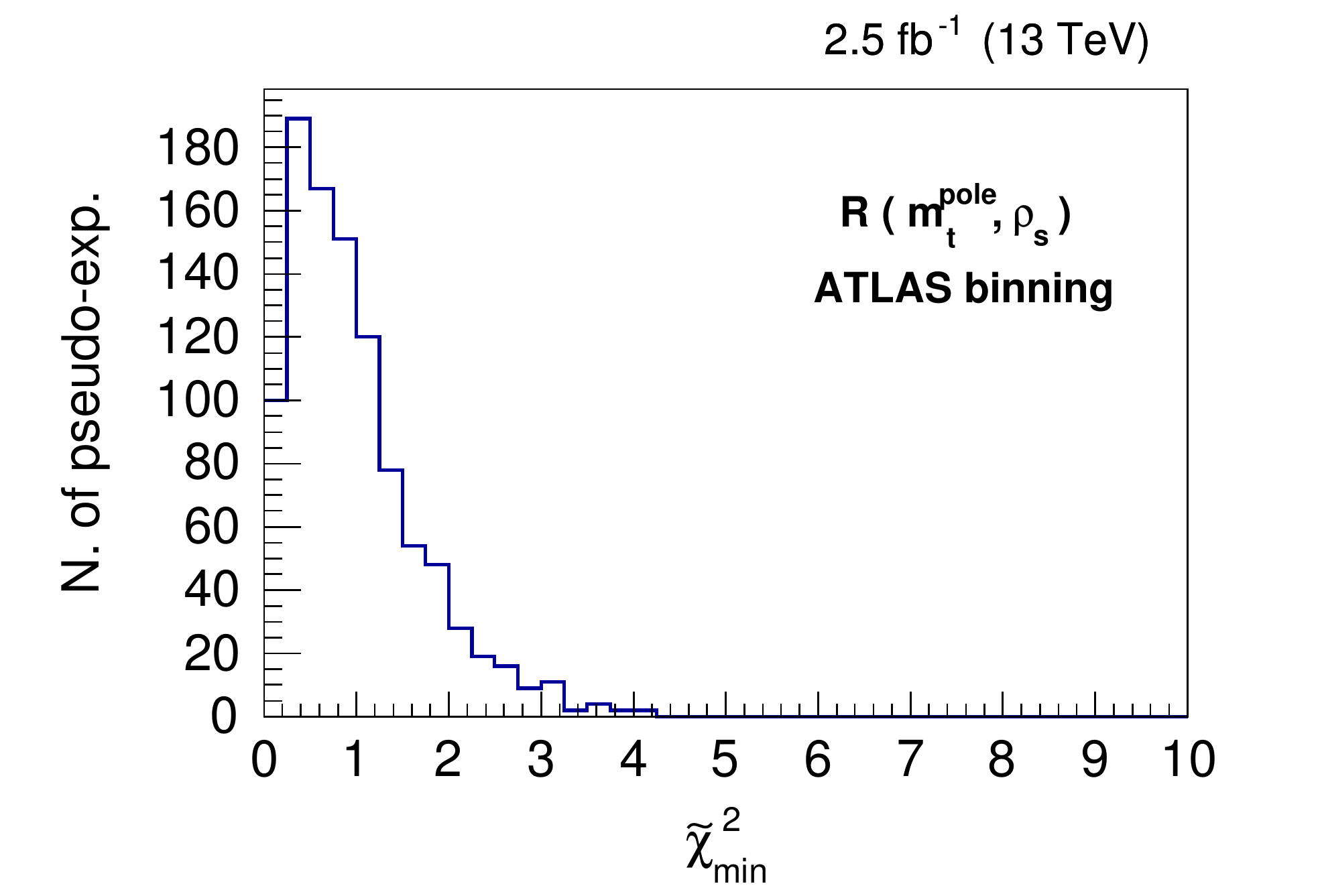}
\includegraphics[width=0.47\textwidth]{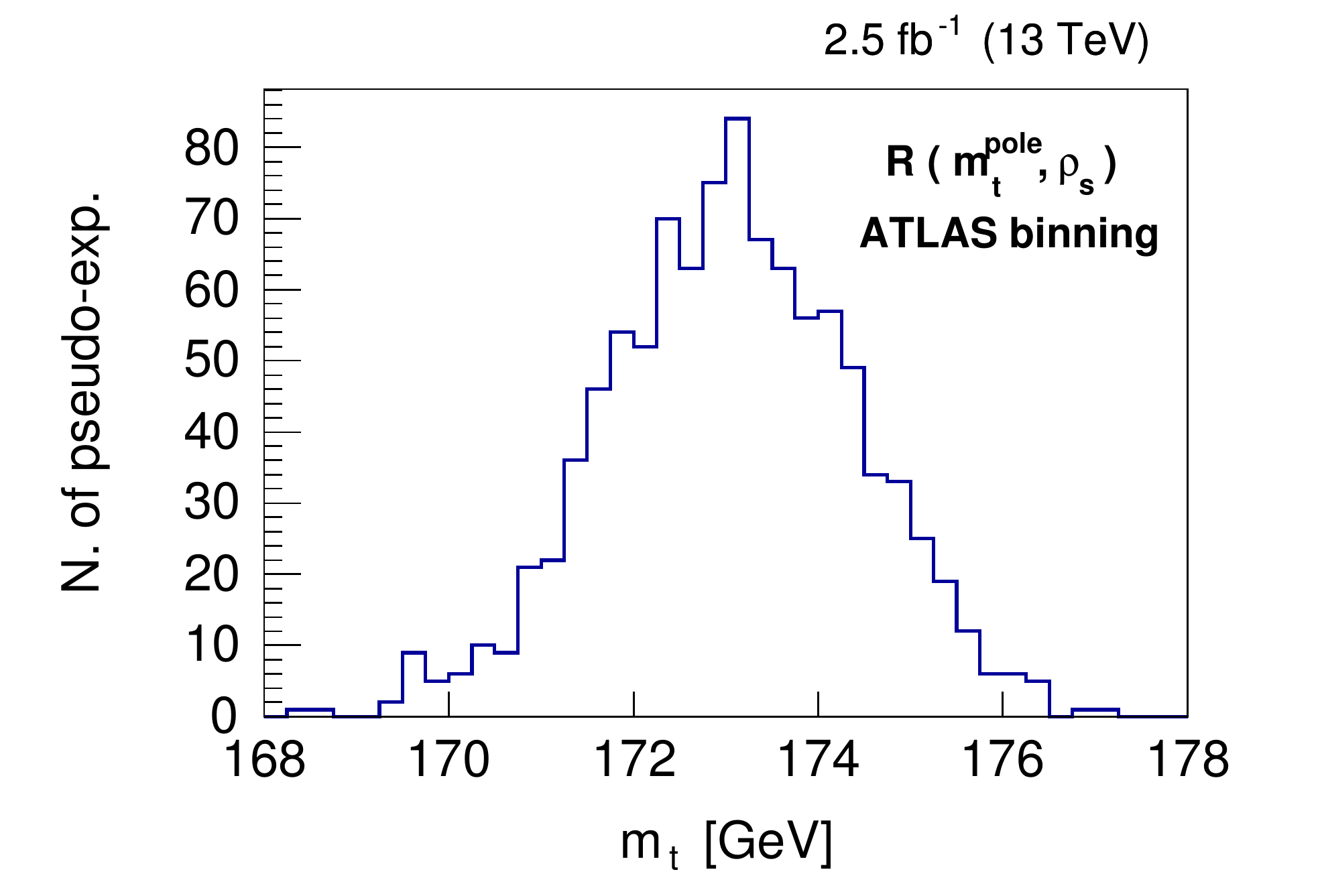}
\includegraphics[width=0.47\textwidth]{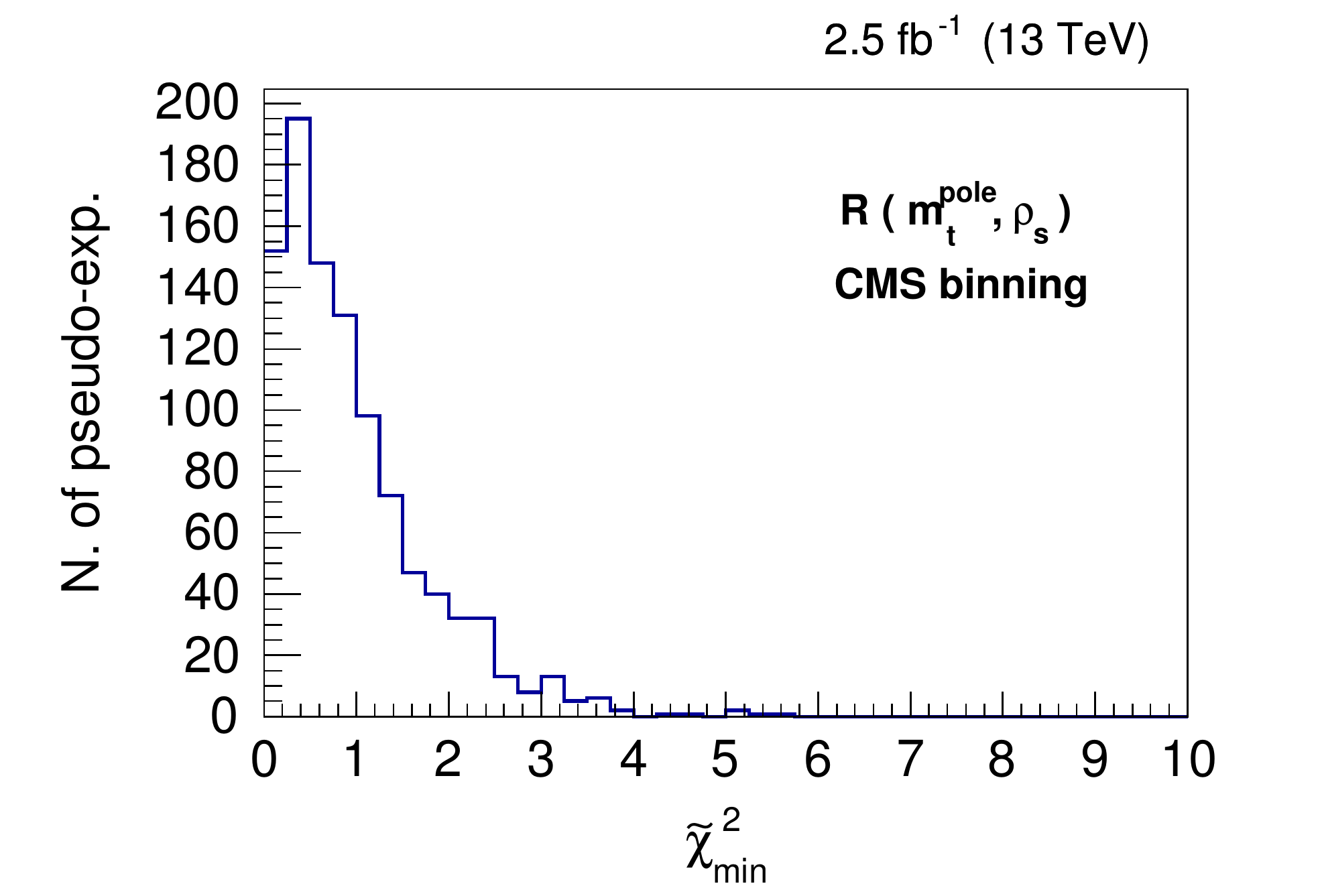}
\includegraphics[width=0.47\textwidth]{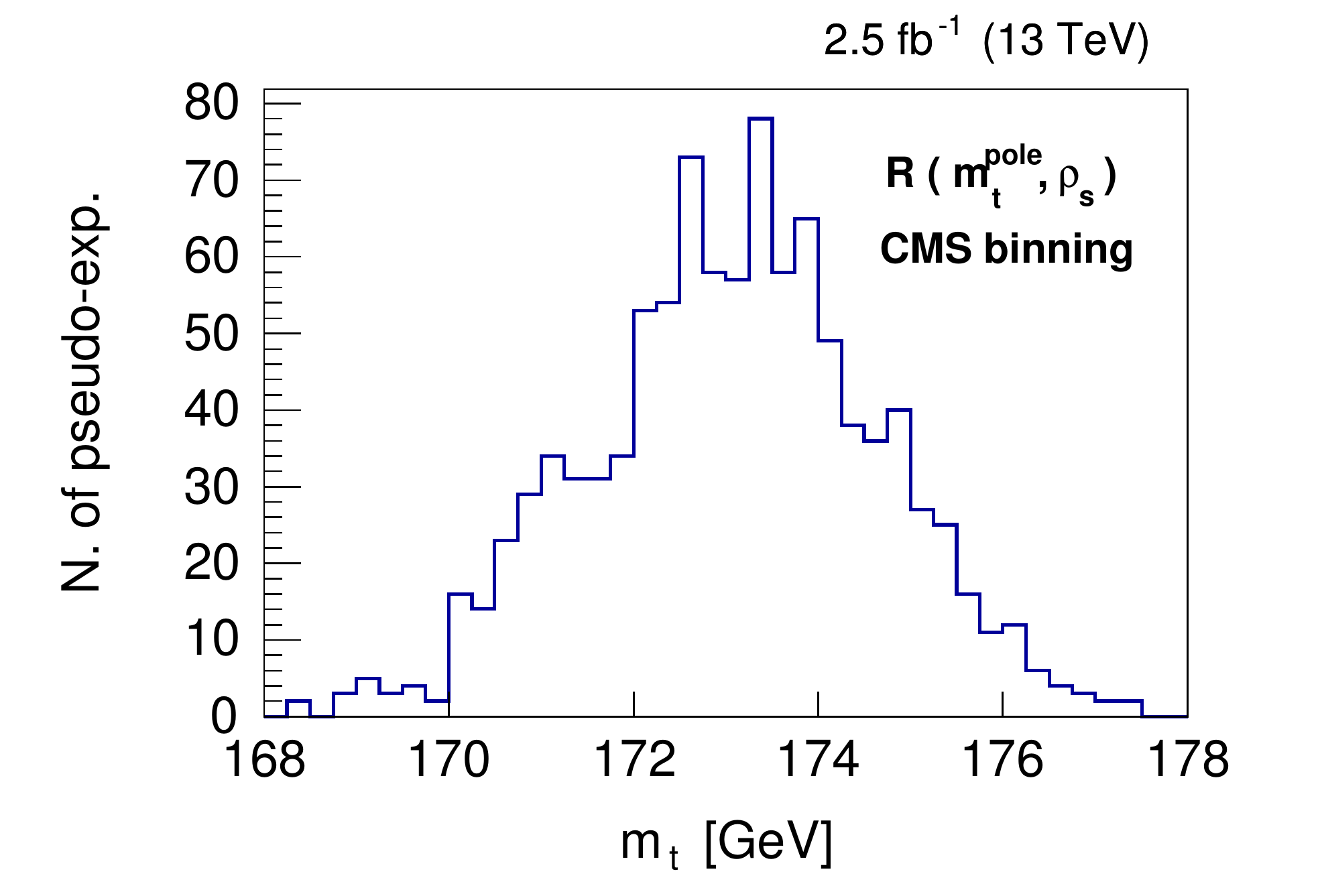}
\caption{\it
Distribution of minimum $\tilde{\chi}^2 \: ( \: \equiv \chi^2/d.o.f.)$
and of the corresponding top quark mass from 1000 pseudo-experiments.
Luminosity of $\mathcal{L} = 2.5 \mbox{ fb}^{-1}$ is assumed. Results
are shown for full theory for the $pp \to e^+\nu_e \mu^- \bar{\nu}_\mu
b\bar{b} j+ X$ production process at the LHC with $\sqrt{s} =$ 13
TeV. The CT14 PDF set and $\mu_0 = H_T/2$ are used.}
\label{psedoexperiment2.5}
\end{center}
\end{figure}
\begin{figure}[h!]
\begin{center}
\includegraphics[width=0.47\textwidth]{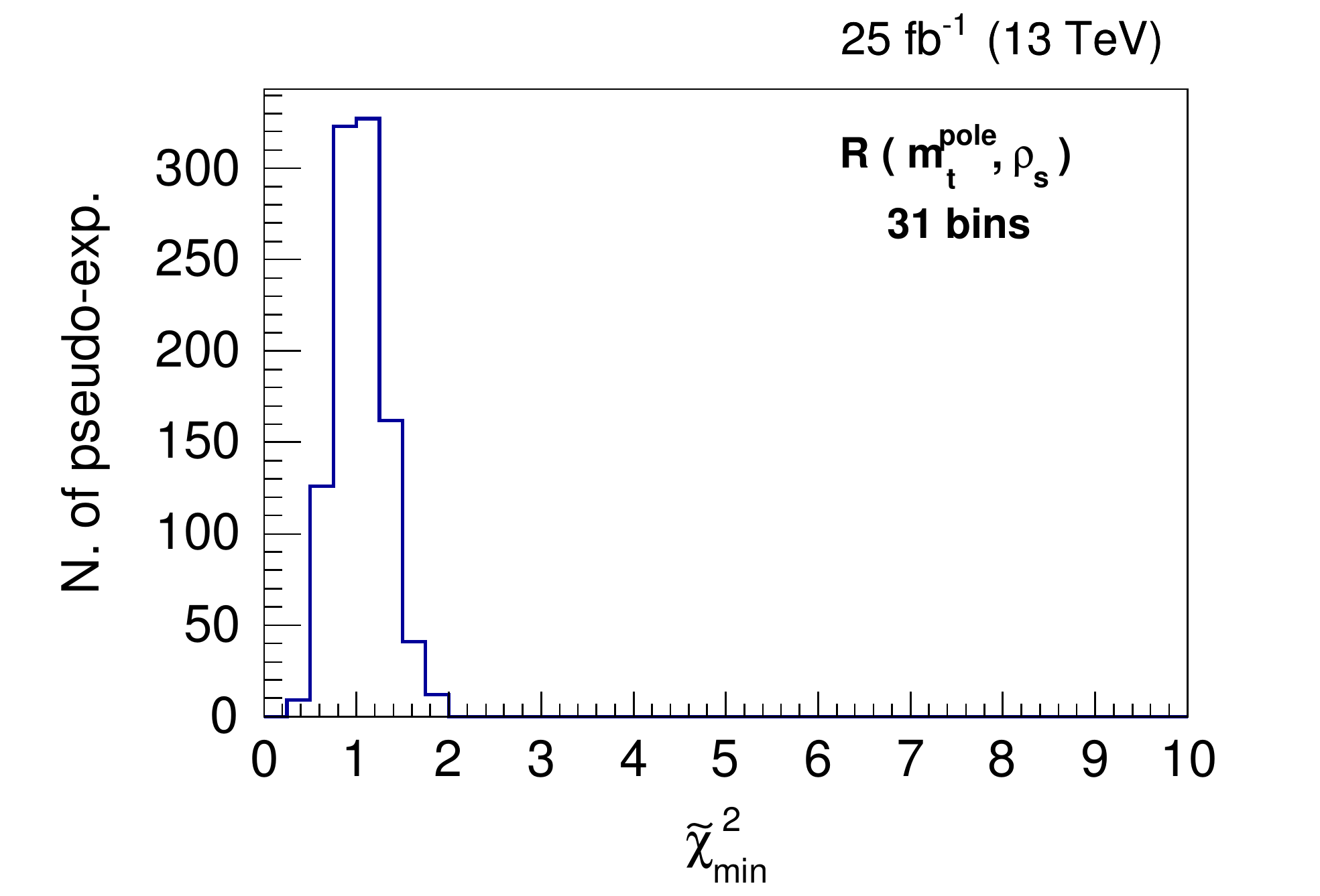}
\includegraphics[width=0.47\textwidth]{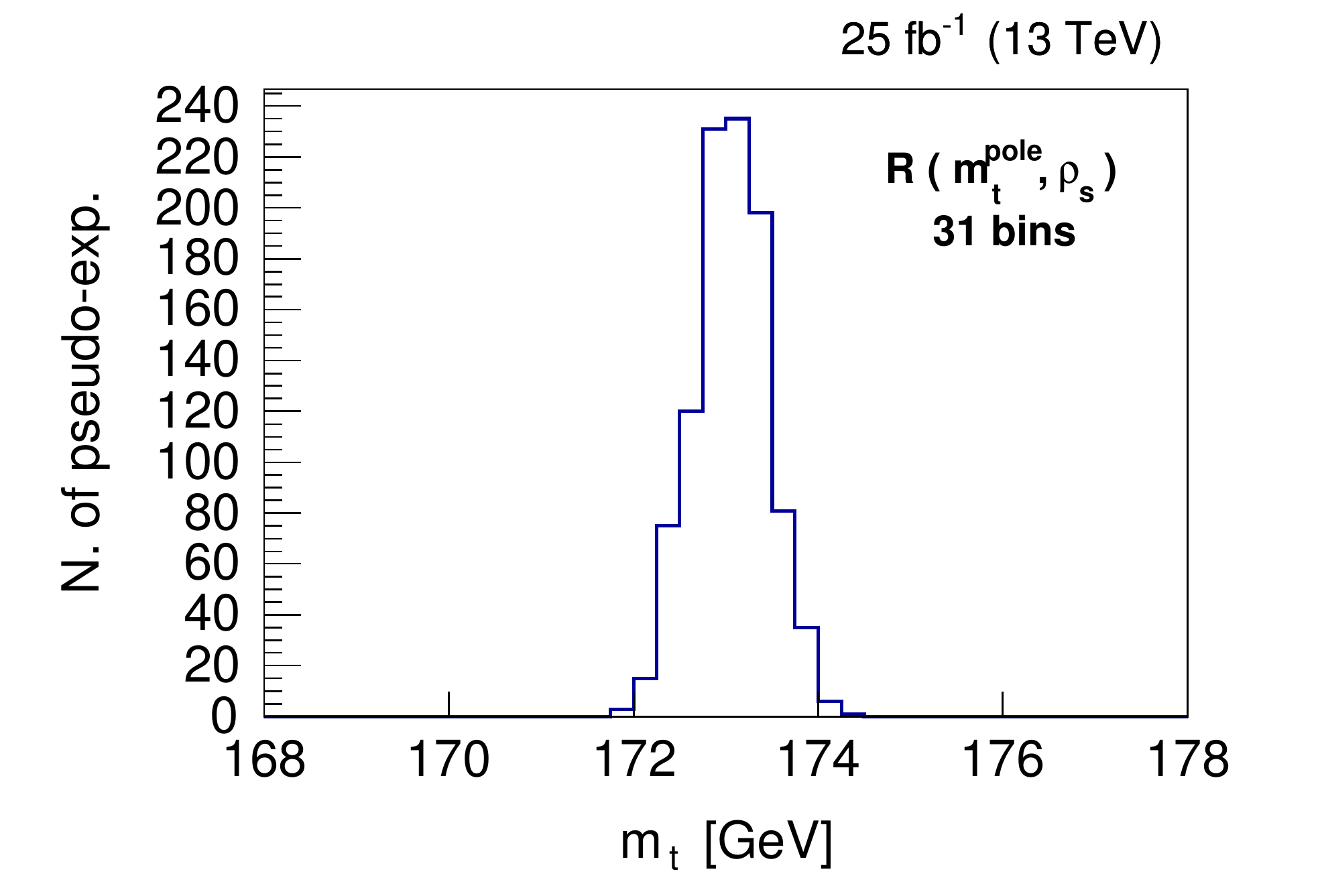}
\includegraphics[width=0.47\textwidth]{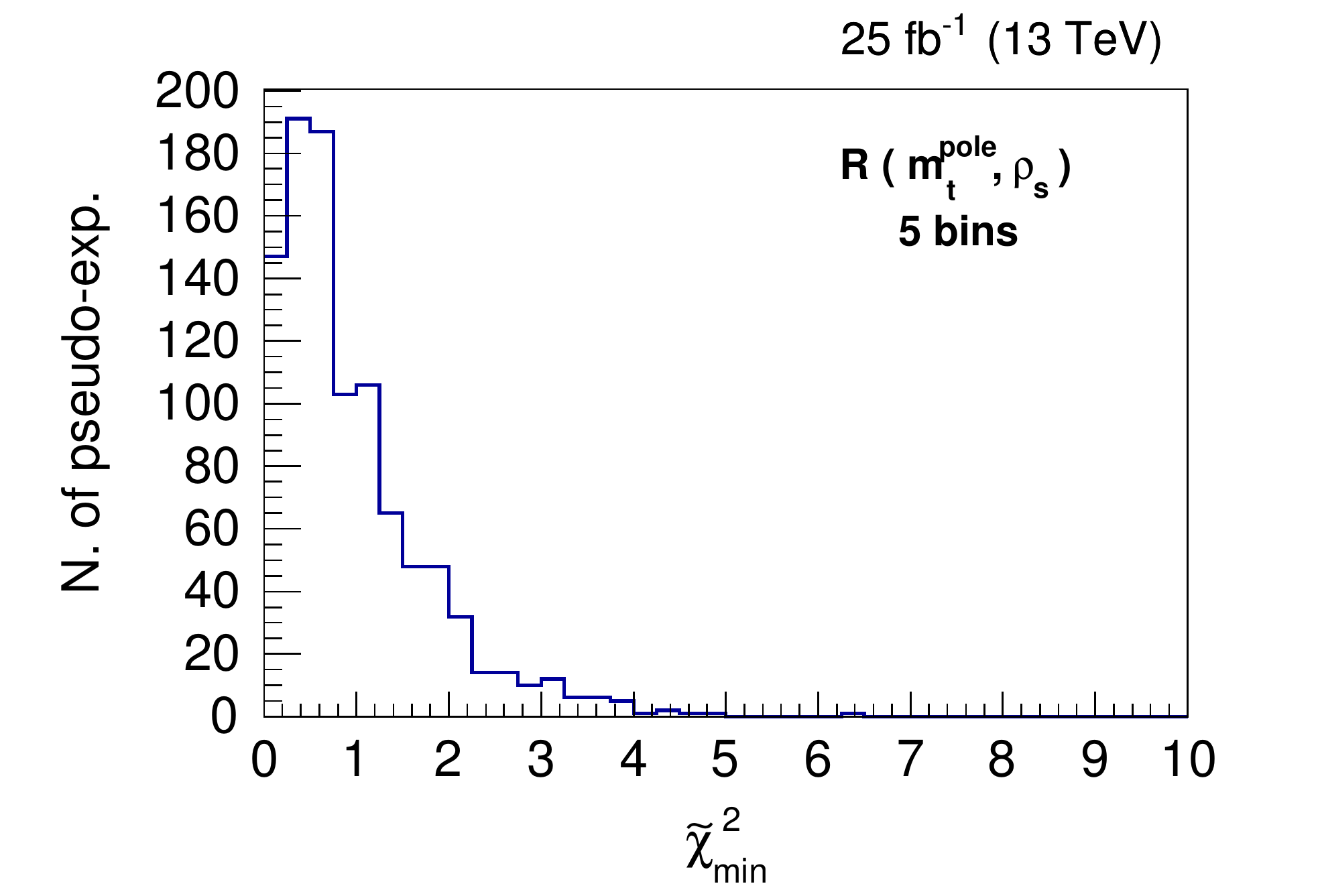}
\includegraphics[width=0.47\textwidth]{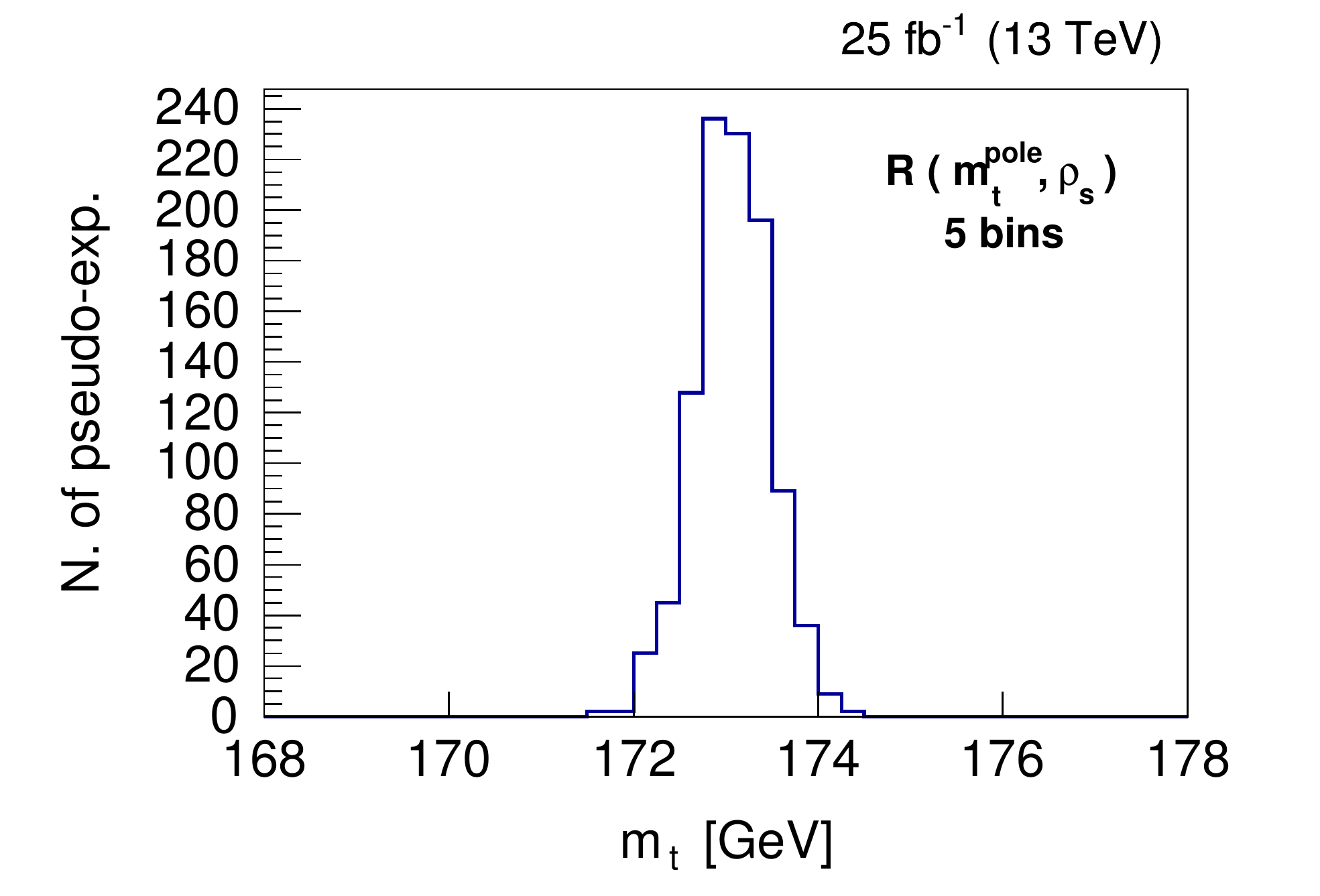}
\includegraphics[width=0.47\textwidth]{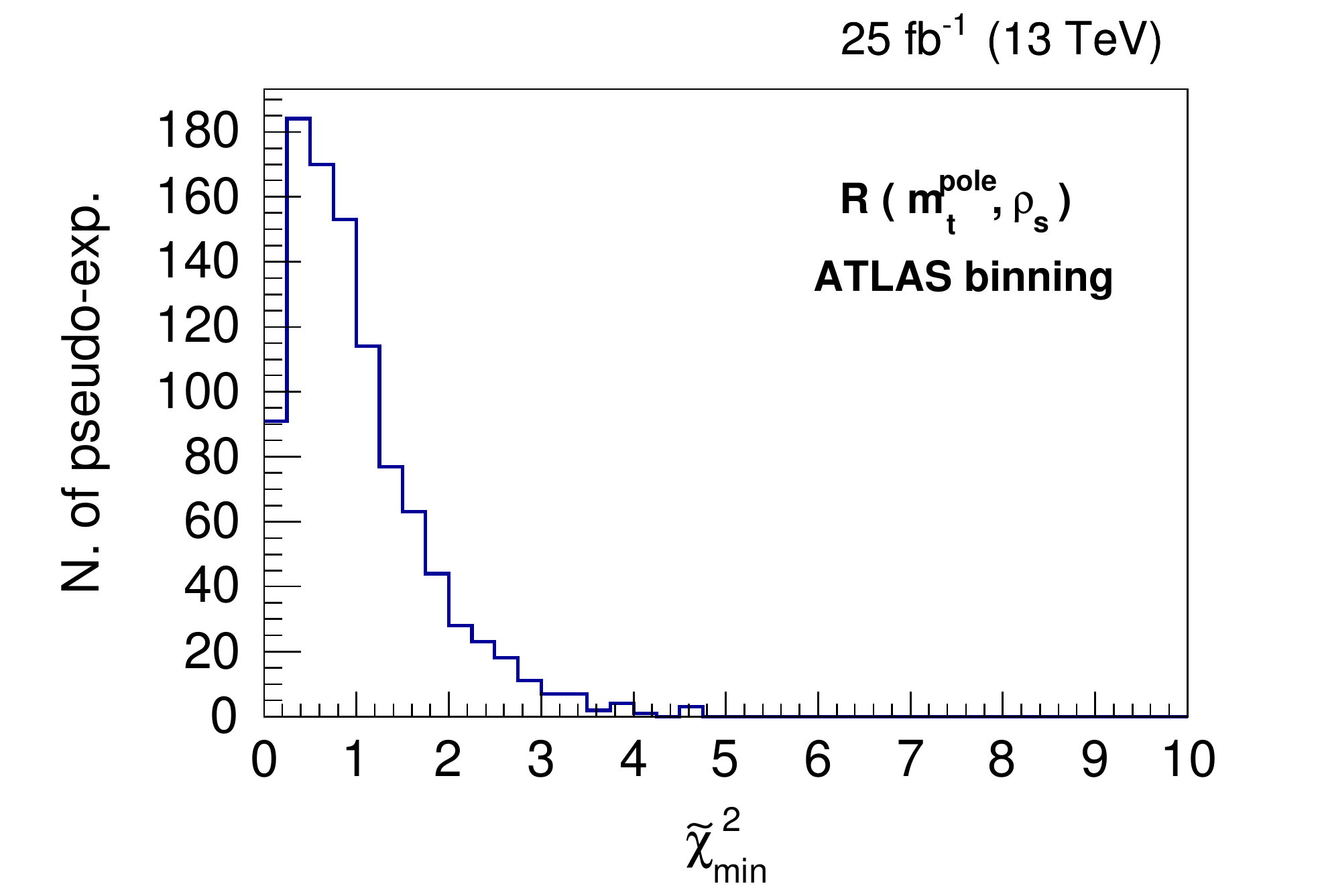}
\includegraphics[width=0.47\textwidth]{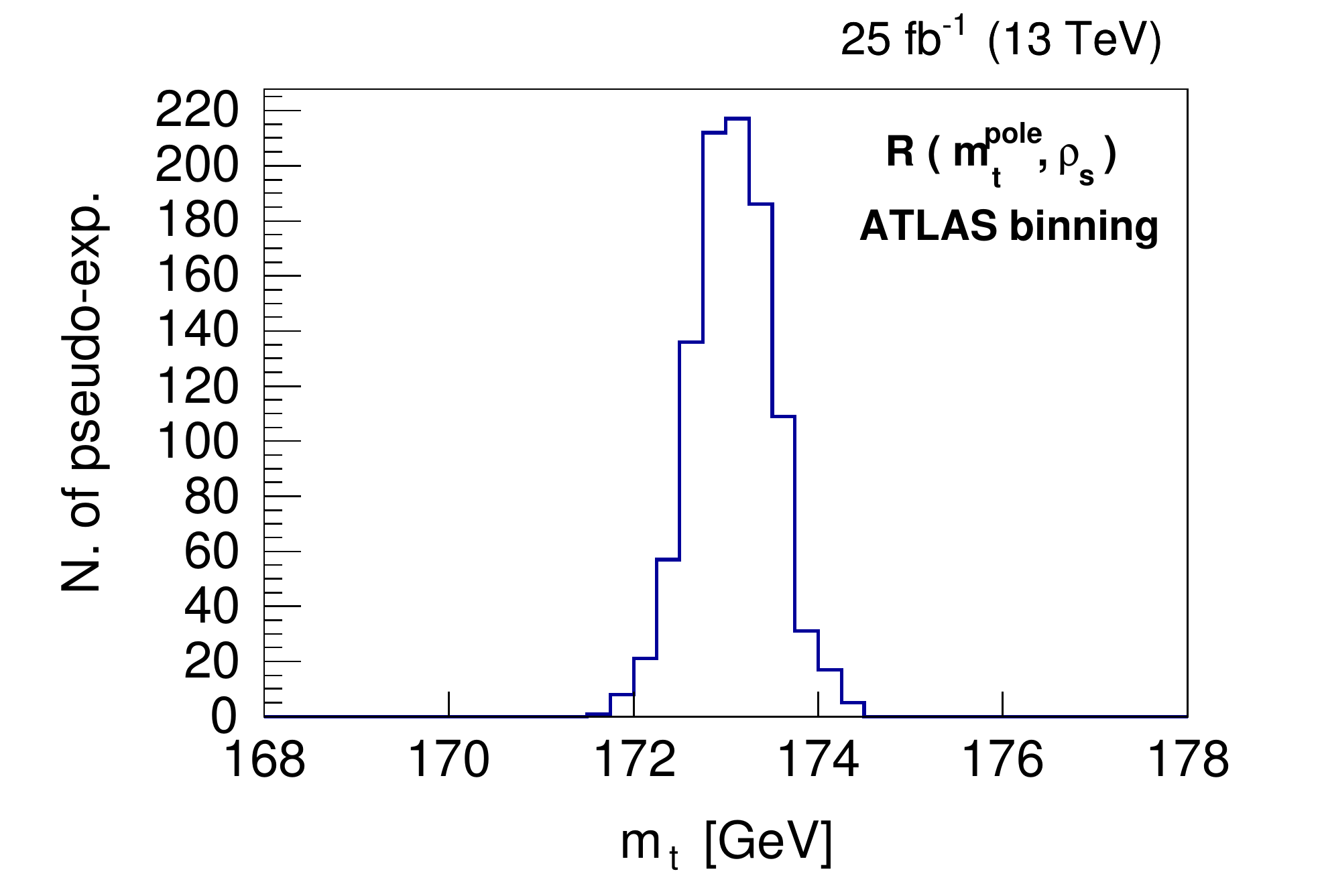}
\includegraphics[width=0.47\textwidth]{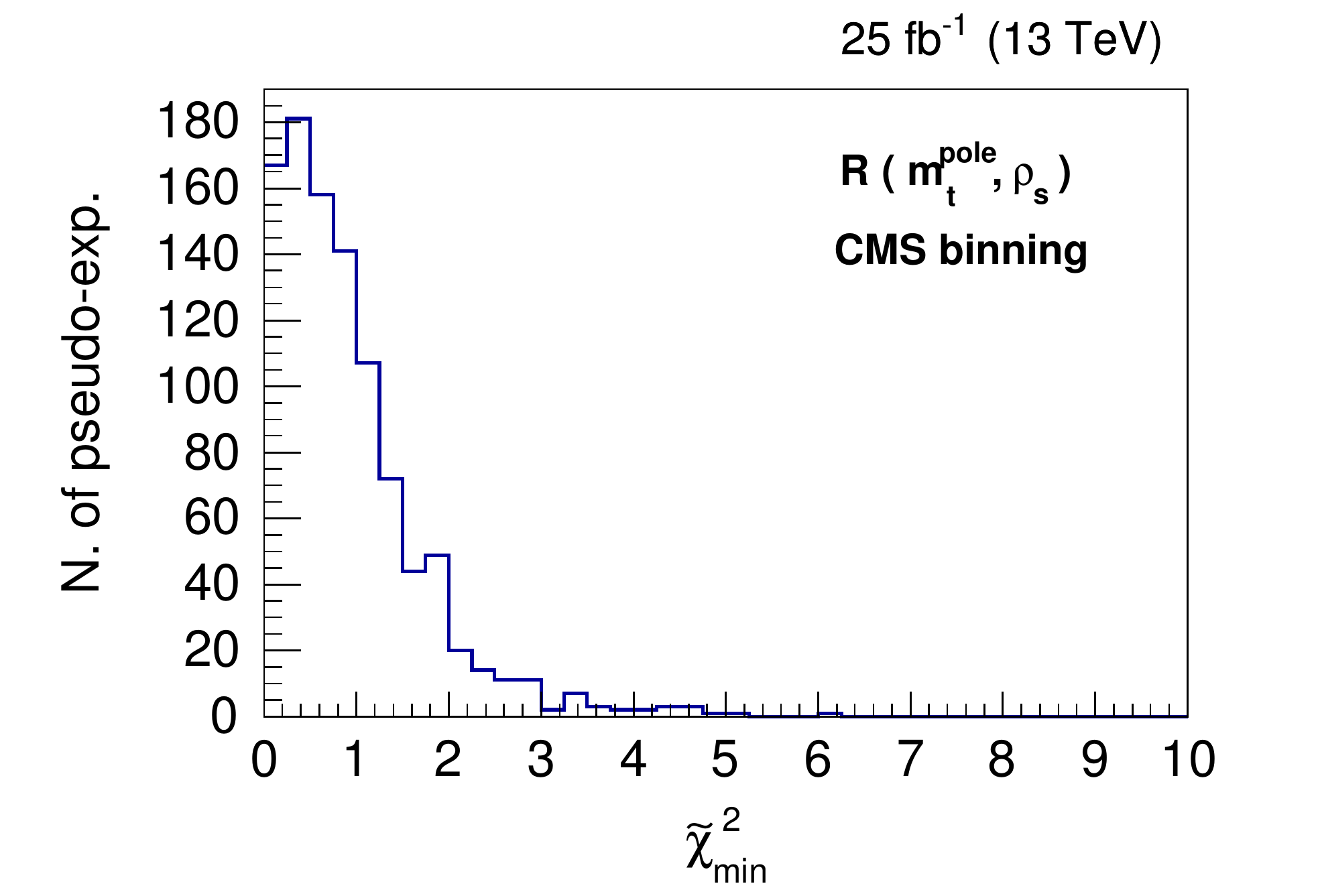}
\includegraphics[width=0.47\textwidth]{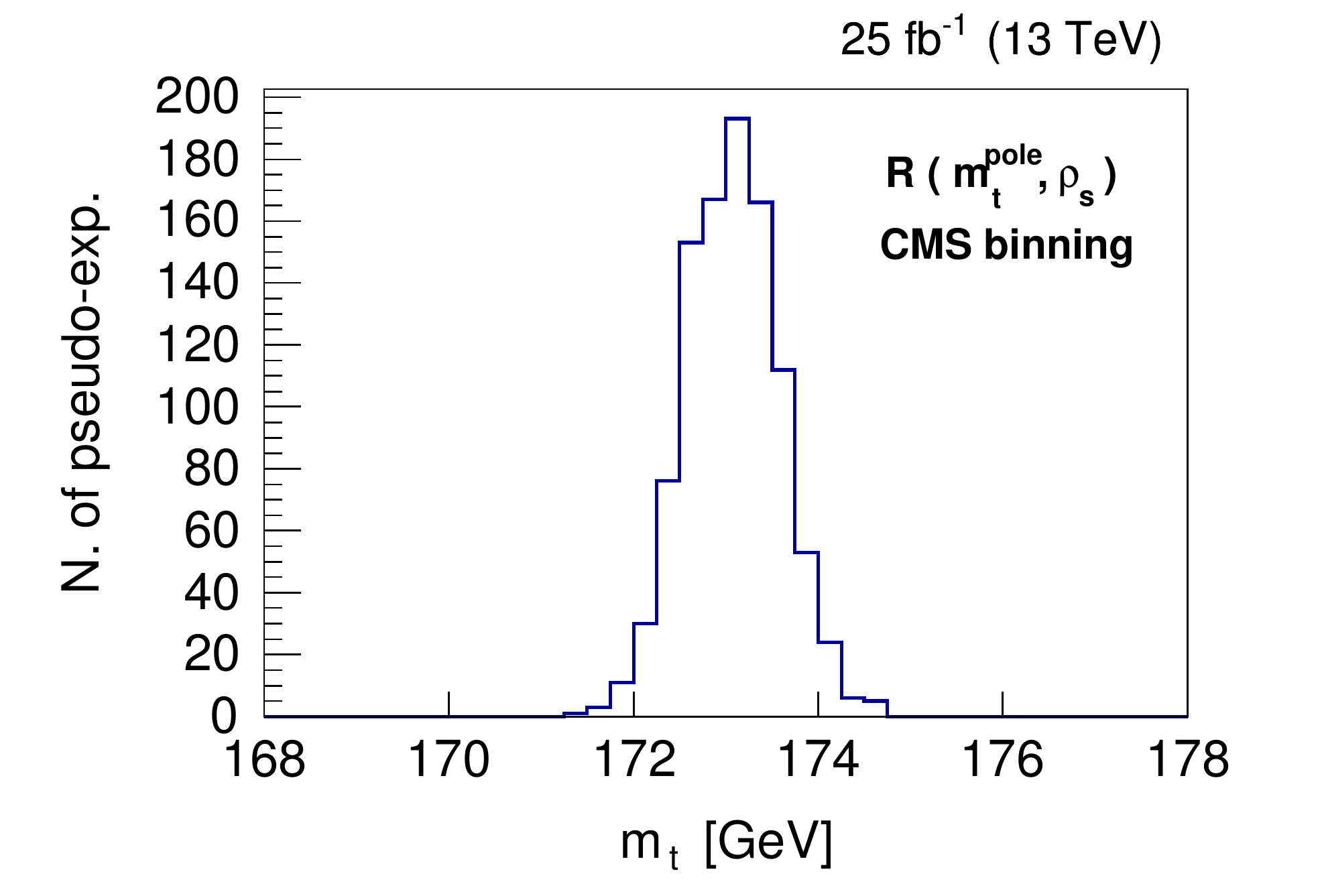}
\caption{\it 
Distribution of minimum $\tilde{\chi}^2 \: ( \: \equiv \chi^2/d.o.f.)$
and of the corresponding top quark mass from 1000 pseudo-experiments.
Luminosity of $\mathcal{L} = 25 \mbox{ fb}^{-1}$ is assumed. Results
are shown for full theory for the $pp \to e^+\nu_e \mu^- \bar{\nu}_\mu
b\bar{b} j+ X$ production process at the LHC with $\sqrt{s} =$ 13
TeV. The CT14 PDF set and $\mu_0 = H_T/2$ are used.}
\label{psedoexperiment25}
\end{center}
\end{figure}
\begin{figure}[t!]
\begin{center}
\includegraphics[width=0.49\textwidth]{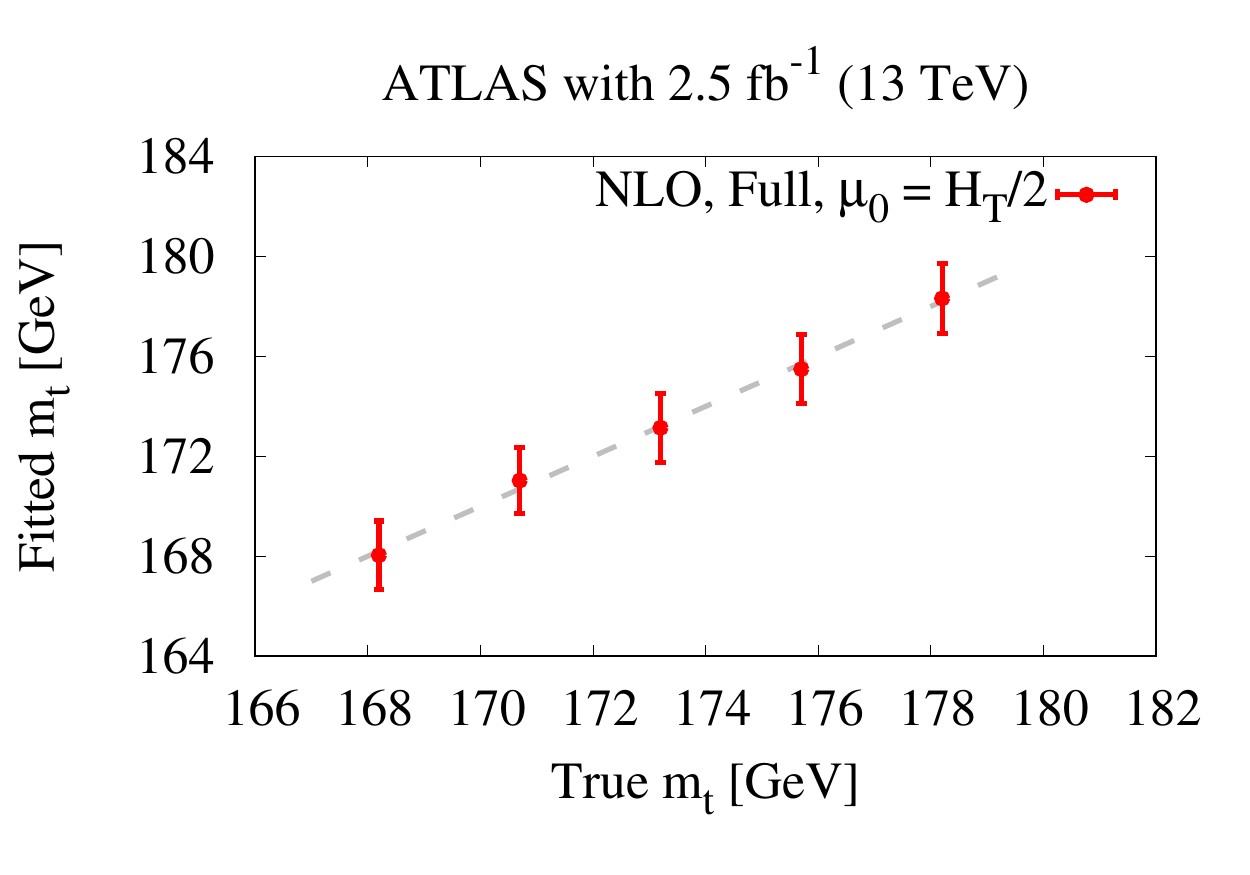}
\includegraphics[width=0.49\textwidth]{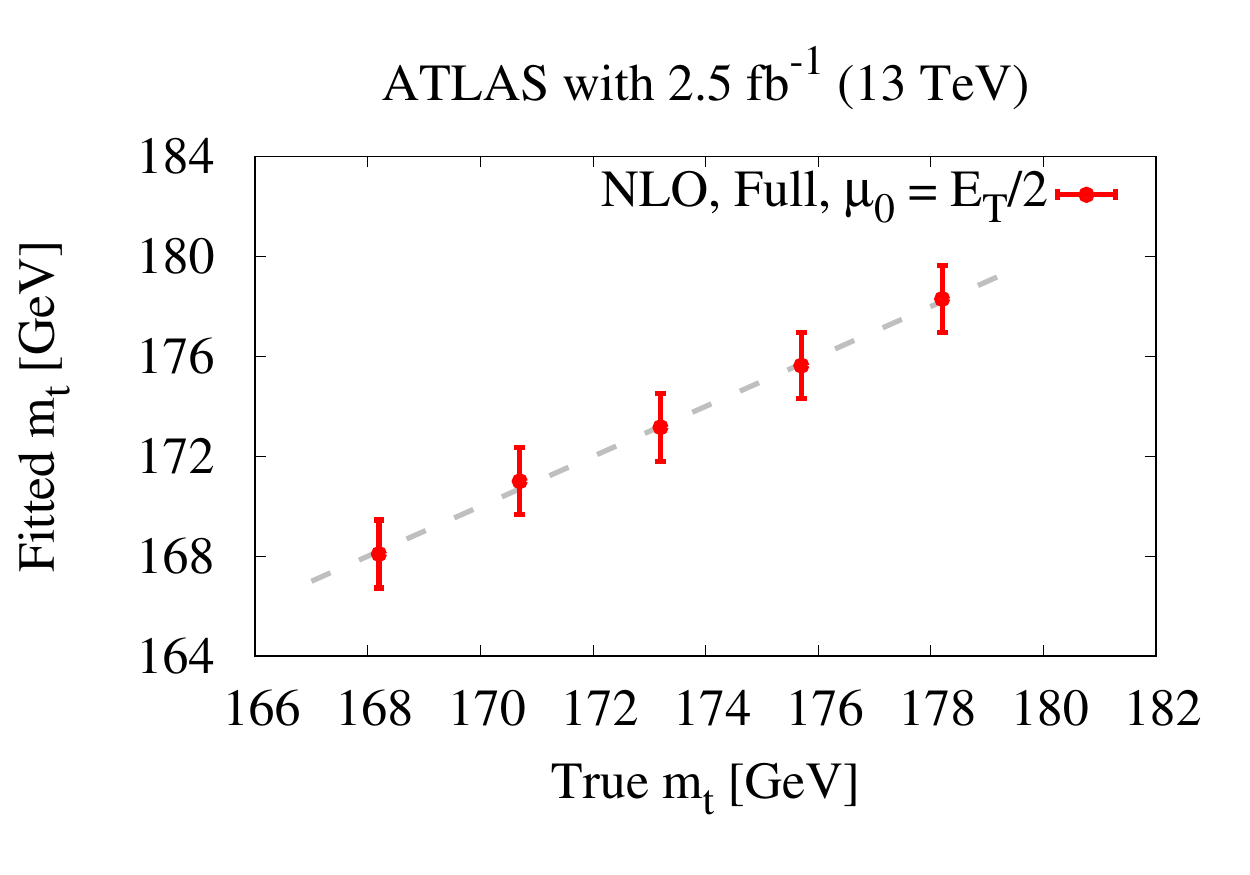}
\includegraphics[width=0.49\textwidth]{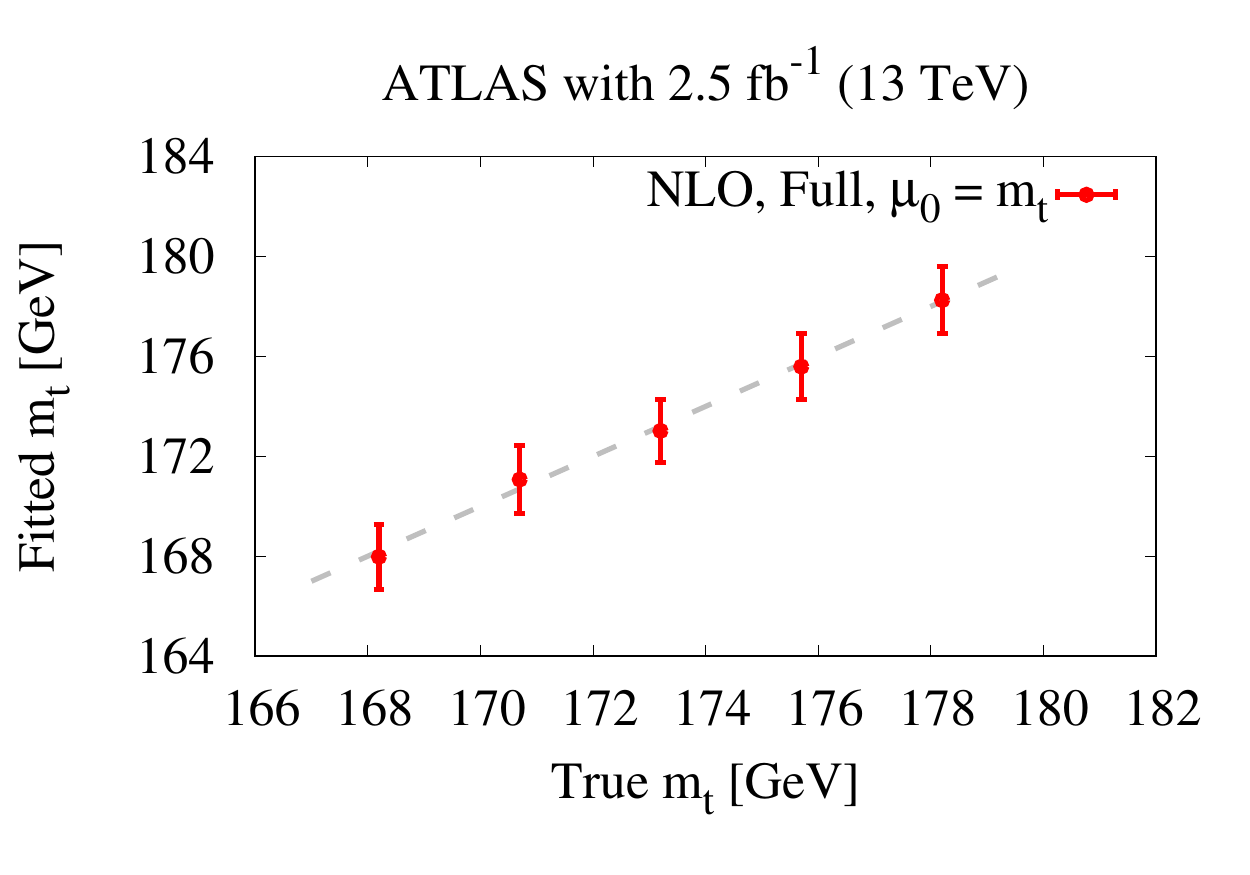}
\includegraphics[width=0.49\textwidth]{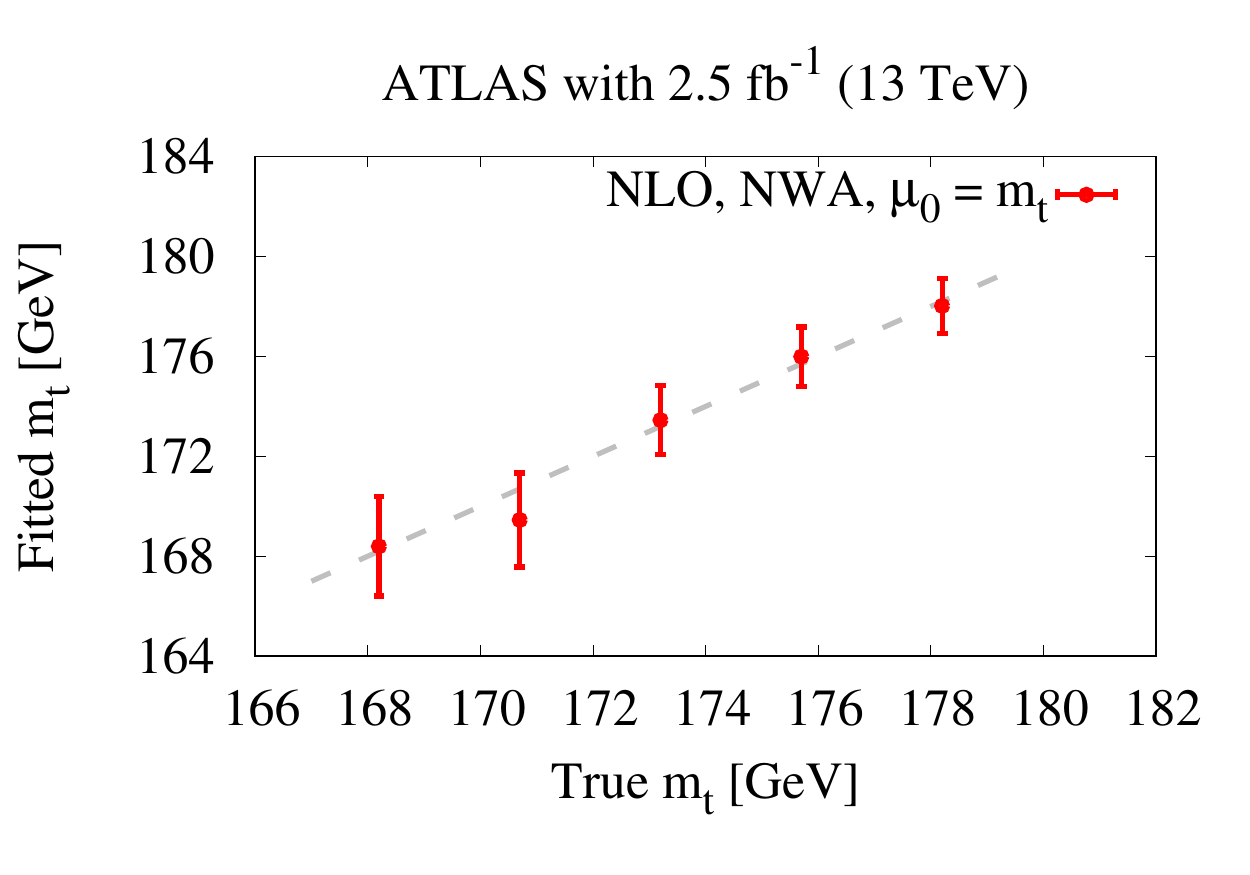}
\includegraphics[width=0.49\textwidth]{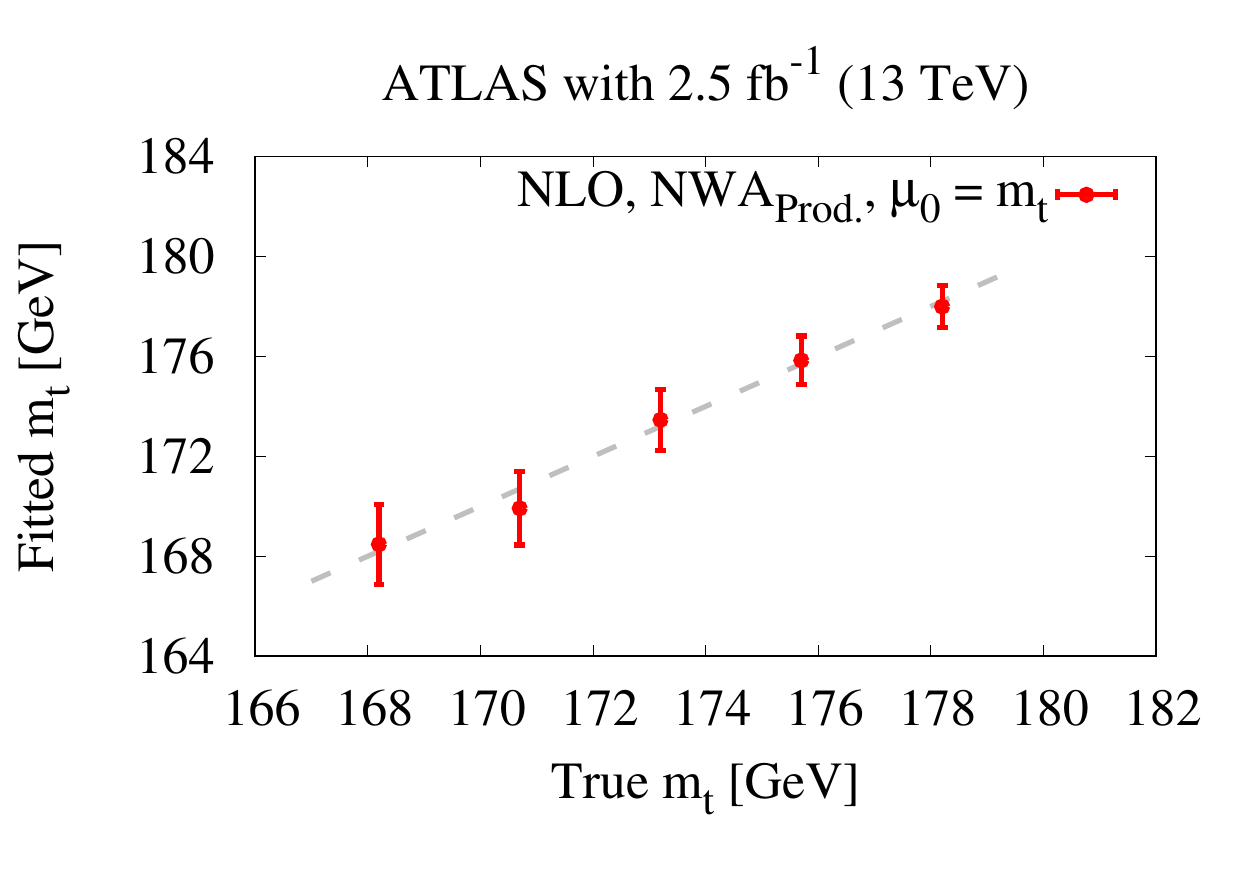}
\end{center}
\caption{\it A difference between the fitted top quark mass and the
top quark mass assumed in the theoretical prediction used for the
generation of the pseudo-data set. The ATLAS binning is
assumed. Luminosity of ${\cal L}=$ 2.5 fb${}^{-1}$ is considered and
the CT14 PDF set is employed. The grey (dashed) line corresponds to
"Fitted $m_t$" $=$ "True $m_t$". }
\label{fig:bias1}
\end{figure}
\begin{figure}[t!]
\begin{center}
\includegraphics[width=0.49\textwidth]{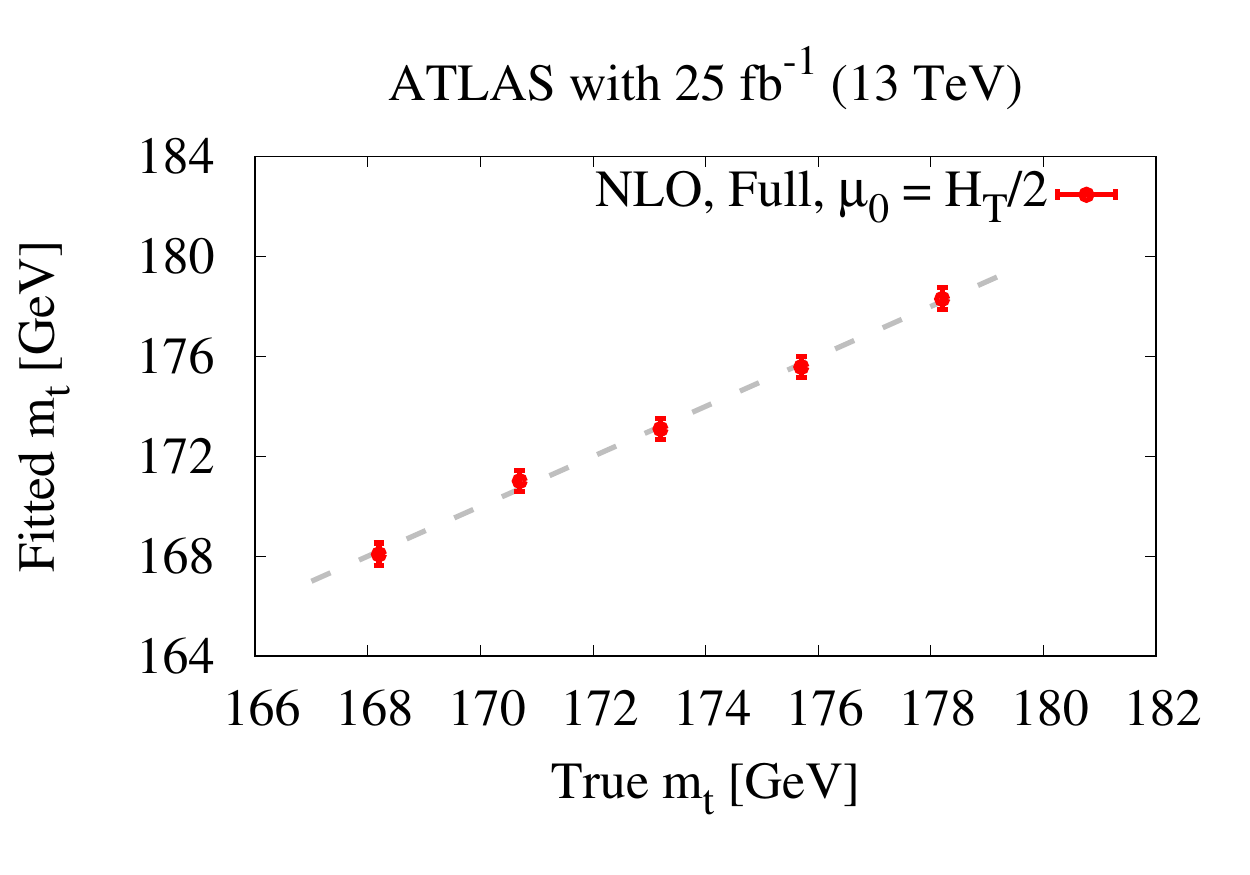}
\includegraphics[width=0.49\textwidth]{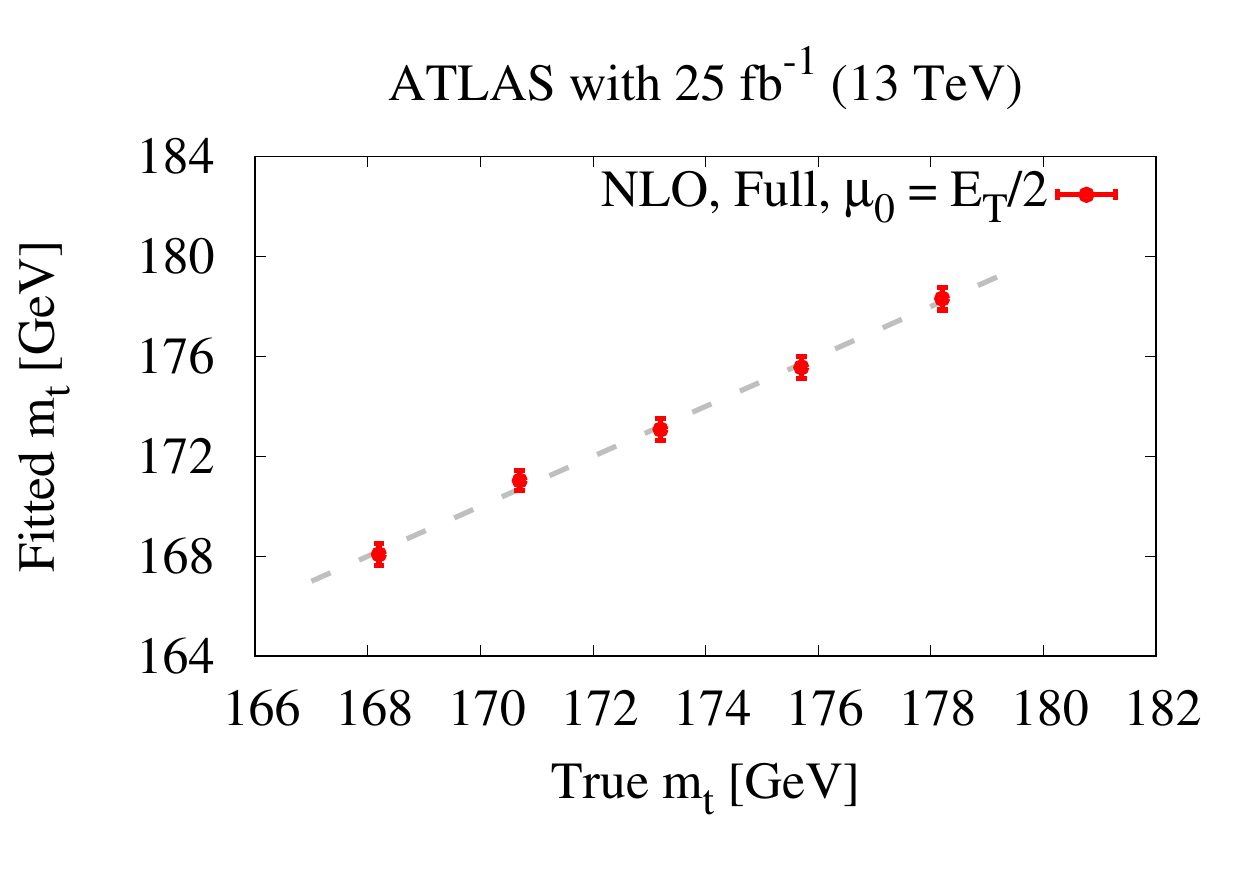}
\includegraphics[width=0.49\textwidth]{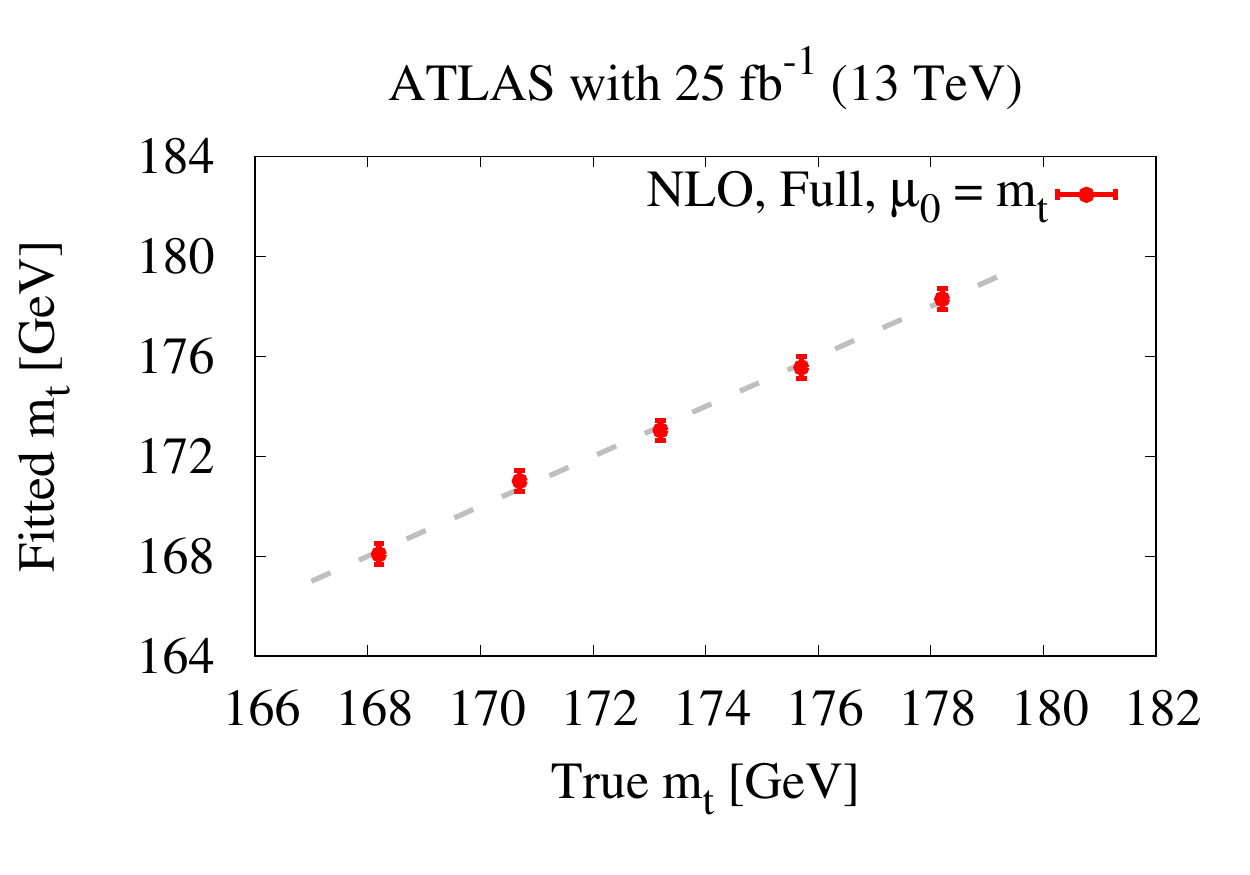}
\includegraphics[width=0.49\textwidth]{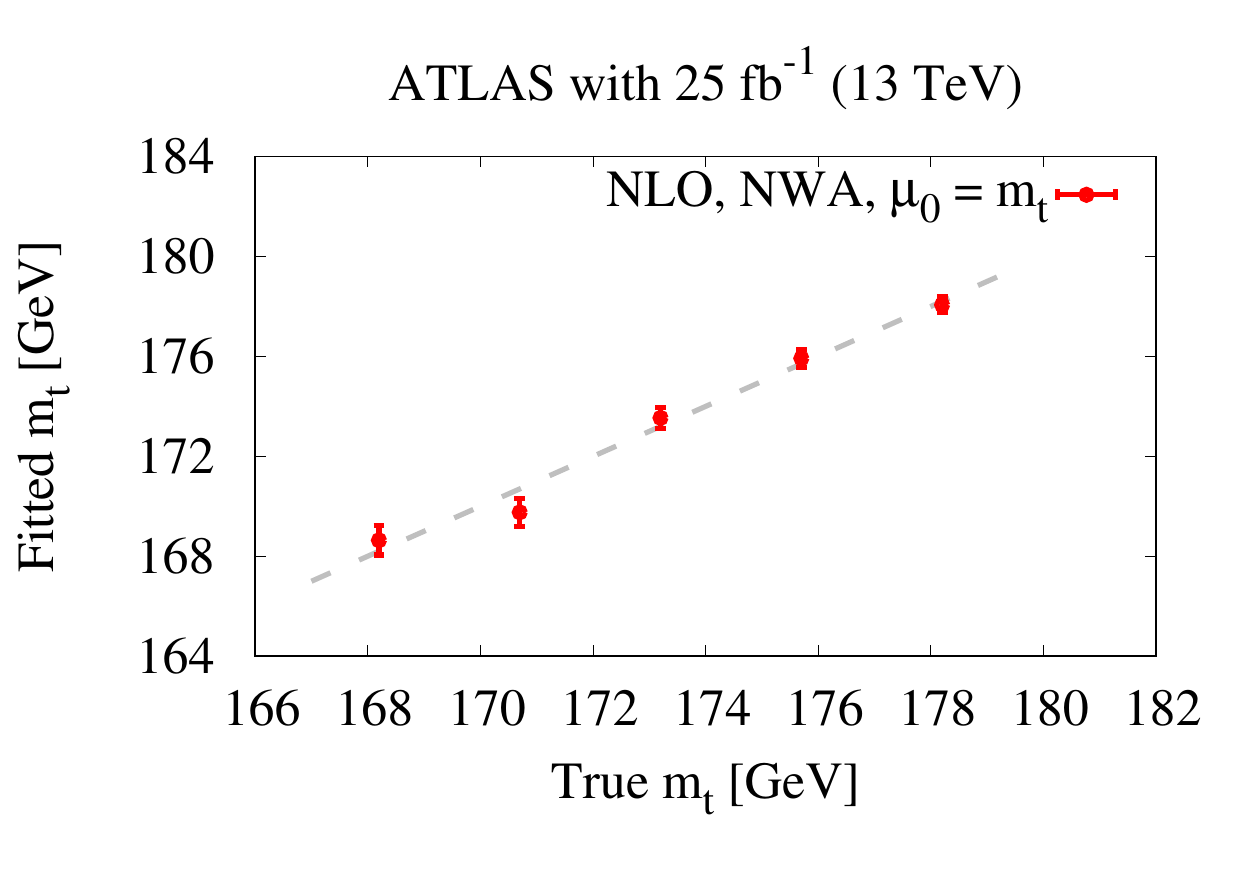}
\includegraphics[width=0.49\textwidth]{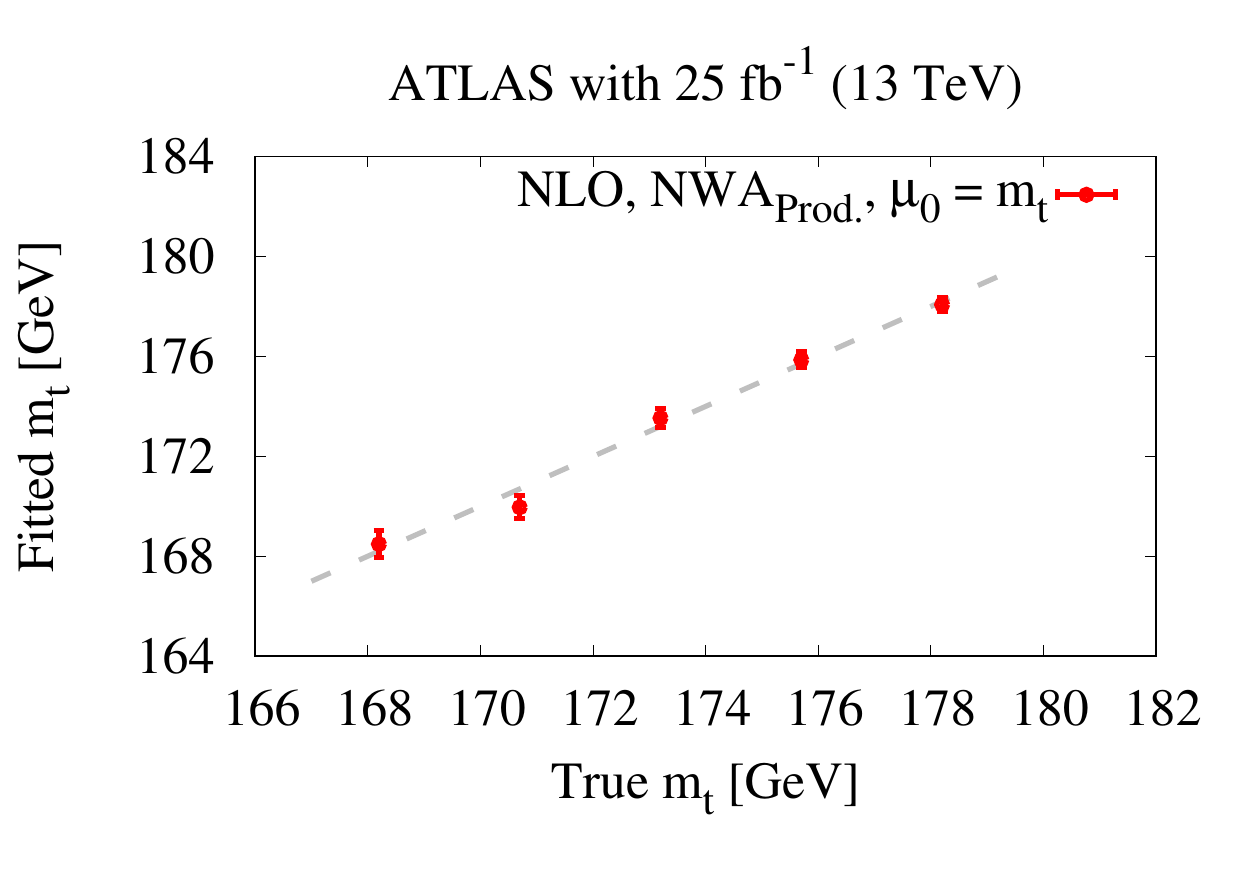}
\end{center}
\caption{\it A difference  between the fitted top quark mass and the
top quark mass assumed in the theoretical prediction used for the
generation of the pseudo-data set. The ATLAS binning is
assumed. Luminosity of ${\cal L}=$ 25 fb${}^{-1}$ is considered and
the CT14 PDF set is employed. The grey (dashed) line corresponds to
"Fitted $m_t$" $=$ "True $m_t$". }
\label{fig:bias2}
\end{figure}
%

In the next step, the ${\cal R}(m_t^{pole},\rho_s)$ differential
distributions shall be used to obtain the top quark mass.  To this end
a set of pseudo-data is compared to ${\cal R}(m_t^{pole},\rho_s)$ as
generated with five different top quark masses and with three
different theoretical descriptions of the $pp\to e^+\nu_e \mu^-
\bar{\nu}_\mu b\bar{b}j +X$ production process. The pseudo-data set is
generated randomly according to the best theoretical prediction at
hand, i.e. the {\it Full} prediction at NLO in QCD as generated with
$m_t=173.2$ GeV and $\mu_R=\mu_F=\mu_0=H_T/2$.  Unless explicitly
mentioned this particular setup with the CT14 PDF set will always be
employed for the generation of the pseudo-data sets for all considered
observables. For completeness, in Figure \ref{fig:1observable_ht} the
normalised $\rho_s$ observable is plotted again, however, this time
the {\it Full} case (red solid line) is shown for $\mu_R=\mu_F=\mu_0
=H_T/2$.  When comparing the differential ${\cal
K}-$factor for the {\it Full} case in Figure \ref{fig:1observable} and
Figure  \ref{fig:1observable_ht} we find that the large corrections in
the region $\rho_s \le 0.3$ are removed. The effect can be attributed
to the scale choice  made. The kinematic tail of
$M_{t\bar{t}j}$ only shows perturbative convergence when dynamic 
scales are employed, and for the $\rho_s$ observable the high energy
kinematic tail corresponds to low values of $\rho_s$. Thus, in
addition to differences for large values of $\rho_s$ present in Figure
\ref{fig:1observable}, there are now only differences up to $-15\%$ at
low values of $\rho_s$. Since this region is sensitive to the top
quark mass we expect to see an impact on $m_t$.  Moreover, even though
we have only simulated decays of the weak bosons to different lepton
generations, i.e. $W^+W^-\to e^+\nu_e \mu^- \bar{\nu}_\mu$ omitting
same generation lepton interference effects as occurring in $W^+W^-\to
e^+\nu_e e^- \bar{\nu}_e$ we adjust the counting factor to correspond
to the production of all combinations of charged leptons of the first
two generations.  The interference effects can be safely neglected
because they are at the per-mille level for our inclusive selection
cuts as has been directly checked using LO results. The complete cross
section with $\ell = e^\pm, \mu^\pm$ is, thus, obtained by multiplying
the result for $pp\to e^+\nu_e \mu^- \bar{\nu}_\mu b\bar{b}j +X$ with
a lepton flavour factor of $4$. In this way, two cases of integrated
luminosity $2.5 \, {\rm fb}^{-1}$ and $25 \, {\rm fb}^{-1}$, that we
shall consider in the following, correspond, assuming perfect detector
efficiency, approximately to $5400$ and $54000$ events
respectively. Errors on the pseudo-data are calculated according to
the Bernoulli distribution. Notice that the theoretical predictions
are calculated with such high statistics that the Monte Carlo errors
in each bin are negligible compared to the errors of the pseudo-data
samples with the chosen luminosities. Examples of the pseudo-data sets
for both cases, $2.5$ fb${}^{-1}$ and $25$ fb${}^{-1}$, are shown in
Figure \ref{fig:1pseudodata}.  We shall consider various choices for
the number of bins and the bin size for the $\rho_s$ observable to
check whether there is any effect on $m_t$. More precisely, we
consider $31$ and $5$ bins of equal size as well as the bin intervals
as proposed by ATLAS \cite{Aad:2015waa} and CMS \cite{CMS:2016khu}
collaborations in their studies at the LHC with $\sqrt{s}=7$ and
$\sqrt{s}=8$ TeV using the $\ell+$jets and the di-lepton top quark
decay channels respectively. The latter three cases are summarised in
Table \ref{binning}. In Figure \ref{5-ATLAS-CMS}, templates for the
{\it Full} case for five different top quark masses with the different
bin size are given assuming $\mu_0 = H_T /2$ and the CT14 set for
PDFs.   To emphasise the regions with the largest
sensitivity to the top quark mass the ratio to the result with the
default value of the top quark mass, $m_t=$ 173.2 GeV, is also shown. 
The top quark mass is determined by a comparison of the pseudo-data
with the theoretical predictions for different values of $m_t$ in
individual bins of the normalised $\rho_s$ distribution. The most
probable value of the top quark mass is extracted by means of the
$\chi_i^2$ distribution for each bin $i$. To be more precise, for each
bin the predicted theoretical values (cross sections) for different
$m_t$ are fitted using a second order polynomial function $f_i(m_t)$
in order to obtain a continuous distribution as a function of the top
quark mass. Example of such functions for the ATLAS and CMS intervals
are shown in Figure \ref{ATLASbinning} and Figure \ref{CMSbinning},
where $R_i(m_t^{pole})$ is defined as
\begin{equation}
R_i(m_t^{pole})= \int_{\rho_{i-1}}^{\rho_i} d\rho_s
\frac{1}{\sigma_{t\bar{t}j}} \frac{d\sigma_{t\bar{t}j}}{d\rho_s}
(m_t^{pole},\rho_s)\,, \quad \quad   \quad \quad \quad 
 i=1,\dots \,,
\end{equation}
with 
\begin{equation}
\begin{array}{lll}
\rho_i^{\rm ATLAS}= \left( 0, 0.25, 0.325, 0.425,0.525,
0.675,1\right)\,, && \quad \quad \quad 
i=0,\dots,6\,,\\[0.2cm]\\
\rho_i^{\rm CMS} = \left( 0, 0.2, 0.3,
0.45,0.6,1\right)\,, &&\quad \quad \quad 
 i=0,\dots,5\,.
\end{array}
\end{equation}
 Afterwards the $\chi_i^2$ distribution is
constructed according to the following formula
\begin{equation} \chi_i^2(m_t) = \frac{\left(N^{pseudo-data}_i
-f_i(m_t)\right)^2}{\left(\delta N_i^{pseudo-data}\right)^2}\,,
\end{equation}
where $f_i(m_t)$ represents the fit to the given theoretical
predictions in the bin $i$, $N^{pseudo-data}_i$ is the number of
the selected pseudo-data events in that bin and $\delta N_i^{pseudo-data}$
stands for statistical uncertainty of the pseudo-data in the bin $i$. The
$\chi^2_i$ distribution does not take into account the theoretical
uncertainties stemming from the scale variation and from the PDF
uncertainties, which are going to be treated as external variations as
described below. The global $\chi^2$ is calculated by simply summing
all bins since individual bins are not correlated
\begin{equation}
\chi^2 = \sum_{i=1}^{N-1} \chi^2_i(m_t)\,,
\end{equation}
where $N$ is the number of bins. The number of degrees of freedom is
reduced since one degree of freedom is used by the normalisation of
the theoretical distributions. As usual we expect that the numerator
of each term will be of the order of $\delta N_i^{pseudo-data}$, so
that each term in the sum will be of the order of unity. Hence a
sample value of $\tilde{\chi}^2 \equiv \chi^2/{d.o.f}$ should be
approximately equal to $1$.  If this is the case, we shall conclude
that our pseudo-data are well described by the values we have chosen
for the $f_i(m_t)$ functions. If our sample value of $\tilde{\chi}^2$
turns out to be much larger than $1$ we may conclude the opposite.
The resulting representative $\tilde{\chi}^2$ distributions with the
binning as proposed by the ATLAS and the CMS collaborations for both
cases of the integrated luminosity, i.e. $2.5$ fb${}^{-1}$ and $25$
fb${}^{-1}$, are shown in Figure \ref{chi2}. The position of the
minimum of the $\chi^2$ distribution is taken as the extracted top
quark mass value, $m_t^{out}$. The statistical uncertainty on the top
quark mass $\delta m_t^{out}$ is calculated in the standard way, i.e
as the $\pm1\sigma$ deviation from the minimum by applying the
$\chi^2+1$ variation. The sensitivity to the theoretical assumptions
and their uncertainties is assessed by using  one thousand
pseudo-data sets. Afterwards, the averaged $\tilde{\chi}^2$ and
$m_t^{out}$ are inferred and $\delta m^{out}_t$ is taken as $\pm 1\sigma$
deviation from the averaged $m_t^{out}$ by applying $68.3\%$ C.L. to
the following (sorted) spread $\left| m_t^{out}-m^i_t \right|$, where
$i=1,\dots,1000$ counts the pseudo-experiments.  Distributions of
the minimum $\tilde{\chi}^2$ and of the corresponding top quark mass from
the 1000 pseudo-experiments are presented in Figure
\ref{psedoexperiment2.5} and Figure \ref{psedoexperiment25}. Results
are shown for ${\cal L} =$ $2.5$ fb${}^{-1}$ and ${\cal L} =$ $25$
fb${}^{-1}$ respectively, and for the {\it Full} theory with
$\mu_R=\mu_F=\mu_0 = H_T /2$ and the CT14 PDF set. For a given
luminosity,  a higher number of bins corresponds to the better top quark
mass resolution and to a substantial decrease of the spread of the
$\tilde{\chi}^2_{min}$ values. Once the luminosity is increased, see
Figure \ref{psedoexperiment25}, the top quark mass resolution is also
improved as of course anticipated.  The improved resolution can be
used to make a more accurate determination of the top quark mass.

Theoretical uncertainties stemming from the scale
variation and various PDF parameterisations are included in the
following manner.  For each source of uncertainty the normalised
$\rho_s$ differential distribution with various top quark masses are
prepared replacing old template distributions with default setup, i.e.
with the $\mu_R=\mu_F=\mu_0$ and the CT14 PDF set. Thus, for each top
quark mass value considered we generate the following additional
normalised $\rho_s$ distributions
\begin{equation}
\rho_s(2\mu_0, {\rm CT14})\,, \quad \quad\rho_s(\mu_0/2, {\rm
  CT14})\,, \quad \quad
\rho_s(\mu_0, {\rm NNPDF3})\,, \quad \quad
\rho_s(\mu_0, {\rm MMHT14})\,.
\end{equation}
For each case, the $\chi^2$ distribution is calculated and the
corresponding top quark mass is inferred. The difference between the
central values of the new extracted top quark masses and $m_t^{out}$
as obtained from the default case, $\rho_s(\mu_0, {\rm CT14})$, is
taken as the systematic uncertainty. To be more precise, the
theoretical uncertainties are estimated according to the following
formulae
\begin{equation}
\begin{split}
\left( \Delta m_t^{out}\right)_\mu &=  \pm \max \Big\{ 
 \left|m_t^{out}\left(\frac{\mu_0}{2}, {\rm CT14}\right) 
-m_t^{out}(\mu_0, {\rm CT14})
 \right| , \\[0.2cm]
& \quad \quad \quad \quad \quad 
\left|m_t^{out}(2\mu_0, {\rm CT14})
 -m_t^{out}(\mu_0, {\rm CT14}) \right| \Big\}\,,
\end{split}
\end{equation}
\begin{equation}
\begin{split}
\left( \Delta m_t^{out}\right)_{\rm PDF} &= 
 \pm
\max \Big\{ 
 \left|m_t^{out}(\mu_0, {\rm MMHT14}) 
-m_t^{out}(\mu_0, {\rm CT14}) \right| ,\\[0.2cm]
&\quad \quad \quad \quad \quad 
 \left|m_t^{out}(\mu_0,{\rm NNPDF3})
 -m_t^{out}(\mu_0,{\rm CT14}) \right| \Big\}\,.
\end{split}
\end{equation}
To be more conservative the highest value from the two obtained is
chosen and symmetrisation is not utilised. Let us note
that the simultaneous variation of the renormalisation and
factorisation scales up and down by a factor of 2 around the central
value $\mu_0$ is motivated by our previous findings. In
Ref.~\cite{Bevilacqua:2016jfk} we have shown that the scale variation
for the process under consideration is fully driven by the changes in
$\mu_R$ independently of the scale choice. Let us additionally add
that scale variations are applied to the numerator and denominator of
the normalised distributions in a correlated way. 

At last, possible biases on the $m_t$ extraction have also been
examined by employing all theoretical descriptions to obtain the
pseudo-data sets. Thus, the {\it Full} case for three different scale
choices $\mu_0=m_t$, $\mu_0=H_T/2$ and $\mu_0=E_T/2$ as well as {\it
NWA} and {\it NWA}${}_{Prod.}$ for $\mu_0=m_t$ have been used not only
as templates but also for the pseudo-data generation.  In the end, the
value of the top quark mass obtained from the global $\chi^2$
distribution has been compared to the top quark mass of the
theoretical sample used as an input. The final results, assuming the
ATLAS binning, are presented in Figure \ref{fig:bias1} and Figure
\ref{fig:bias2} separately for ${\cal L}=$ $2.5$ fb${}^{-1}$ and 
${\cal L}=$ $25$ fb${}^{-1}$. A good agreement within the
corresponding statistical errors has been found for each top quark
mass, for all theoretical predictions and for both luminosity
cases. Neither a favoured value of the top quark mass nor a bias
towards a higher or a lower $m_t$ has been observed. Thus, we can
conclude that the method used is indeed unbiased.

%
%
\section{Numerical Results for $\boldsymbol{m_t}$ Based 
on the  Normalised  $\boldsymbol{\rho_s}$  Distribution}
%
%

%
\begin{table}[t!]
\begin{center}
\begin{tabular}{ccccc}
&&&&\\
\text{Theory, NLO  QCD}& $m^{out}_t \pm \delta m^{out}_t$
& Averaged & Probability &
$m_t^{in} -m_t^{out}$ \\ \text{CT14 PDF} & [GeV] &
$\chi^2/{\rm d.o.f.}$
& {\it p-value} & [GeV] \\
&&&&\\
\hline\hline
&&&&\\
&\quad \quad \quad \quad 
{\it 31 bins} &&&\\
\hline\hline
{\it Full}, $\mu_0=H_T/2$ &173.38 $\pm$ 1.34 
& 1.04 & 0.40 (0.8$\sigma$) & $-0.18$\\
{\it Full}, $\mu_0=E_T/2$ & 172.84 $\pm$ 1.33 & 1.05 
& 0.39 (0.9$\sigma$) & $+0.36$\\
{\it Full},  $\mu_0=m_t$ &  174.11 $\pm$ 1.39 & 1.07 
& 0.37 (0.9$\sigma$) & $-0.91$\\
\hline
{\it NWA},  $\mu_0=m_t$ & 175.70 $\pm$ 0.96 & 1.17 
& 0.24 (1.2$\sigma$)&$-2.50$\\
{\it NWA}${}_{Prod.}$, $\mu_0=m_t$ & 169.93 $\pm$ 0.98 & 1.20 &
 0.20  (1.3$\sigma$)& $+3.27$\\
\hline\hline
&&&&\\
&\quad \quad \quad \quad   {\it 5 bins}&&&\\
\hline\hline
{\it Full}, $\mu_0=H_T/2$ & 173.15 $\pm$ 1.32   &0.93  
& 0.44 (0.8$\sigma$)& $+0.05$ \\
{\it Full}, $\mu_0=E_T/2$ & 172.55 $\pm$ 1.18 & 1.07  
& 0.37 (0.9$\sigma$) & $+0.65$ \\
{\it Full},  $\mu_0=m_t$ &  173.92 $\pm$ 1.38 &1.48 
& 0.20  (1.3$\sigma$) & $-0.72$ \\
\hline
{\it NWA},  $\mu_0=m_t$ & 175.54 $\pm$ 0.97 & 1.38 
& 0.24 (1.2$\sigma$) & $-$2.34\\
{\it NWA}${}_{Prod.}$, $\mu_0=m_t$ & 169.37  $\pm$ 1.43  &1.16  
& 0.33  (1.0$\sigma$)& $+$3.83\\
\hline\hline
&&&&\\
& \quad \quad \quad \quad  {\it ATLAS binning}&&&\\
\hline\hline
{\it Full}, $\mu_0=H_T/2$& 173.05 $\pm$ 1.31 & 0.99 
&0.42 (0.8$\sigma$)& $+0.15$\\
{\it Full}, $\mu_0=E_T/2$ &  172.19 $\pm$ 1.34 & 1.05
 & 0.39 (0.9$\sigma$) &$+1.01$\\
{\it Full}, $\mu_0=m_t$ & 173.86 $\pm$ 1.39   & 1.42 
& 0.21 (1.2$\sigma$) & $-0.66$\\
\hline
{\it NWA},  $\mu_0=m_t$ & 175.22 $\pm$ 1.15 &
1.38 & 0.23 (1.2$\sigma$)&  $-2.02$\\
{\it NWA}${}_{Prod.}$, $\mu_0=m_t$ & 169.39  $\pm$ 1.46  &
1.12 & 0.35 (0.9$\sigma$) &  $+3.81$\\
\hline
\hline
&&&&\\
&\quad \quad \quad \quad  {\it CMS binning}&&&\\
\hline\hline
{\it Full}, $\mu_0=H_T/2$&173.09 $\pm$ 1.53 & 0.94 
& 0.44 (0.8$\sigma$) & $+0.11$\\
{\it Full}, $\mu_0=E_T/2$ & 172.20 $\pm$ 1.54 &0.96 
& 0.43 (0.8$\sigma$) & $+1.00$\\
{\it Full}, $\mu_0=m_t$ & 173.94 $\pm$ 1.49  &1.42 
& 0.22 (1.2$\sigma$) & $-0.74$\\
\hline
{\it NWA},  $\mu_0=m_t$ & 175.66 $\pm$ 1.10 &
1.42 & 0.22 (1.2$\sigma$)& $-2.46$\\
{\it NWA}${}_{Prod.}$, $\mu_0=m_t$ & 169.96  $\pm$ 1.80  &
1.00 & 0.41  (0.8$\sigma$)& $+3.24$\\
\end{tabular}
 \end{center}
\caption{\it 
Mean value of the top quark mass, $m_t^{out}$, from 1000
pseudo-experiments as obtained from the normalised $\rho_s$
differential distribution for the $pp\to e^+\nu_e \mu^-\bar{\nu}_\mu
b\bar{b}j +X$ production process at the LHC with $\sqrt{s} =$ 13 TeV.
Also shown is 68$\,\%$ C.L. (1$\sigma$) statistical error of the top
quark mass, $\delta m_t^{out}$, together with the averaged minimal  
$\chi^2/d.o.f$ and the p-value.  The number of standard deviations
corresponding to each p-value is presented in parentheses. In the last
column the top quark mass shift, defined as $m_t^{in} -m_t^{out}$,
with $m_t^{in} =$ 173.2 GeV, is also given.  Luminosity of 2.5
$fb^{-1}$ is assumed.}
\label{tab:1}
\end{table}
\begin{table}[t!]
\begin{center}
\begin{tabular}{ccccc}
&&&&\\ \text{Theory, NLO  QCD}& $m^{out}_t \pm \delta m^{out}_t$
& Averaged & Probability &
$m_t^{in} -m_t^{out}$ \\ \text{CT14 PDF} & [GeV] &
$\chi^2/{\rm d.o.f.}$
& {\it p-value} & [GeV] \\
&&&&\\
\hline\hline 
&&&&\\
&\quad \quad \quad  \quad {\it 31 bins}&&&\\
\hline\hline
{\it Full}, $\mu_0=H_T/2$ & 173.09 $\pm$ 0.42 & 1.04 
& 0.41 (0.8$\sigma$) & $+0.11$\\
{\it Full}, $\mu_0=E_T/2$ &  172.45  $\pm$ 0.39  & 1.12 
& 0.30 (1.0$\sigma$) & $+0.75$\\
{\it Full}, $\mu_0=m_t$ & 173.76 $\pm$ 0.40  &1.87 
&0.003 (3.0$\sigma$) & $-0.56$\\
\hline
{\it NWA},  $\mu_0=m_t$ & 175.65 $\pm$ 0.31 & 2.99 
& $7\cdot 10^{-8}$ (5.4$\sigma$) & $-2.45$\\
{\it NWA}${}_{Prod.}$, $\mu_0=m_t$ & 169.59 $\pm$ 0.30 
& 3.10 &$2\cdot 10^{-8}$ (5.6$\sigma$)& $+3.61$\\
\hline\hline
&&&&\\
&\quad \quad \quad \quad  {\it 5 bins} &&&\\
\hline\hline
{\it Full}, $\mu_0=H_T/2$ & 173.08 $\pm$ 0.40 &0.94  
&  0.44  (0.8$\sigma$)& $+0.12$\\
{\it Full}, $\mu_0=E_T/2$ & 172.48  $\pm$ 0.38 &1.58 
&  0.18  (1.3$\sigma$) &$+0.72$ \\
{\it Full}, $\mu_0=m_t$ & 173.75  $\pm$ 0.40  &6.76 
&  $2\cdot 10^{-5}$  (4.3$\sigma$) &  $-0.55$\\
\hline
{\it NWA},   $\mu_0=m_t$ & 175.49 $\pm$ 0.30 &
5.31 &  $2\cdot 10^{-4}$ (3.7$\sigma$) & $-2.29$\\
{\it NWA}${}_{Prod.}$, $\mu_0=m_t$ & 169.39  $\pm$ 0.47 &
3.42 &  $8\cdot 10^{-3}$ (2.6$\sigma$) & $+$3.81\\
\hline\hline
&&&&\\
&\quad \quad \quad \quad {\it ATLAS binning} &&&\\
\hline\hline
{\it Full}, $\mu_0=H_T/2$ &  173.06 $\pm$ 0.44 & 0.97 
& 0.44 (0.8$\sigma$) & $+0.14$\\
{\it Full}, $\mu_0=E_T/2$ & 172.36 $\pm$ 0.44 & 1.38 
& 0.23 (1.2$\sigma$) & $+0.84$\\
{\it Full}, $\mu_0=m_t$ & 173.84 $\pm$ 0.42 & 5.12 
& $1\cdot 10^{-4}$ (3.9$\sigma$)& $-0.64$\\
\hline
{\it NWA},  $\mu_0=m_t$ & 175.23 $\pm$ 0.37 &
5.28 & $7\cdot 10^{-5}$ (4.0$\sigma$) &   $-2.03$ \\
{\it NWA}${}_{Prod.}$, $\mu_0=m_t$ & 169.43  $\pm$ 0.50 &
2.61 & 0.02 (2.3$\sigma$) &  $+3.77$\\
\hline\hline
&&&&\\
&\quad \quad \quad  \quad  {\it CMS binning} &&&\\
\hline\hline
{\it Full}, $\mu_0=H_T/2$ & 173.09 $\pm$ 0.50 & 0.96 
& 0.43 (0.8$\sigma$)& $+0.11$\\
{\it Full}, $\mu_0=E_T/2$ & 172.22 $\pm$ 0.48 & 1.32 
& 0.26 (1.1$\sigma$)& $+0.98$\\
{\it Full}, $\mu_0=m_t$ & 174.02 $\pm$ 0.46 &6.57 
& $3 \cdot 10^{-5}$ (4.2$\sigma$) & $-0.82$\\
\hline
{\it NWA},  $\mu_0=m_t$ & 175.74 $\pm$ 0.34 &
6.00 & $8\cdot 10^{-5}$ (3.9$\sigma$) & $-2.54$\\
{\it NWA}${}_{Prod.}$, $\mu_0=m_t$ & 170.22  $\pm$ 0.53 &
2.19 & 0.07 (1.8$\sigma$) & $+2.98$\\
\end{tabular}
 \end{center}
\caption{\it 
Mean value of the top quark mass, $m_t^{out}$, from 1000
pseudo-experiments as obtained from the normalised $\rho_s$
differential distribution for the $pp\to e^+\nu_e \mu^-\bar{\nu}_\mu
b\bar{b}j +X$ production process at the LHC with $\sqrt{s}=$ 13
TeV. Also shown is 68$\,\%$ C.L. (1$\sigma$) statistical error of the
top quark mass, $\delta m_t^{out}$, together with the averaged minimal 
$\chi^2/d.o.f$ and the p-value.  The number of standard deviations
corresponding to each p-value is presented in parentheses. In the last
column the top quark mass shift, defined as $m_t^{in} -m_t^{out}$,
with $m_t^{in} =$ 173.2 GeV, is also given.  Luminosity of 25
$fb^{-1}$ is assumed.}
\label{tab:2}
\end{table}

Our findings for the top quark mass, as determined from the normalised
$\rho_s$ distribution using the methods described in the previous
section, are summarised in Table \ref{tab:1} and Table
\ref{tab:2}. They are obtained for the integrated luminosity of ${\cal
L} =2.5$ fb${}^{-1}$ and ${\cal L} =25$ fb${}^{-1}$ at the LHC with 
$\sqrt{s}=13$ TeV. We show the mean value of the top quark mass as
collected from the 1000 pseudo-experiments, $m_t^{out}$, the $68\,\%$
C.L. (1$\sigma$) statistical error on the top quark mass, $\delta
m_t^{out}$, and the averaged minimal $\chi^2/d.o.f$.  The significance
of a discrepancy between the pseudo-data and what one expects under
the assumption of particular theoretical description is quantified by
giving the probability value, the {\it p-value}. The latter is defined
as the probability to find $\chi^2$ in the region of equal or lesser
compatibility with the theory in question than the level of
compatibility observed with the pseudo-data. Thus, in Table
\ref{tab:1} and Table \ref{tab:2} the {\it p-value} is also provided
together with the corresponding number of standard deviations, which
is shown in parentheses. Let us note at this point, that the smaller
the {\it p-value} the larger the significance because it tells us that
the theoretical description under consideration might not adequately
describe the pseudo-data.  We would normally start to question the
theoretical description employed only if we were to have found the
{\it p-value} smaller than $0.0455$ (larger than $2\sigma$).  If the
{\it p-value} is larger than $0.0455$ (smaller than $2\sigma$) then we
assume that the pseudo-data are consistent with the theoretical
approach used to model the process under consideration. Results with
{\it p-value} smaller than $0.0027$ (larger than $3\sigma$) can be
considered to be disfavoured by the pseudo-data. Finally, in Table
\ref{tab:1} and Table \ref{tab:2} we also give the top quark mass
shift, defined as $m_t^{in} -m_t^{out}$, where $m_t^{in}=173.2$ GeV.

We start with results for ${\cal L=}$ 2.5 fb${}^{-1}$ that are
collected in Table \ref{tab:1}. The first thing that we can notice is
an overall agreement, within $0.8\sigma-1.3\sigma$, between various
theoretical descriptions and the pseudo-data. Moreover, for all
considered cases, the averaged minimal $\chi^2/d.o.f$ is of the order
of $1$. However, depending on the theory at hand, various mass shifts
are observed. For the {\it Full} case, independently of the bin size
and the scale choice, a difference from $m_t^{in}$ up to $1$ GeV can
be identified. On the other hand, a shift of $2.0-2.5$ GeV is visible
for {\it NWA}. Should we use the {\it Full} case with the fixed scale
$\mu_0=m_t$ for the generation of the pseudo-data instead, the shift
of $1.2-2.0$ GeV would rather be seen for {\it NWA}. However, the
statistical uncertainty $\delta m^{out}_t$ is still quite high for
this case, that is of the order of $1$ GeV.   For the
higher luminosity case, that we shall present in the next step,
despite the diminished quality of the $\chi^2$ fit the mass shifts will
persist. They are again up to $2.5$ GeV $(2.0 ~{\rm GeV})$ for {\it
  NWA} with $\mu_0=m_t$ when pseudo-data are generated from the  {\it
  Full} case with $\mu_0=H_T/2$ $(\mu_0=m_t)$. In that case $\delta
m^{out}_t$ is of the order of $0.3-0.4$ GeV only. Thus, the off-shell
effects and non-resonant contributions of the top quark and the $W$
gauge boson are not negligible for the top quark mass extraction from
the ${\cal R}(m_t^{pole},\rho_s)$ observable.  For the last case
considered, that is {\it NWA}${}_{Prod.}$, a substantial deviation
from $m_t^{in}$ of the order of $3.2-3.8$ GeV is observed, which can
be explained by substantial shape differences of the normalised
$\rho_s$ distribution in the regions sensitive to $m_t$.  The large
mass shifts for two NWA cases suggest that for ${\cal
R}(m_t^{pole},\rho_s)$ the full theory description for the $pp\to
\ell^+ \nu_\ell \ell^- \bar{\nu}_\ell b\bar{b}j$ production process is
indeed required.  Additionally, when examining {\it Full} and {\it
NWA} cases closer, for example for the same scale choice, that is for
$\mu_0=m_t$, we can see that the statistical uncertainty of
$m_t^{out}$ is always higher in the former case. This suggests
underestimation of $\delta m_t^{out}$ by about $20\%-45\%$ in the case
of {\it NWA}.  Let us remind here, that the {\it NWA}${}_{Prod.}$ case
is far from complete theory since only higher order corrections to
on-shell top quark pair production with one hard jet are
incorporated. Thus, we mostly show this case for reasons of comparison
and to underline the importance of QCD corrections and jet radiation
in top quark decays. Moreover, let us stress here, that in ATLAS and
CMS experimental analyses \cite{Aad:2015waa,CMS:2016khu}, the on-shell
$t\bar{t}j$ production process calculated at NLO in QCD is combined
with a parton shower. Top quark decays are treated in the parton
shower approximation omitting $t\bar{t}$ spin correlations. However,
the shower programs include higher-order corrections to the hard
subprocess in an approximate way by including the leading-logarithmic
contributions to all orders.  These dominant contributions are
associated with collinear parton splittings or soft gluon
emissions. Additionally, the parton shower approximation takes into
account not only the collinear enhanced real parton emissions at each
order in perturbation theory but also, by unitarity, virtual effects
of the same order. Such effects are included in the probability of not
splitting during evolution from one scale to the other encoded in the
Sudakov form factor. Finally, top quark decays in standard shower
programs are not based on a strict NWA, but rather obey a Breit-Wigner
distribution that should account for the dominant off-shell effects.
Therefore, NLO plus parton shower results are better approximations of
{\it NWA} rather than of {\it NWA}${}_{Prod.}$.  Nevertheless, in
Ref. \cite{Aad:2015waa,CMS:2016khu}, such predictions are first tuned
to data and afterwards unfolded back to the parton level to obtain the
on-shell top quarks, that are used to construct ${\cal
R}(m^{out}_t,\rho_s)$.

In the next step, we concentrate on results obtained for increased
integrated luminosity of ${\cal L}=$ 25 fb${}^{-1}$, which are
summarised in Table \ref{tab:2}. First, as expected, the statistical 
uncertainty $\delta m_t^{out}$ decreases with the square root of
luminosity. Secondly, our conclusions about the top quark mass shift
derived for ${\cal L}=$ 2.5 fb${}^{-1}$ are not altered. 
Thirdly, underestimation of the statistical uncertainties
on $m_t$ by the {\it NWA} case can still be observed. Here this effect amounts
to $15\%-35\%$.  However, unlike  the case of low integrated
luminosity, for ${\cal L=}$ 25 fb${}^{-1}$ sensitivity to the various
theoretical predictions is clearly visible. This can be best observed
in the changes of ${\chi^2/d.o.f}$ and the {\it p-value}.  The
pseudo-data are properly described only by {\it Full} either with
$\mu_R=\mu_F= \mu_0=H_T/2$ ($0.8\sigma$) or with
$\mu_R=\mu_F=\mu_0=E_T/2$ ($1.0\sigma-1.3\sigma$). The best 
agreement in the former case is rather trivial. {\it Full} with
$\mu_0=H_T/2$ will always work since it is used to obtain our
pseudo-data sets. Less trivial is the fact that also {\it Full} with
$\mu_0=E_T/2$ performs very well at least with the integrated
luminosity at hand. This is due to the fact that these two scales
provide very similar results. On the other hand, independently of the
bin size the {\it NWA} case is disfavoured at the $4\sigma-5\sigma$
level and {\it NWA}${}_{Prod.}$ at the $2\sigma-5\sigma$ level.  Even
for the {\it Full} case with $\mu_0=m_t$ discrepancies at the level of
$3\sigma-4\sigma$ are observed. The latter finding underlines the fact
that when differential cross sections are employed not only the full
off-shell effects and non-resonant background contributions of the top
quark and $W$ gauge boson but also the scale choices play an important
role. We note here, that a higher number of bins that corresponds to
increased sensitivity to $m_t$, helps to clearly distinguish between
the case where the theory (still) agrees with the pseudo-data and the
one where the theory is disfavoured by such pseudo-data.
 
In the following, systematic uncertainties on $m_t^{out}$ are
examined. They are estimated based on the full theory because
ultimately only this description should be used for the normalised
$\rho_s$ distribution.  Our findings are luminosity independent.
However, as expected, in the case of the scale dependence they depend
on the scale choice. Additionally they vary with the bin size
used. For $\mu_0=H_T/2$ and $\mu_0=E_T/2$ theoretical uncertainties
stemming from the scale variation have been estimated to be of the
order of $0.6\,{\rm GeV}-1.2\,{\rm GeV}$, whereas for $\mu_0=m_t$ they
are larger of the order of $2.1\, {\rm GeV}-2.8\,{\rm GeV}$. The
smallest values are obtained for the case of the largest number of
bins of equal size. As mentioned before the theoretical uncertainties,
as obtained from the scale dependence of the templates, are not the
only source of systematic uncertainties. Another source comes from
various PDF parameterisations. Here, quite uniform uncertainties in
the range of $0.4 \,{\rm GeV} - 0.7\, {\rm GeV}$ have been
obtained. Thus, PDF uncertainties on $m_t$ for the process under
scrutiny are well below the theoretical uncertainties due to scale
dependence, which remain the dominant source of the theoretical
systematics on the top quark mass extraction.

%
\section{Comparison to the $\boldsymbol{\rho^\prime_s}$ 
Observable}
%

%
\begin{figure}[t!]
\begin{center}
\includegraphics[width=1.0\textwidth]{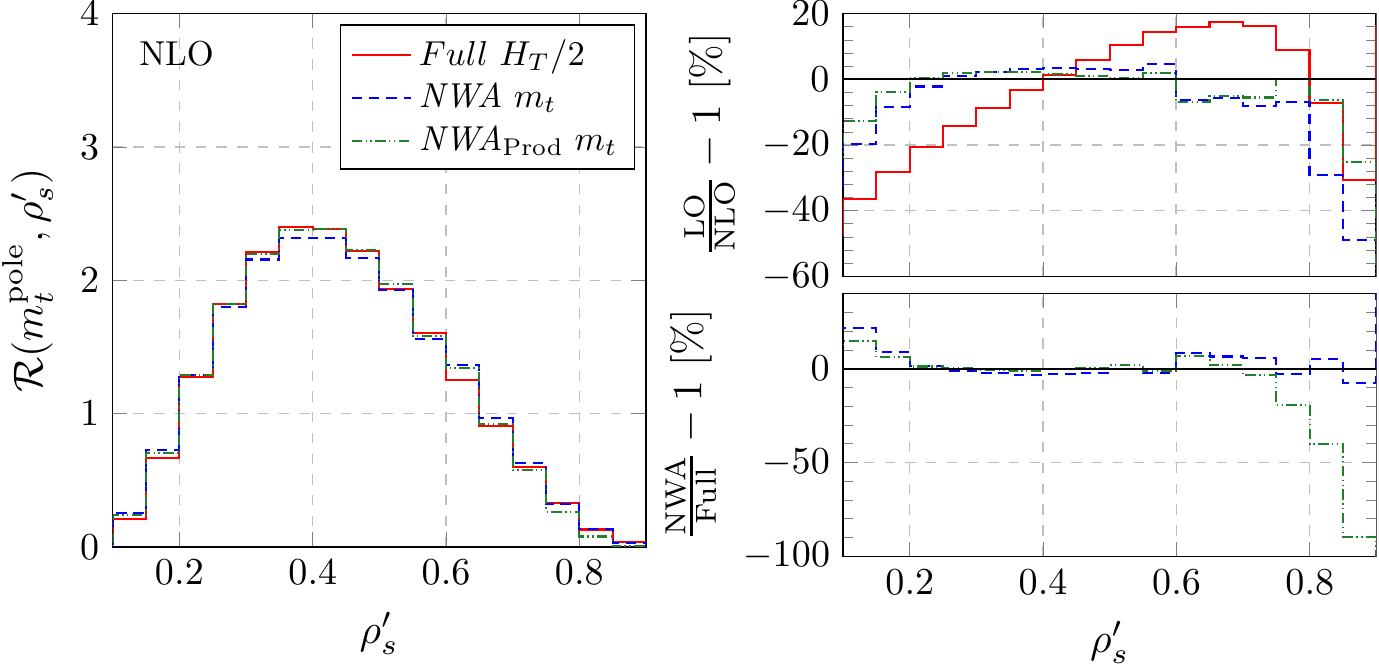}
\includegraphics[width=0.7\textwidth]
{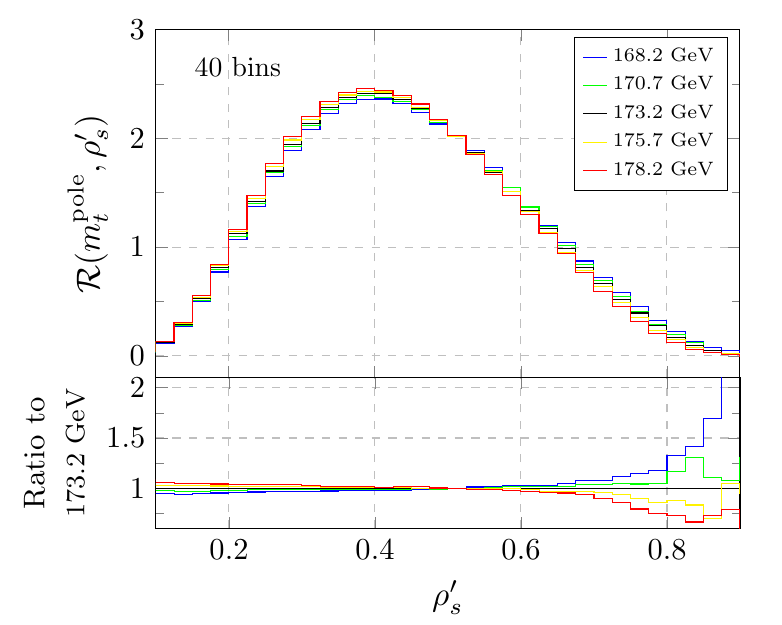}
\end{center}
\caption{\it Normalised $\rho^\prime_s$ distribution at NLO QCD for
the $pp\to e^+\nu_e \mu^-\bar{\nu}_\mu b\bar{b}j +X$ production
process at the LHC with $\sqrt{s} =$ 13 TeV. The CT14 PDF set is
used. Upper panel: three different theoretical descriptions with
$m_t=$ 173.2 GeV and $\mu_R =\mu_F =\mu_0$, where $\mu_0=m_t$ for both
NWA cases and $\mu_0=H_T/2$ for the Full case. Also given are the
relative size of NLO QCD corrections and the combined relative size of
finite-top-width and finite-W-width effects for the normalised
$\rho^\prime_s$ observable.  Lower panel: the Full case for five
different top quark masses for $\mu_0=H_T/2$. Also
plotted is the ratio to the result with the default value of $m_t$,
i.e. $m_t =$ 173.2 GeV.} 
\label{fig:2observable}
\end{figure}

We also examine a slightly modified version of the normalised $\rho_s$
distribution dubbed $\rho_s^\prime$. The difference between $\rho_s$
and $\rho_s^\prime$ comes from the second hard jet. Namely, if a
second (leading-order) jet is resolved, it is added to the invariant
mass of the $t\bar{t}j$ system. The upper part of Figure
\ref{fig:2observable} presents the ${\cal
R}(m_t^{pole},\rho^\prime_s)$ differential distribution for three
considered cases, {\it Full}, {\it NWA} and {\it NWA}${}_{Prod.}$. The
renormalisation and factorisation scales are set to the common value
$\mu_0$, where $\mu_0=m_t$ for both {\it NWA} cases and $\mu_0=H_T/2$
for the {\it Full} case. Also shown are the relative NLO QCD
corrections $\sigma^{\rm LO}/\sigma^{\rm NLO} -1$ and the relative
deviation of the NWA results from the full calculation.  On the other
hand, in the lower part of Figure \ref{fig:2observable} the dependence
on $m_t$ of ${\cal R}(m_t^{pole},\rho^\prime_s)$ is provided. We show
the best theoretical prediction, {\it Full}, with
$\mu_R=\mu_F=\mu_0=H_T/2$ for five different top quark masses. We can
see that the beginning of the spectrum is mostly affected, because the
additional hard jet essentially modifies the tails of
$M_{t\bar{t}j}$. Moreover, the peak of the distribution is shifted
towards smaller values of $\rho^\prime_s$. The magnitude and sign of
higher order corrections for {\it NWA} and {\it NWA}$_{Prod.}$ have
also changed for $\rho_s^\prime < 0.3$.  For the {\it Full} case we
have obtained shape differences from $-40\%$ to $+20\%$, which once
again underline the importance of the inclusion of higher-order
corrections. Nevertheless, a large impact on the top quark mass extraction
is not expected since the highest sensitivity falls in the range of
high values of $\rho^\prime_s$ as can be deduced from the lower part
of Figure \ref{fig:2observable} where a similar dependence on $m_t$ as
in the case of $\rho_s$ is visible. For low values of $\rho_s^\prime$
we observe differences up to $+20\%$ for {\it NWA} and $+15\%$ for
{\it NWA}$_{Prod.}$, however, they are present only around $\rho_s^\prime
\approx 0.1$ where the dependence on $m_t$ is diminished. Our findings on
$m_t^{out}$ can be summarised as follows. For the low luminosity case
all extracted top quark masses are at most $1\sigma$ away from the
corresponding values obtained with the help of $\rho_s$. As a
consequence, a similar size of the top quark mass shifts is noted.  We
should stress here that the quality of $\chi^2/d.o.f.$ is worsened for
all but the {\it Full} case with $\mu_0=H_T/2$ and $\mu_0=E_T/2$. For
${\cal L=}$ 25 fb${}^{-1}$ only these two cases should be employed
since other theoretical approaches are disfavoured beyond $5\sigma$
level. The main reason for which the previously defined variable
$\rho_s$ should be used instead of $\rho_s^\prime$, however, is the
size of theoretical uncertainties. Once $\rho_s^\prime$ is used the
theoretical uncertainties due to the scale dependence are driven by
the leading order scale dependence of the second hard jet and a
significant increase is observed. Namely, they amount to $3 \,{\rm
GeV}-4\,{\rm GeV}$ for the {\it Full} case with $\mu_0=H_T/2$ or
$\mu_0=E_T/2$. On the other hand, the magnitude of the PDF
uncertainties is the same.

%
\section{Comparison to Other Observables}
\label{section:otherobservables}
%

\begin{figure}[t!]
\begin{center}
\includegraphics[width=1.0\textwidth]{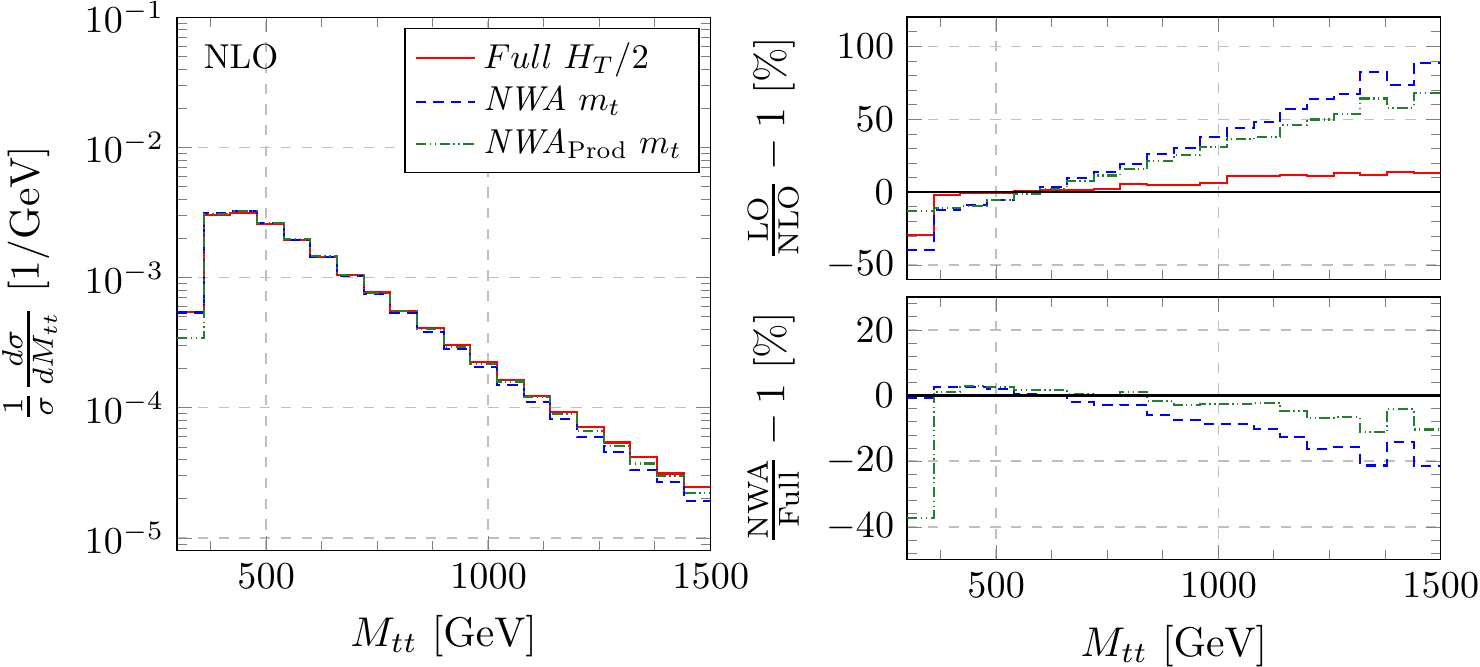}
\includegraphics[width=0.7\textwidth]
{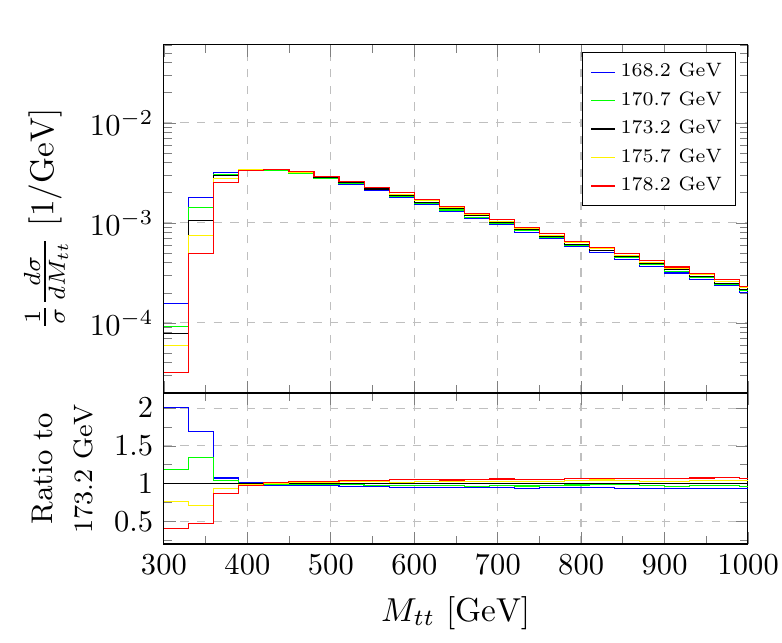}
\end{center}
\caption{\it  
Normalised $M_{t\bar{t}}$ distribution at NLO QCD for the $pp\to
e^+\nu_e \mu^-\bar{\nu}_\mu b\bar{b}j +X$ production process at the
LHC with $\sqrt{s} =$ 13 TeV. The CT14 PDF set is used. Upper panel:
three different theoretical descriptions with $m_t=$ 173.2 GeV and
$\mu_R =\mu_F =m_0$, where $\mu_0 = m_t$ for both NWA cases and $\mu_0
= H_T /2$ for the Full case. Also given are the relative size of NLO
QCD corrections and the combined relative size of finite-top-width and
finite-W-width effects for the normalises $M_{t\bar{t}}$ observable.
Lower panel: the Full case for five different top quark masses for
$\mu_0 = H_T/2$. Also plotted is the ratio to the result
with the default value of $m_t$, i.e. $m_t =$ 173.2 GeV.} 
\label{fig:3observable}
\end{figure}

In the following, we shall focus on examining the sensitivity of the
normalised $\rho_s$ distribution by comparing it to other observables
sensitive to $m_t$ in the $pp\to t\bar{t}j$ production process.  Since
$\rho_s$ is defined as the (inverse) invariant mass of the $t\bar{t}$
plus additional hard jet system the natural observable to start with
is the invariant mass of the top anti-top pair alone. In such a way we
can assess the impact of the additional hard jet on the $m_t$
extraction. The normalised $M_{t\bar{t}}$ differential distribution is
presented in Figure \ref{fig:3observable} together with its dependence
on $m_t$. For the top quark mass study a range up to $1$ TeV has only
been used, the reason being the minimised difference between {\it
Full} and {\it NWA} in this region.  Furthermore, high energy regions
are not only sensitive to electroweak corrections but also could be
potentially diluted by new, not yet discovered, heavy resonances
decaying to $t\bar{t}$ final states. Our findings are summarised in
Table \ref{tab:mtt}. For the same luminosity, the $M_{t\bar{t}}$
differential distribution seems to perform better than $\rho_s$
yielding statistical uncertainties a factor of $2$ to $2.4$
smaller. More importantly, the shift of the top quark mass,
$m_t^{in}-m_t^{out}$, is greatly reduced. For {\it Full} and {\it NWA}
it is below or of the order of $0.5$ GeV and for the {\it
NWA}${}_{Prod.}$ case it is equal to $1.8$ GeV. For a measurement not
only the sensitivity is important but also the reliability of the
observable used. For example in the case of $M_{t\bar{t}}$ a good
sensitivity is achieved mostly due to a few first bins.  In this
extreme threshold regime, however, theoretical predictions would
require to go beyond fixed order perturbation theory resumming 
threshold effects and soft gluon emissions. Such studies have been
carried out for the invariant mass distribution of the on-shell top
quarks in the $t\bar{t}$ production process at the LHC with
$\sqrt{s}=14$ TeV \cite{Kiyo:2008bv}. From that study one can conclude
that in the threshold region the enhancement of the cross section
amounts to roughly a factor $3$, additionally a significant shift of
the threshold is observed. Compared to the inclusive fixed order NLO
total cross section for $t\bar{t}$ production with $\mu_0=m_t$,
however, the increase is relatively small, of the order of $1\%$
only. In principle the shape of the differential distribution
$d\sigma/dM_{t\bar{t}}$ could be distorted in the threshold region,
which in turn can affect the mean value of $m_t^{out}$ and shift it
towards smaller values. In practise, however, one needs to realise
here, that the size of the region where these effects can be visible
is of the order of $1$ GeV only.  Therefore a very fine resolution
would be required to get sensitivity to the threshold effects. In our
studies such effects are incorporated into one bin of $30$ GeV
size. We have even checked a larger bin size of $60$ GeV (smaller
number of bins) and confirmed that our findings on $m_t^{out}$
extraction are unchanged as can be seen from Table
\ref{tab:mtt}. Therefore, we conclude that we are below any
sensitivity to such threshold effects since they are completely
washed  out by our $M_{t\bar{t}}$ resolution. Consequently,
$M_{t\bar{t}}$ can be safely used. We are not aware of similar studies
for the on-shell $t\bar{t}j$ production at the LHC.  Thus, it is still
not clear to which extent the normalised $\rho_s$ distribution can be
affected by initial state radiation as well as bound state corrections
for $\rho_s \approx 1$. The theoretical uncertainties of $m_t^{out}$
based on $M_{t\bar{t}}$ stemming from the scale variation are
estimated to be of the order of $1.3$ GeV, thus slightly larger than
in the case of $\rho_s$. On the other hand, by comparison the PDF
uncertainties are negligible, at the level of $0.1$ GeV. One more
time, we can observe that for ${\cal L=}$ 25 fb${}^{-1}$ only the {\it
Full} theory adequately describes the pseudo-data. We would like to
stress here, that the $M_{t\bar{t}}$ observable, similarly as the
normalised $\rho_s$ distribution, allows to unfold the (real) data to
the perturbative partonic level uniquely linking the $m_t^{out}$ value
to the top quark mass from the SM Lagrangian. 
%
\begin{table}[t!]
\begin{center}
\begin{tabular}{ccccc}
&&&&\\
\text{Theory, NLO  QCD}& $m^{out}_t \pm \delta m^{out}_t$
& Averaged & Probability &
$m_t^{in} -m_t^{out}$ \\ \text{CT14 PDF} & [GeV] &
$\chi^2/{\rm d.o.f.}$
& {\it p-value} & [GeV] \\
&&&&\\
\hline\hline
&&&&\\
&\quad \quad   \quad {\it 25 bins} $@$ 2.5 fb${}^{-1}$
  &&&\\
\hline\hline
{\it Full}, $\mu_0=H_T/2$&  173.24 $\pm$  0.55 &
1.03  & 0.42 (0.8$\sigma$) & $-0.04$\\
{\it Full}, $\mu_0=E_T/2$& 173.11 $\pm$ 0.54 &
1.04  & 0.41 (0.8$\sigma$) & $+0.09$\\
{\it Full}, $\mu_0=m_t$& 173.52 $\pm$ 0.57 
& 1.08 & 0.36 (0.9$\sigma$) & $-0.32$\\
\hline
{\it NWA},  $\mu_0=m_t$& 173.72 $\pm$ 0.51  & 
1.17 &  0.25 (1.1$\sigma$) & $-0.52$\\
{\it NWA}$_{\rm Prod.}$, $\mu_0=m_t$& 171.36 $\pm$ 0.48 &
1.62 &  0.03 (2.2$\sigma$) & $+1.84$\\
\hline\hline
&&&&\\
&\quad \quad \quad  {\it 25 bins} $@$ 25 fb${}^{-1}$  
&&&\\
\hline\hline
{\it Full}, $\mu_0=H_T/2$& 173.18  $\pm$ 0.18 &
1.03  & 0.42 (0.8$\sigma$) & $+0.02$\\
{\it Full}, $\mu_0=E_T/2$& 173.10 $\pm$ 0.17 &
1.11 & 0.32 (1.0$\sigma$) & $+0.10$\\
{\it Full}, $\mu_0=m_t$& 173.50 $\pm$ 0.16 &
1.87 & 0.006 (2.7$\sigma$) & $-0.30$\\
\hline
{\it NWA},  $\mu_0=m_t$& 173.73 $\pm$ 0.17 &
3.03 & $8\cdot 10^{-7}$ (4.9$\sigma$) & $-0.53$\\
{\it NWA}$_{\rm Prod.}$, $\mu_0=m_t$& 171.39 $\pm$ 0.15 &
7.84 & 0 ($\gg 5\sigma$) & $+1.81$\\
\hline\hline
&&&&\\
&\quad \quad  \quad   {\it 12 bins} $@$ 2.5 fb${}^{-1}$  
&&&\\
\hline\hline
{\it Full}, $\mu_0=H_T/2$&173.21 $\pm$ 0.52  & 1.01 
& 0.43 $(0.8\sigma)$ & $-0.01$\\
{\it Full}, $\mu_0=E_T/2$&  173.10 $\pm$ 0.55 & 1.02 
& 0.42 $(0.8\sigma)$ & $+0.10$\\
{\it Full}, $\mu_0=m_t$& 173.48 $\pm$ 0.54 & 1.13 
& 0.33 $(1.0\sigma)$ & $-0.28$\\
\hline
{\it NWA},  $\mu_0=m_t$&  173.67 $\pm$ 0.51 & 1.23
& 0.26 $(1.1\sigma)$ & $-0.47$\\
{\it NWA}$_{\rm Prod.}$, $\mu_0=m_t$& 170.92 $\pm$ 0.47 & 1.31
& 0.21  $(1.2\sigma)$ & $+2.28$\\
\hline\hline
&&&&\\
&\quad \quad \quad   {\it 12 bins} $@$ 25 fb${}^{-1}$  
&&&\\
\hline\hline
{\it Full}, $\mu_0=H_T/2$& 173.17 $\pm$ 0.17  & 1.00 
& 0.45 $(0.7\sigma)$ & $+0.03$\\
{\it Full}, $\mu_0=E_T/2$& 173.10 $\pm$ 0.17 & 1.18 
& 0.30 $(1.0\sigma)$ & $+0.10$\\
{\it Full}, $\mu_0=m_t$& 173.51 $\pm$ 0.17 & 2.72 
& 0.002 $(3.1\sigma)$& $-0.31$\\
\hline
{\it NWA},  $\mu_0=m_t$& 173.67 $\pm$ 0.16 & 3.69 
& $3\cdot 10^{-5}$  $(4.2\sigma)$& $-0.47$\\
{\it NWA}$_{\rm Prod.}$, $\mu_0=m_t$&170.96 $\pm$ 0.16 & 4.36 &
$1\cdot  10^{-6}$ $(4.9\sigma)$ & $+2.24$\\
\end{tabular}
 \end{center}
\caption{\it 
Mean value of the top quark mass, $m_t^{out}$, from 1000
pseudo-experiments as obtained from the normalised $M_{t\bar{t}}$
distribution for the $pp\to e^+\nu_e \mu^-\bar{\nu}_\mu b\bar{b}j +X$
production process at the LHC with $\sqrt{s} =$ 13 TeV.  Also shown is
68$\,\%$ C.L. (1$\sigma$) statistical error of the top quark mass,
$\delta m_t^{out}$, together with the averaged minimal $\chi^2/d.o.f$
and the p-value.  The number of standard deviations corresponding to
each p-value is presented in parentheses. In the last column the top
quark mass shift, defined as $m_t^{in} -m_t^{out}$, with $m_t^{in}=$
173.2 GeV, is also given. Luminosity of 2.5 $fb^{-1}$ and 25 $fb^{-1}$
is assumed.}
\label{tab:mtt}
\end{table}

Similar performance as in the case of $M_{t\bar{t}}$ can been obtained
with the help of the more exclusive observable, $M_{b\ell}$, defined as
the invariant mass of a $b$-jet and a charged lepton. This observable
is frequently used for top quark mass measurements by both ATLAS and
CMS experimental collaborations in the di-lepton top quark decay
channel, see e.g. Ref. \cite{Aaboud:2016igd, Sirunyan:2017idq}. We employ
the invariant mass of the positron and a $b$-jet, keeping
in mind that experimentally one cannot uniquely determine which
$b$-jet should be taken into account to build the observable. If the
$be^+$ pair that returns the smallest invariant mass will be chosen,
however, then the probability that both final states come from the
decay cascade initiated by the same top quark increases
\cite{Beneke:2000hk}. Thus, we define $M_{be^+}$ as
%
\begin{figure}[t!]
\begin{center}
\includegraphics[width=1.0\textwidth]{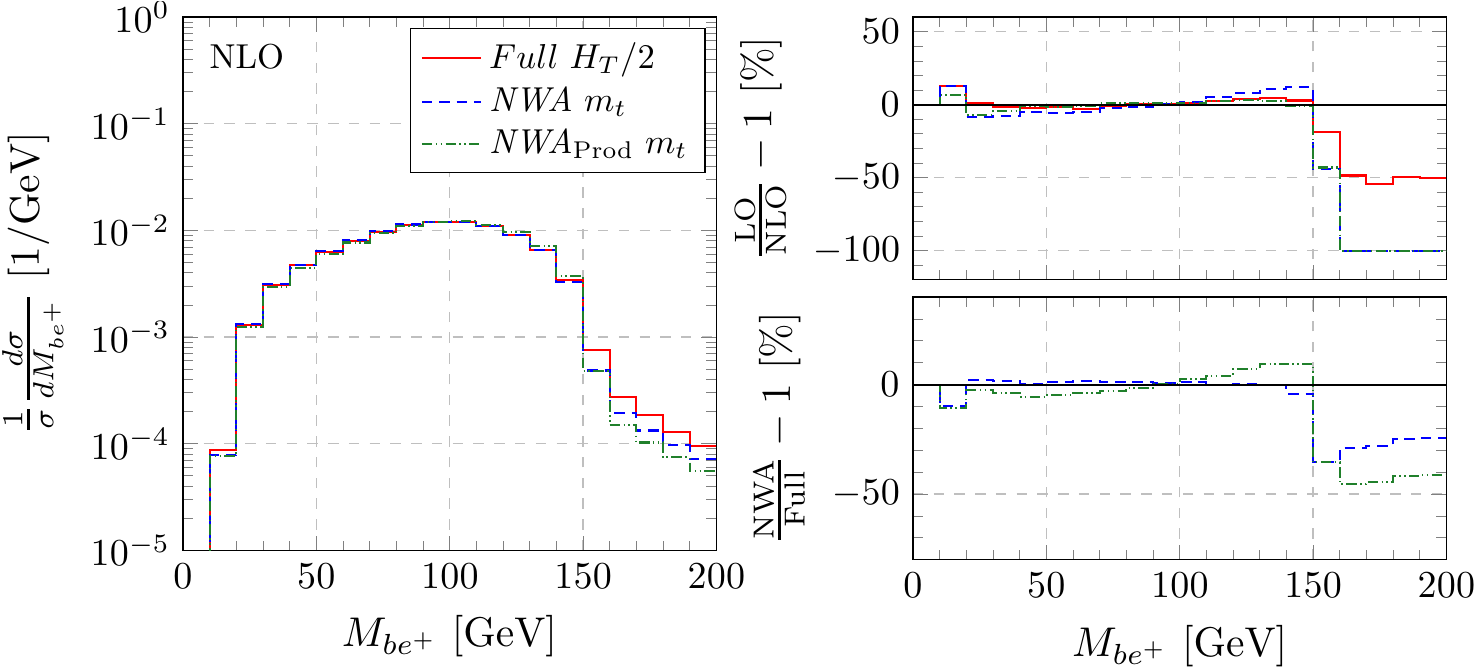}
\includegraphics[width=0.7\textwidth]{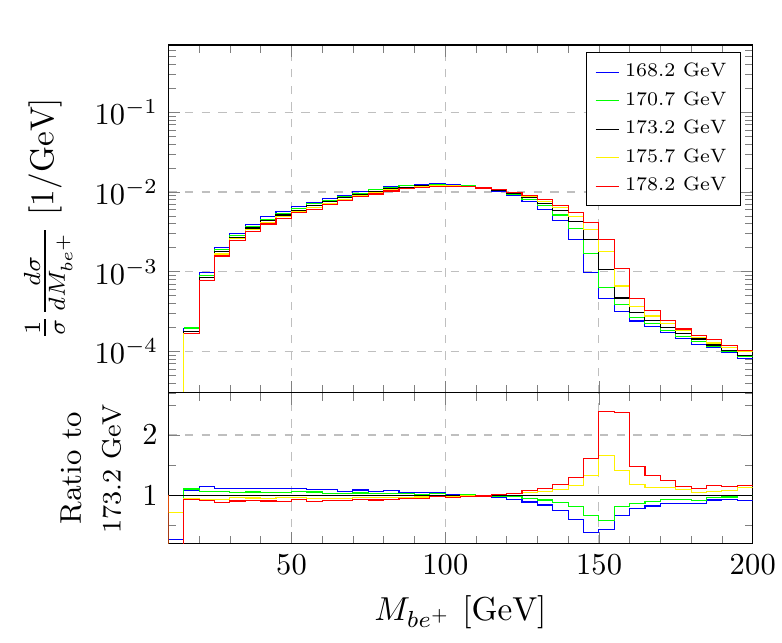}
\end{center}
\caption{\it 
Normalised $M_{b\ell}$ distribution at NLO QCD for the $pp\to e^+\nu_e
\mu^-\bar{\nu}_\mu b\bar{b}j +X$ production process at the LHC with
$\sqrt{s} =$ 13 TeV.  The CT14 PDF set is used. Upper panel: three
different theoretical descriptions with $m_t=$ 173.2 GeV and $\mu_R
=\mu_F =\mu_0$, where $\mu_0=m_t$ for both NWA cases and $\mu_0=H_T/2$
for the Full case.  Also given are the relative size of NLO QCD
corrections and the combined relative size of finite-top-width and
finite-W-width effects for the normalised $M_{b\ell}$ observable.
Lower panel: the Full case for five different top quark masses for
$\mu_0 =H_T/2$. Also plotted is the ratio to the result
with the default value of $m_t$, i.e. $m_t =$ 173.2 GeV.} 
\label{fig:4observable}
\end{figure}
\begin{table}[t!]
\begin{center}
\begin{tabular}{ccccc}
&&&&\\
\text{Theory, NLO  QCD}& $m^{out}_t \pm \delta m^{out}_t$
& Averaged & Probability &
$m_t^{in} -m_t^{out}$ \\ \text{CT14 PDF} & [GeV] &
$\chi^2/{\rm d.o.f.}$
& {\it p-value} & [GeV] \\
&&&&\\
\hline\hline
&&&&\\
&\quad \quad  {\it 31 bins} $@$ 2.5 fb${}^{-1}$  &&&\\
\hline\hline
{\it Full}, $\mu_0=H_T/2$&173.09 $\pm$ 0.48 & 1.05 
& 0.38 (0.9$\sigma$) & $+$0.11\\
{\it Full}, $\mu_0=E_T/2$& 173.01 $\pm$ 0.50 & 1.06
& 0.37 (0.9$\sigma$) & $+$0.19\\
{\it Full}, $\mu_0=m_t$& 173.07 $\pm$ 0.49 & 1.22 
& 0.18 (1.3$\sigma$) &$+$0.13\\
\hline
{\it NWA},  $\mu_0=m_t$&173.90 $\pm$ 0.50& 1.11
& 0.30 (1.0$\sigma$) &$-$0.70 \\
{\it NWA}$_{\rm Prod.}$, $\mu_0=m_t$& 172.56 $\pm$ 0.54& 1.64
& 0.01 (2.6$\sigma$) &$+$0.64 \\
\hline\hline
&&&&\\
&\quad  \quad  {\it 31 bins} $@$ 25 fb${}^{-1}$  &&&\\
\hline\hline
{\it Full}, $\mu_0=H_T/2$& 173.18 $\pm$ 0.15 &
1.02  & 0.42 (0.8$\sigma$) & $+$0.02 \\
{\it Full}, $\mu_0=E_T/2$& 173.23 $\pm$ 0.15 & 1.03
& 0.41  (0.8$\sigma$)&$-$0.03\\
{\it Full}, $\mu_0=m_t$& 173.22 $\pm$ 0.16 & 1.78
& 0.005  (2.8$\sigma$) &$-$0.02\\
\hline
{\it NWA},  $\mu_0=m_t$& 173.98  $\pm$ 0.16 & 2.56 
& 5 $\cdot 10^{-6}$ (4.6$\sigma$) & $-$0.78\\
{\it NWA}$_{\rm Prod.}$, $\mu_0=m_t$&  172.62 $\pm$ 0.17 & 8.23
& 0 ($\gg 5\sigma$) & $+$0.58\\
\end{tabular}
 \end{center}
\caption{\it 
Mean value of the top quark mass, $m_t^{out}$, from 1000
pseudo-experiments as obtained from the normalised $M_{b\ell}$
differential distribution for the $pp\to e^+\nu_e \mu^-\bar{\nu}_\mu
b\bar{b}j +X$ production process at the LHC with $\sqrt{s} =$ 13
TeV.  Also shown is 68$\,\%$ C.L. (1$\sigma$) statistical error of
the top quark mass, $\delta m_t^{out}$, together with the averaged
minimal $\chi^2/d.o.f$ and the p-value.  The number of standard
deviations corresponding to each p-value is presented in
parentheses. In the last column the top quark mass shift, defined as
$m_t^{in} -m_t^{out}$, with $m_t^{in} =$ 173.2 GeV, is also
given. Luminosity of 2.5 $fb^{-1}$ and 25 $fb^{-1}$ is assumed.}
\label{tab:mbl}
\end{table}
%
\begin{equation}
 M_{be^+} = \min   \left\{ M_{b_1 e^+} , M_{b_2 e^+} 
\right\}\,.
\end{equation}
The $M_{be^+}$ observable possesses a kinematic endpoint that can be
derived from the on-shell top-quark decay into $t\to W^+b \to e^+
\nu_e b$. Since we have $m^2_t = p^2_t = m^2_W + 2p_bp_{e^+} +
2p_bp_{\nu_e}$ the invariant mass of the positron and the bottom quark
is given by $M_{be^+}=\sqrt{2p_bp_{e^+}}$ and in the massless case
should be smaller or equal to $\sqrt{m^2_t-m^2_W}$. When both the $t$
quark and the $W$ gauge boson are treated as on-shell particles at the
lowest order this strict kinematic limit amounts to
$M^{max.}_{be^+}=153.4$ GeV. Additional radiation, for example from
parton showers or the real emission part of the higher order
corrections, as well as off-shell effects and non-resonant
contributions of the top quark and the $W$ gauge boson introduce a
smearing of $M^{max.}_{be^+}$. In Figure \ref{fig:4observable},
$M_{be^+}$ is shown together with its top quark mass dependence.  A
sharp fall of the cross section around the value of $153$ GeV is
clearly observed. In the range below the kinematical cut-off the size
of off-shell effects is negligible. Above the $M^{max.}_{be^+}$ value,
however, these effects are large, of the order of $-35\%$ or even
$-45\%$ if only LO top quark decays are incorporated. In spite of
that, a substantial impact on $m_t^{out}$ is not expected if the whole
range of $M_{b\ell}$ is to be taken into account, the reason being a
drop of the cross section by one or even two orders of magnitude for
$M_{be^+} \gtrsim M^{max.}_{be^+}$. Our findings on $m_t^{out}$ are
recapitulated in Table \ref{tab:mbl}. We  first observe that for
the same case of the integrated luminosity a similar size of
statistical uncertainties is obtained as for $M_{t\bar{t}}$. Next, for
the {\it Full} case with ${\cal L}=$ 2.5 fb${}^{-1}$ the top quark
mass shift is much smaller of the order of $0.2$ GeV only. When
luminosity is increased it is even further reduced down to $0.03$
GeV. At the same time for {\it NWA} and {\it NWA}${}_{Prod.}$ we have
obtained a change by $0.6-0.8$ GeV.  Comparing the {\it NWA} case 
with pseudo-data generation from {\it Full} with $\mu_R=\mu_F=\mu_0
=m_t$ we find a shift in the top quark mass extraction of $-0.70$
GeV. We note that this result is compatible with the shift of $-0.83$
GeV observed for the process $pp \to e^+ \nu_e \mu^- \bar{\nu}_\mu b
\bar{b}$ in Ref.~\cite{Heinrich:2017bqp}, that was generated for the
similar setup. Also for ${\cal L}=$ 2.5 fb${}^{-1}$ all theoretical
descriptions can be employed, whereas for the case of ${\cal L}=$ 25
fb${}^{-1}$ only the {\it Full} approach provides {\it p-value} larger
than $0.0027$ (below $3\sigma$). Lastly, theoretical uncertainties are
very small, i.e. of the order of $0.05$ GeV for the {\it Full} case
with dynamical scale choice and $1$ GeV for the {\it Full} case with a
fixed scale. The PDF uncertainties are independent of the scale choice
and yield $0.02-0.03$ GeV. Overall, considering all aspects,
i.e. statistical uncertainties on $m_t^{out}$, the top quark mass
shift, the quality of the $\chi^2$ fit as well as theoretical
uncertainties, the $M_{b\ell}$ observable provides the best
sensitivity to the top quark mass when the {\it Full} case is
employed.
%
\begin{figure}[t!]
\begin{center}
\includegraphics[width=1.0\textwidth]{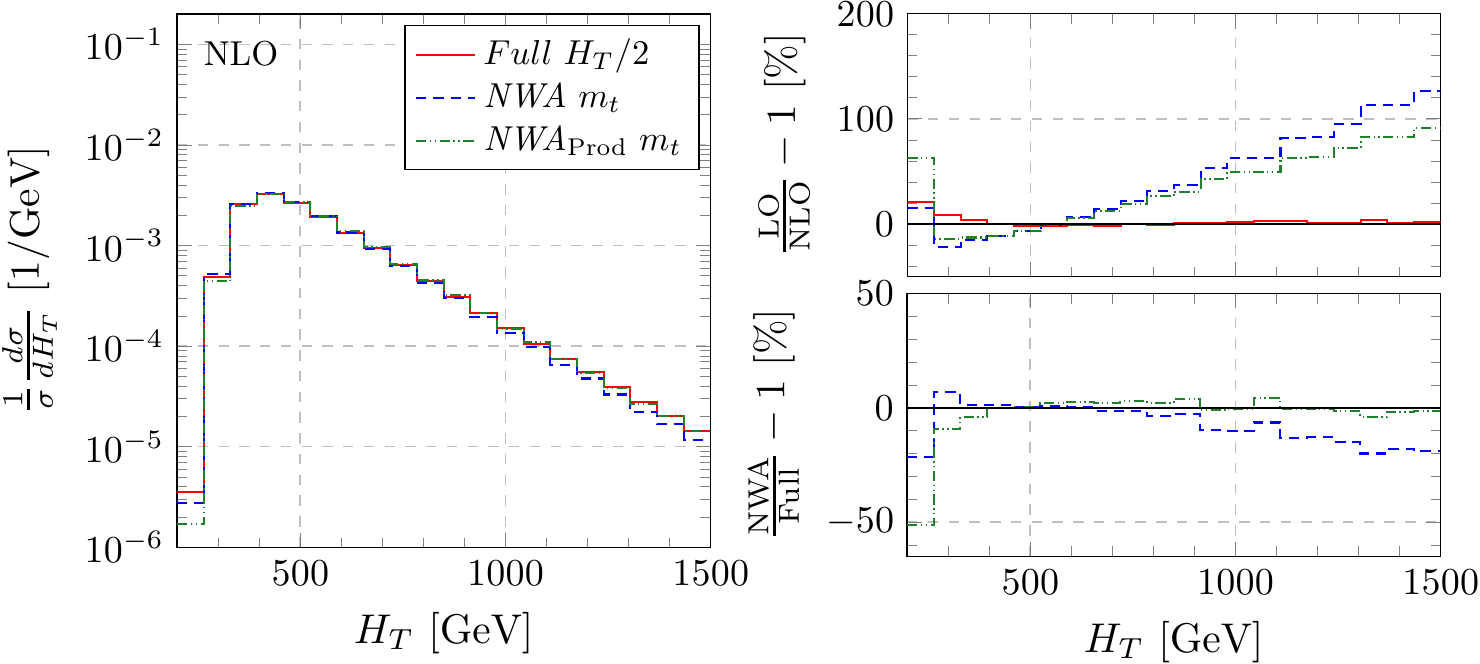}
\includegraphics[width=0.7\textwidth]{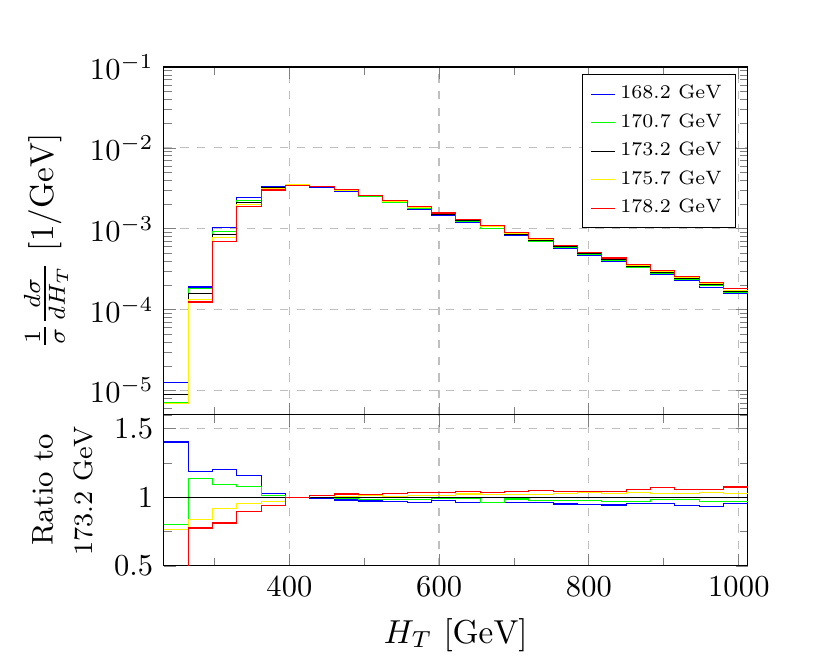}
\end{center}
\caption{\it 
Normalised $H_T$ distribution at NLO QCD for the $pp\to e^+\nu_e
\mu^-\bar{\nu}_\mu b\bar{b}j +X$ production process at the LHC with
$\sqrt{s} =$ 13 TeV. The CT14 PDF set is used. Upper panel: three
different theoretical descriptions with $m_t=$ 173.2 GeV and $\mu_R
=\mu_F =\mu_0$, where $\mu_0=m_t$ for both NWA cases and $\mu_0=H_T/2$
for the Full case.  Also given are the relative size of NLO QCD
corrections and the combined relative size of finite-top-width and
finite-W-width effects for the normalised $H_T$ observable.  Lower
panel: the Full case for five different top quark masses for $\mu_0
=H_T/2$. Also plotted is the ratio to the result with the
default value of $m_t$, i.e. $m_t =$ 173.2 GeV.} 
\label{fig:5observable}
\end{figure}
\begin{table}[t!]
\begin{center}
\begin{tabular}{ccccc}
&&&&\\
\text{Theory, NLO  QCD}& $m^{out}_t \pm \delta m^{out}_t$
& Averaged & Probability &
$m_t^{in} -m_t^{out}$ \\ \text{CT14 PDF} & [GeV] &
$\chi^2/{\rm d.o.f.}$
& {\it p-value} & [GeV] \\
&&&&\\
 \hline\hline 
&&&&\\
&\quad \quad {\it 22 bins} $@$ 2.5 fb${}^{-1}$  &&&\\
\hline\hline
{\it Full}, $\mu_0=H_T/2$ & 173.14 $\pm$ 1.18 & 1.01 
& 0.44 (0.8$\sigma$) & $+0.06$ \\
{\it Full}, $\mu_0=E_T/2$ & 172.49 $\pm$ 1.19 & 1.05 
& 0.39 (0.8$\sigma$) & $+0.71$ \\
{\it Full}, $\mu_0=m_t$ & 173.39 $\pm$ 1.23 & 1.07 
& 0.37 (0.9$\sigma$) & $-0.19$ \\
\hline
{\it NWA},  $\mu_0=m_t$ & 174.47 $\pm$ 1.19 & 1.06 
& 0.38 (0.9$\sigma$) & $-1.27$ \\
{\it NWA}$_{\rm Prod.}$, $\mu_0=m_t$ & 171.21 $\pm$ 1.15 
& 1.06 & 0.37 (0.9$\sigma$) & $+1.99$\\
\hline\hline
&&&&\\
&\quad \quad {\it 22 bins} $@$  25 fb${}^{-1}$  &&&\\
\hline\hline
{\it Full}, $\mu_0=H_T/2$ & 173.27 $\pm$ 0.39 & 1.03 
& 0.41 (0.8$\sigma$) & $-0.07$\\
{\it Full}, $\mu_0=E_T/2$ & 172.55 $\pm$ 0.37  & 1.17 
& 0.26 (1.1$\sigma$) & $+0.65$ \\
{\it Full}, $\mu_0=m_t$ & 173.45 $\pm$ 0.40  & 1.83 
& 0.01 (2.6$\sigma$) & $-0.25$\\
\hline
{\it NWA},  $\mu_0=m_t$ & 174.62 $\pm$ 0.36  & 1.89 
& 0.008 (2.6$\sigma$)  & $-1.42$ \\
{\it NWA}$_{\rm Prod.}$, $\mu_0=m_t$ & 171.22 $\pm$ 0.34 & 1.48 
& 0.07 (1.8$\sigma$) & $+1.98$ \\
\hline\hline
&&&&\\
&\quad \quad {\it 22 bins} $@$ 50  fb${}^{-1}$  &&&\\
\hline\hline
{\it Full}, $\mu_0=H_T/2$ & 173.27 $\pm$ 0.27 & 1.02 
& 0.43 ($0.8\sigma$) &  $-0.07$\\
{\it Full}, $\mu_0=E_T/2$ & 172.56 $\pm$ 0.27 & 1.40 
& 0.11 ($1.6\sigma$) & $+0.64$\\
{\it Full}, $\mu_0=m_t$ & 173.45 $\pm$ 0.28 & 2.79 
& $2\cdot 10^{-5}$ ($4.3\sigma$) & $-0.25$\\
\hline
{\it NWA},  $\mu_0=m_t$ &174.63 $\pm$ 0.25  & 2.82 
& $2\cdot 10^{-5}$ ($4.3\sigma$) & $-1.43$\\
{\it NWA}$_{\rm Prod.}$, $\mu_0=m_t$ & 171.25 $\pm$ 0.25  & 2.00 
&0.004 $(2.9\sigma)$ & $+1.95$\\
\end{tabular}
 \end{center}
\caption{\it 
Mean value of the top quark mass, $m_t^{out}$, from 1000
pseudo-experiments as obtained from the normalised $H_T$ differential
distribution for the $pp\to e^+\nu_e \mu^-\bar{\nu}_\mu b\bar{b}j +X$
production process at the LHC with $\sqrt{s} =$ 13 TeV. Also shown is
68$\,\%$ C.L. (1$\sigma$) statistical error of the top quark mass,
$\delta m_t^{out}$, together with the averaged minimal $\chi^2/d.o.f$
and the p-value.  The number of standard deviations corresponding to
each p-value is presented in parentheses. In the last column the top
quark mass shift, defined as $m_t^{in} -m_t^{out}$, with $m_t^{in} =$
173.2 GeV, is also given. Luminosity of 2.5 $fb^{-1}$, 25 $fb^{-1}$
and 50 $fb^{-1}$ is assumed.}
\label{tab:ht}
\end{table}
%

Our last (exclusive) observable, that we would like to examine, is the
total transverse momentum of the top anti-top plus one hard jet
system, $H_T$, defined as
\begin{equation}
H_T= p_T(e^+)+p_T(\mu^-)+p_T(j_{b_1})+p_T(j_{b_{2}}) +p_T(j_1) +
p_T^{miss}\,.
\end{equation}
Let us remind  that in the case of two resolved jets the one with
highest transverse momentum is chosen. In Figure \ref{fig:5observable},
this observable is presented, again together with its dependence on
$m_t$. For normalised distributions a shape difference between {\it
Full} and {\it NWA}${}_{Prod.}$ is noticeable, which will be
definitely reflected on the mean value of $m_t^{out}$. By applying the
same arguments as for $M_{t\bar{t}}$ also here a range only up to $1$
TeV is used in the top quark mass studies. Our results on $m_t^{out}$
are provided in Table \ref{tab:ht}. This observable has a similar
performance in terms of statistical uncertainties as the normalised
$\rho_s$ distribution. The top quark mass shift for the {\it Full}
case is also comparable, i.e. between $0.1-0.7$ GeV independently of
the luminosity considered and reduced in the case of {\it NWA} and {\it
NWA}${}_{Prod.}$.  In the latter cases it amounts to $1.4$ GeV and $2$
GeV respectively. What seems to be different, however, is the good
quality of the $\chi^2$ fit independently of the theory applied and
luminosity examined.  To be more precise, for ${\cal L=}$ 2.5
fb${}^{-1}$ all theoretical descriptions are within $1\sigma$ with the
pseudo-data whereas for the ${\cal L=}$ 25 fb${}^{-1}$ case the same
applies to the {\it Full} case with $\mu_0=H_T/2$ and
$\mu_0=E_T/2$. Nevertheless, for {\it Full}, {\it NWA} and {\it
NWA}${}_{Prod.}$ with $\mu_0=m_t$ we have an agreement within
$2.6\sigma$ with the pseudo-data. This suggests that a larger
integrated luminosity is required to clearly differentiate among various
theoretical approaches used in the calculation of higher order QCD
corrections to the $pp\to t\bar{t}j$ production process in the
di-lepton channel at the LHC. Indeed, already for ${\cal L=}$ 50
fb${}^{-1}$, which corresponds approximately to $10800$ events, again
only the {\it Full} case with a dynamical scale choice, either
$\mu_0=H_T/2$ or $\mu_0= E_T/2$, reproduces the pseudo-data adequately
as can be seen from Table \ref{tab:ht}. The remaining cases, {\it
Full} and {\it NWA} with $\mu_0=m_t$, are disfavoured beyond the
$4\sigma$ level. In the former case the $4\sigma$ difference can be
simply attributed to the fixed scale choice used for the description
of the $H_T/2$ differential distribution, which not sufficiently
describes tails of the distribution.  As to the theoretical
uncertainties the contribution related to unknown higher-order
corrections is estimated to be of the order of $0.5$ GeV $-$ $1.8$ GeV
for a dynamical scale choice and $2$ GeV for a fixed scale.  We have
also analysed the theoretical error arising from different
parametrisation of PDFs, being able to quantify it at the level of
$0.4$ GeV, thus well below the uncertainty associated with the scale
dependence.

%
\section{Summary and Conclusions}
%

In this paper we have studied the normalised $\rho_s$ differential
distribution including the leptonic top quark decays.  We focused on
fixed order NLO QCD calculations at the LHC with $\sqrt{s} = 13$ TeV.
Three different theoretical descriptions of the top quark decay chain
have been investigated. In the first approach we included all
interferences, off-shell effects and non-resonant backgrounds. In the
second case top quark decays in the narrow width approximation have
been considered. To be more precise two cases have been employed: NLO
QCD corrections to the $pp \to t\bar{t}j$ production process with
leading order decays and the more sophisticated case with QCD
corrections and jet radiation present also in top quark decays. We
have used these various theoretical prescriptions to investigate their
impact on the extraction of the top quark mass. We have compared them
to the pseudo-data sets, that have been generated from the best
theoretical description, i.e. the {\it Full} prediction at NLO in QCD
as generated with $m_t=173.2$ GeV and
$\mu_R=\mu_F=\mu_0=H_T/2$. Moreover, we have quantified associated
theoretical uncertainties. For the low integrated luminosity case with
${\cal L} = 2.5 ~{\rm fb}^{-1}$ that corresponded in our case to
approximately $5400$ events assuming perfect detector efficiency and
to the statistical uncertainty on the top quark mass of the order of
$\delta m^{out}_t=1 ~{\rm GeV} -1.5 ~{\rm GeV}$, all theoretical
prescriptions seemed to be in agreement with the pseudo-data sets. The
largest discrepancy amounted to $1.3\sigma$ only. Additionally, the
averaged minimal $\chi^2/d.o.f$ was always around $1$. However,
substantial mass shifts, even up to $2.5$ GeV and $3.8$ GeV, have been
observed in the case of {\it NWA} and {\it NWA}$_{Prod.}$
respectively. We have checked that generating the pseudo-data sets
with the {\it Full} case but for $\mu_R=\mu_F=\mu_0=m_t$ does not
change the situation, namely mass shifts up to $2$ GeV for {\it NWA}
and $3.8$ GeV for {\it NWA}$_{Prod.}$ are still obtained. Thus, they
cannot be ascribed only to effects of the scale choice used in the
generation of the pseudo-data sets. For the higher luminosity case,
that corresponded to $54 000$ events and $\delta m_t^{out}=0.3 ~{\rm
GeV} -0.5 ~{\rm GeV}$, despite the diminished quality of the $\chi^2$
fit these mass shifts remained unchanged. Taking into account the size
of the statistical uncertainty on the top quark mass and the
negligible statistical errors of theoretical predictions as compared
to pseudo-data errors we conclude that independently of the integrated
luminosity case only the {\it Full} prediction with either
$\mu_R=\mu_F=\mu_0=H_T/2$ or $\mu_R=\mu_F=\mu_0=E_T/2$ should be used
to extract the top quark mass from the normalised $\rho_s$
differential distribution once top quark decays are included. Using
the best theoretical description at hand, we have established that
theoretical uncertainties stemming from the scale variation were
luminosity independent and of the order of $0.6 ~{\rm GeV} - 1.2 ~{\rm
GeV}$. The smallest value has been obtained for the normalised
$\rho_s$ observable with the largest number of bins. Once a fixed
scale has been used instead, they increased to $2.1 ~{\rm GeV} - 2.8
~{\rm GeV}$. Thus, additionally, the importance of the proper scale
choice for the description of the differential cross sections has been
shown here.  Another source of theoretical uncertainties on the top
quark mass extraction coming from various PDF parameterisations has
been estimated to be in the range of $0.4 ~{\rm GeV} - 0.7 ~{\rm
GeV}$.

In the next step we examined a slightly modified version of the
normalised $\rho_s$ differential distribution. Namely, if the second
resolved jet was present it has been included in the definition of the
observable. We have found similar performance as in the $\rho_s$ case
for all aspects but theoretical uncertainties.  The theoretical errors
from the scale dependence increased to $3 ~{\rm GeV} - 4 ~{\rm GeV}$
for the {\it Full} case either with $\mu_0=H_T/2$ or
$\mu_0=E_T/2$. The latter raise has been driven by the leading order
nature of the second resolved jet.

Finally, to check the sensitivity of the $\rho_s$ observable we have
made a comparison to the invariant mass of the $t\bar{t}$ system and
to two other more exclusive observables like the minimal invariant
mass of the charged lepton and $b$-jet as well as the total transverse
momentum of the $e^+\nu_e \mu^- \bar{\nu}_\mu b\bar{b}j$ system. In
terms of the statistical errors on the extraction of $m_t$ and the
mass shift the normalised invariant mass of the top anti-top pair has
performed better than $\rho_s$.  For the same integrated luminosity
case, $\delta m_t^{out}$ was lower by a factor of $2-2.4$. The quality
of the $\chi^2$ fit seemed similar, however, the $m_t^{in}-m_t^{out}$
shift was below $0.1$ GeV for the {\it Full} case with the dynamical
scale choice, $0.3$ GeV for the fixed scale and $0.5$ GeV for the {\it
NWA} case. In the case of {\it NWA}$_{Prod.}$ a somewhat higher value
of the $m_t^{in}-m_t^{out}$ shift, around $2$ GeV, has been
obtained. Thus, in the chosen range, i.e. up to $1$ TeV, and for the
low integrated luminosity case, the off-shell effects and non-resonant
contributions of the top quark and $W$ gauge boson 
were not very crucial. It turned out that the inclusion
of the higher order corrections to the top quark decays was more
important. Both {\it Full} and {\it NWA} cases could be employed for
the $m_t$ extraction.  Generally speaking, the case of the low
integrated luminosity has shown lack of a sensitivity to the details
of the top quark decays. Once increased luminosity was considered,
however, the {\it NWA} case has been disfavoured at the
$4\sigma-5\sigma$ level considering only the statistical
uncertainties.   The performance of the normalised
$M_{t\bar{t}}$ observable was similar to the performance of the more
exclusive and very well known observable used in the alternative $m_t$
measurements, i.e.  the (normalised) minimal invariant mass of the
bottom jet and the charged lepton, $M_{b\ell}$, which has also been
examined.  The last observable that we have studied was the normalised
$H_T$ differential cross section. This exclusive observable proved to
be similar to ${\cal R}(m_t^{pole},\rho_s)$ in terms of $\delta
m^{out}_t$ and the quality of the $\chi^2$ fit, however, the observed
mass shifts were smaller, of the order of $0.7$ GeV for {\it Full},
$1.4$ GeV for {\it NWA} and $2$ GeV for {\it NWA}$_{Prod.}$. In
addition, in order to disfavour the {\it NWA} approach beyond the
$3\sigma-4\sigma$ level the integrated luminosity had to be increased
$20$ times unlike for all other cases where the smaller change from
$2.5 ~{\rm fb}^{-1}$ to $25 ~{\rm fb}^{-1}$ has been sufficient to
obtain the $5\sigma$ level. Overall, among all studied
normalised differential cross sections, $\rho_s$ has shown  the
highest  sensitivity to the top quark and $W$ gauge boson off-shell
effects and non-resonant background contributions.

Let us note here that this is a theoretical study and additional
systematic uncertainties need to be addressed. Among others the impact
of the parton shower on the shape of $\rho_s$, $M_{t\bar{t}}$,
$M_{b\ell}$ and $H_T$ observables should be carefully examined as well
as non-perturbative effects together with the $b$-tagging and neutrino
reconstruction efficiencies should be estimated. These uncertainties
are, however, beyond the scope of this paper. We plan to study them in
a separate publication. Even though we can not quantify the size of
systematic uncertainties on the experimental side we can make the
following general statement. If, for the particular observable that we
have scrutinised for which large mass shifts have not been present,
the systematic uncertainties are larger or of the same order as our
statistical uncertainty $\delta m_t^{out}$ for ${\cal L}=2.5$
fb${}^{-1}$, various theoretical descriptions at NLO in QCD, that have
been investigated in the paper, can be employed to simulate the $pp\to
t\bar{t}j$ production process in the di-lepton top quark decay
channel. This is possible since we do not have sufficient sensitivity
to see differences in the various descriptions of the top quark
decays. In the case of observables with a large mass shift, e.g.
$\rho_s$ or $H_T$, all these theoretical descriptions may still be used but
one would have to compensate for the shift. If the size of systematic
uncertainties, however, is rather similar to $\delta m_t^{out}$ for
${\cal L}=25$ fb${}^{-1}$ or in the case of $H_T$ to $\delta
m_t^{out}$ for ${\cal L}=50$ fb${}^{-1}$, only the {\it Full}
theoretical description with the dynamical scale choice, either
$\mu_0=H_T/2$ or $\mu_0=E_T/2$, should be used to simulate the $pp \to
e^+ \nu_e \mu^- \bar{\nu}_\mu b\bar{b} j$ production process at the
LHC to extract the top quark mass.

A few additional comments are in order. The ${\cal
R}(m_t^{pole},\rho_s)$ differential observable has already been
employed by the ATLAS and CMS experimental collaborations at the LHC
to determine the top quark mass. In both studies, on-shell top quarks
have been used to build the normalised $\rho_s$ observable. In
practise, various Monte Carlo programs have been used where at most
on-shell $t\bar{t}$ or $t\bar{t}j$ samples at NLO in QCD have been
matched with parton shower programs like PYTHIA or
HERWIG. Nevertheless, such theoretical predictions have been first
tuned to data to account for missing perturbative and non-perturbative
contributions. In the next step they are unfolded back to the
so-called parton level to obtain on-shell top quarks. These
calibrations come with additional uncertainties that the experimental
collaborations need to consider.  Finally, such predictions are
contrasted with the same data to extract the top quark mass. NLO QCD
calculations with complete top quark and $W$ gauge boson off-shell
effects and non-resonant contributions included allow, instead, to
define top quarks using kinematics and selection cuts making them much
closer to the experimental data. Thus, for example the top quark mass
can be measured using the fiducial differential cross section as a
function of $\rho_s$ or $M_{t\bar{t}}$. To summarise, the aim of such
precise theoretical predictions can be twofold. First, they can be
used for a direct comparison with the LHC data at the parton level,
which would lead to the much simplified calibration procedure and
substantial reduction of the systematic uncertainties. Secondly, they
can be utilised by the experimental collaborations at the intermediate
level to test the quality of the tuning and unfolding
procedures. Close collaboration on these issues with experimental
colleagues from ATLAS and CMS is already planned.

%
\acknowledgments
%

We would like to thank Juan Fuster for careful reading of the
manuscript and his valuable comments. Furthermore, we thank Jamie
Tattersall and Peter Uwer for discussions. 

The work of M.W. and H.B.H. was supported in part by the German
Research Foundation (DFG) under Grant No. WO 1900/2 $-$ {\it
Top-Quarks under the LHCs Magnifying Glass: From Process Modelling to
Parameter Extraction}. Furthermore, the work of H.B.H. was supported
by a Rutherford Grant ST/M004104/1. The research of G.B. was supported
by grant K 125105 of the National Research, Development and Innovation
Office in Hungary.

Simulations were performed with computing resources
  granted by RWTH Aachen University under project rwth0165.

\begin{figure}[t!]
\begin{center}
\includegraphics[width=1.0\textwidth]{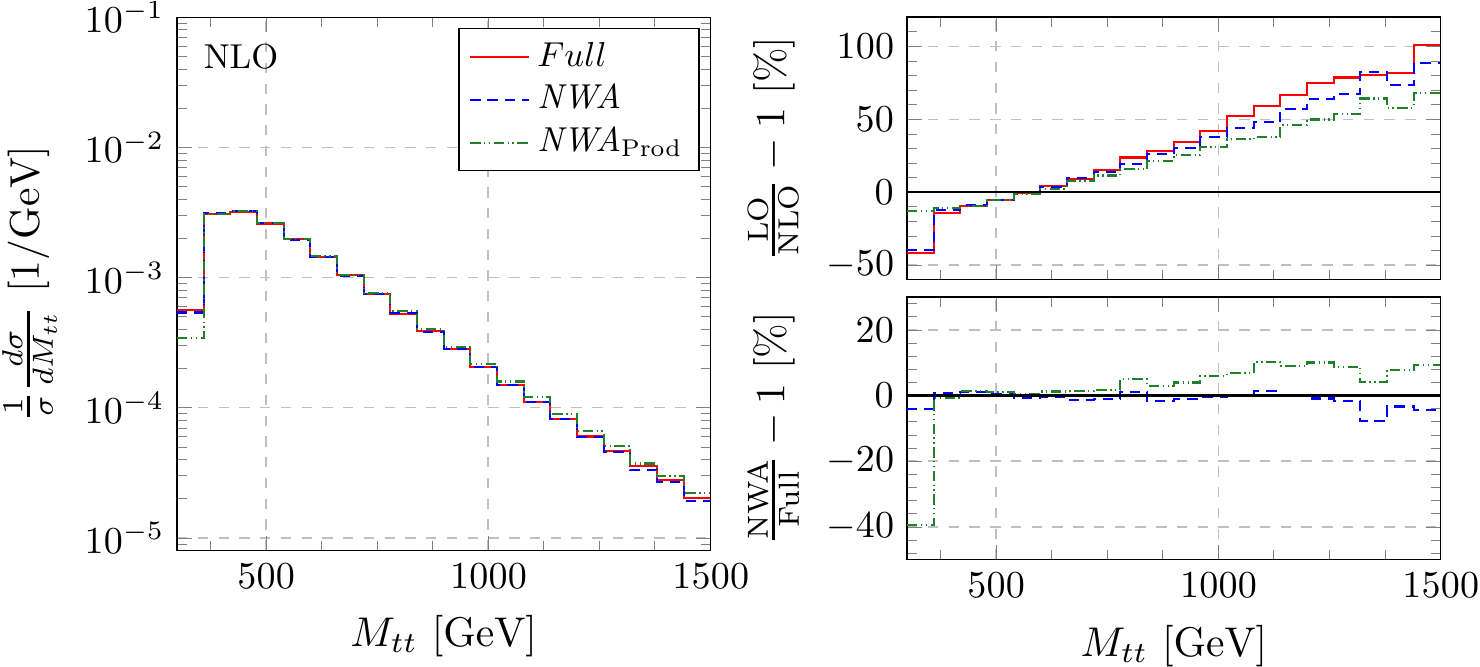}
\end{center}
\caption{\it Normalised $M_{t\bar t}$ differential distribution at NLO
QCD for the $pp \to e^+\nu_e \mu^- \bar{\nu}_\mu b\bar{b} j+ X$
production process at the LHC with $\sqrt{s} =$ 13 TeV. Three
different theoretical descriptions with $m_t=$ 173.2 GeV are shown.
Also given are the relative size of NLO QCD corrections and the
combined relative size of finite-top-width and finite-W-width effects
for the normalised $M_{t\bar t}$ observable.  Renormalisation and
factorisation scales are set to the common value $\mu_R =\mu_F 
=\mu_0$ where $\mu_0 =m_t$.  The CT14 PDF set is employed. }
\label{fig:Mtt_mt}
\end{figure}
\begin{figure}[t!]
\begin{center}
\includegraphics[width=1.0\textwidth]{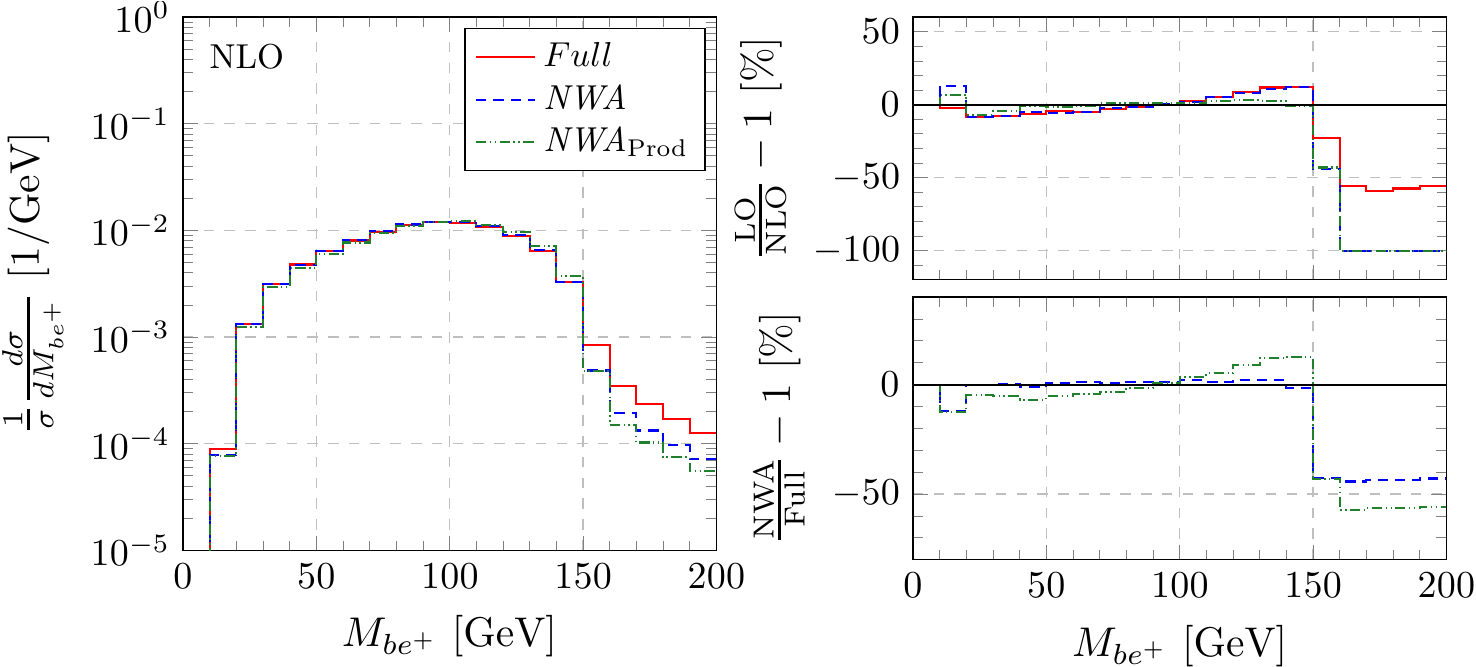}
\end{center}
\caption{\it Normalised $M_{b\ell}$ differential distribution at NLO
QCD for the $pp \to e^+\nu_e \mu^- \bar{\nu}_\mu b\bar{b} j+ X$
production process at the LHC with $\sqrt{s} =$ 13 TeV. Three
different theoretical descriptions with $m_t=$ 173.2 GeV are shown.
Also given are the relative size of NLO QCD corrections and the
combined relative size of finite-top-width and finite-W-width effects
for the normalised $M_{b\ell}$ observable.  Renormalisation and
factorisation scales are set to the common value $\mu_R =\mu_F 
=\mu_0$ where $\mu_0 =m_t$.  The CT14 PDF set is employed. }
\label{fig:Mbl_mt}
\end{figure}
\begin{figure}[t!]
\begin{center}
\includegraphics[width=1.0\textwidth]{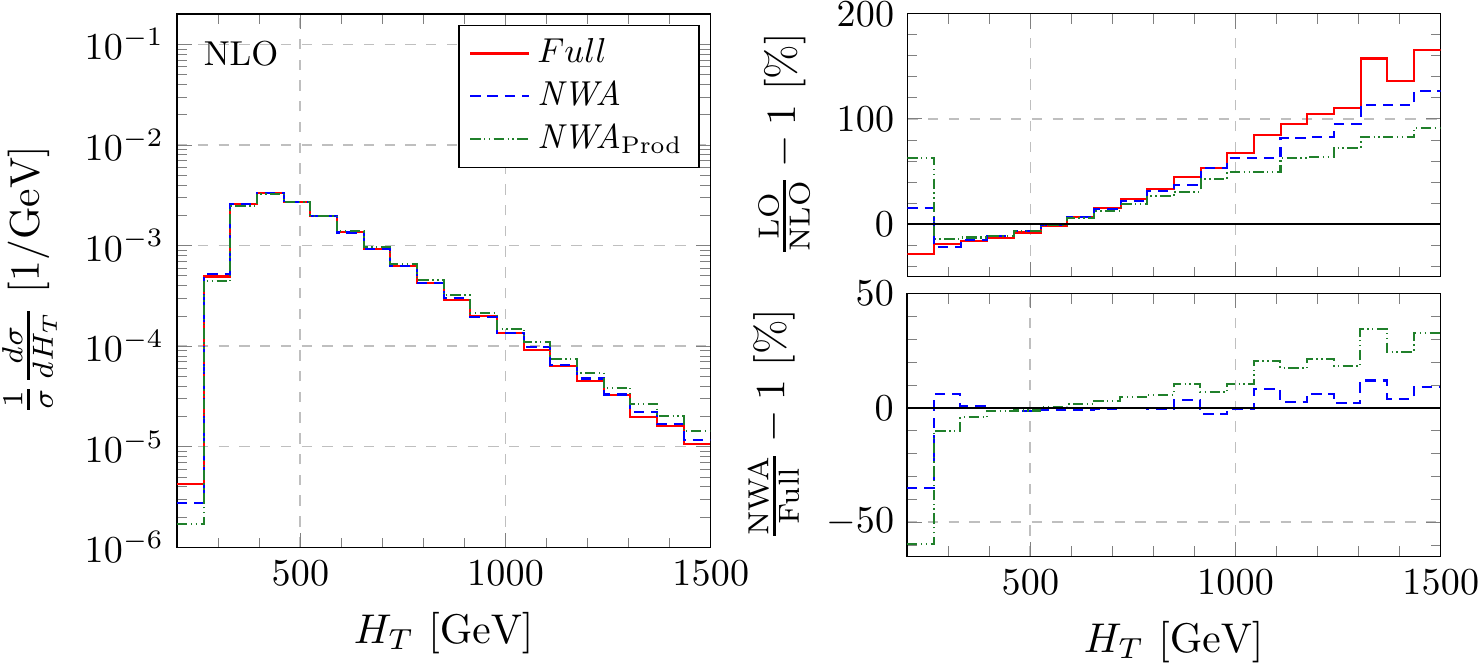}
\end{center}
\caption{\it Normalised $H_T$ differential distribution at NLO QCD for
the $pp \to e^+\nu_e \mu^- \bar{\nu}_\mu b\bar{b} j+ X$ production
process at the LHC with $\sqrt{s} =$ 13 TeV. Three different
theoretical descriptions with $m_t=$ 173.2 GeV are shown.  Also given
are the relative size of NLO QCD corrections and the combined relative
size of finite-top-width and finite-W-width effects for the normalised
$H_T$ observable.  Renormalisation and factorisation scales are set to
the common value $\mu_R =\mu_F =\mu_0$ where $\mu_0 =m_t$.  
The CT14 PDF set is employed. }
\label{fig:HT_mt}
\end{figure}
%

\appendix
\section{Comparison between {\textbf{\textit {Full,}}}
{\textbf{\textit {NWA}}} and 
{\textbf{\textit {NWA}}}$_{\textbf{\textit{Prod.}}}$ for
$\boldsymbol{\mu_0= m_t}$}
\label{App:FullNWAcomparison}

In
Figures~\ref{fig:3observable},~\ref{fig:4observable}~and~\ref{fig:5observable},
we have shown the comparison between three theoretical predictions
corresponding to different modellings of top-quark decays: {\it Full}
with $\mu_0 = H_T/2$, {\it NWA} with $\mu_0 = m_t$ and {\it
NWA}${}_{Prod.}$ with $\mu_0 = m_t$, for $M_{t\bar t}$, $M_{b\ell}$
and $H_T$ observables.  Here, $\mu_0$ is the common value chosen for
the renormalizaton and factorization scales, $\mu_R = \mu_F = \mu_0$.
These comparisons were relevant for our top-quark mass extraction
studies performed in Section~\ref{section:otherobservables}. In this
Appendix, we show the very same comparison albeit with a common scale
choice, $\mu_0 = m_t$, used for all three theoretical predictions.
We believe such comparison will better reflect the finite-top-width
and finite-$W$-width effects in {\it Full} compared to {\it NWA}.  In
Figures~\ref{fig:Mtt_mt},~\ref{fig:Mbl_mt}~and~\ref{fig:HT_mt}, we
show {\it Full}, {\it NWA} and {\it NWA}${}_{Prod.}$ predictions, all
with $\mu_0 = m_t$ for normalised $M_{t\bar t}$, $M_{b\ell}$ and $H_T$
observables.


\begin{thebibliography}{99}

\bibitem{Degrassi:2012ry}
  G.~Degrassi, S.~Di Vita, J.~Elias-Miro, J.~R.~Espinosa,
  G.~F.~Giudice, G.~Isidori and A.~Strumia,
  {\it Higgs mass and vacuum stability in the Standard Model at NNLO}, 
  JHEP {\bf 1208} (2012) 098
   [arXiv:1205.6497 [hep-ph]].

\bibitem{Alekhin:2012py}
  S.~Alekhin, A.~Djouadi and S.~Moch,
  {\it The top quark and Higgs boson masses and the stability of the
    electroweak vacuum},
  Phys.\ Lett.\ B {\bf 716} (2012) 214
    [arXiv:1207.0980 [hep-ph]].

\bibitem{Frixione:2003ei}
  S.~Frixione, P.~Nason and B.~R.~Webber,
 {\it Matching NLO QCD and parton showers in heavy flavor production},
  JHEP {\bf 0308} (2003) 007
    [hep-ph/0305252].

\bibitem{Frixione:2007nw}
  S.~Frixione, P.~Nason and G.~Ridolfi,
  {\it A Positive-weight next-to-leading-order Monte Carlo for heavy
    flavour hadroproduction},
  JHEP {\bf 0709} (2007) 126
    [arXiv:0707.3088 [hep-ph]].

\bibitem{Campbell:2014kua}
  J.~M.~Campbell, R.~K.~Ellis, P.~Nason and E.~Re,
  {\it Top-Pair Production and Decay at NLO Matched with Parton Showers},
  JHEP {\bf 1504} (2015) 114
  [arXiv:1412.1828 [hep-ph]].

\bibitem{Jezo:2015aia}
  T.~Jezo and P.~Nason,
  {\it On the Treatment of Resonances in Next-to-Leading Order
    Calculations Matched to a Parton Shower},
  JHEP {\bf 1512} (2015) 065
   [arXiv:1509.09071 [hep-ph]].

\bibitem{Jezo:2016ujg}
  T.~Jezo, J.~M.~Lindert, P.~Nason, C.~Oleari and S.~Pozzorini,
  {\it An NLO+PS generator for $t\bar{t}$ and $Wt$ production and
    decay including non-resonant and   interference effects},
  Eur.\ Phys.\ J.\ C {\bf 76} (2016) no.12,  691
   [arXiv:1607.04538 [hep-ph]].

\bibitem{ATLAS:2014wva}
  [ATLAS, CDF, CMS and D0 Collaborations],
  {\it First combination of Tevatron and LHC measurements of the
    top-quark mass},
  arXiv:1403.4427 [hep-ex].

\bibitem{Aaboud:2016igd}
  M.~Aaboud {\it et al.} [ATLAS Collaboration],
  {\it Measurement of the top quark mass in the $t\bar{t}\to$ di-lepton
    channel from $\sqrt{s}=8$ TeV ATLAS data},
  Phys.\ Lett.\ B {\bf 761} (2016) 350
    [arXiv:1606.02179 [hep-ex]].

\bibitem{Khachatryan:2015hba}
  V.~Khachatryan {\it et al.} [CMS Collaboration],
  {\it Measurement of the top quark mass using proton-proton data at
    $\sqrt{s}$ = 7 and 8 TeV},
  Phys.\ Rev.\ D {\bf 93} (2016) no.7,  072004
    [arXiv:1509.04044 [hep-ex]].

\bibitem{Czakon:2013goa}
  M.~Czakon, P.~Fiedler and A.~Mitov,
  {\it Total Top-Quark Pair-Production Cross Section at Hadron
    Colliders Through $O(\alpha^4_s)$},
  Phys.\ Rev.\ Lett.\  {\bf 110} (2013) 252004
   [arXiv:1303.6254 [hep-ph]].

\bibitem{Aad:2014kva}
  G.~Aad {\it et al.} [ATLAS Collaboration],
  {\it Measurement of the $t\bar{t}$ production cross-section using
    $e\mu $ events with b-tagged jets in pp collisions at $\sqrt{s}$ =
    7 and 8  $\,\mathrm{TeV}$ with the ATLAS detector},
  Eur.\ Phys.\ J.\ C {\bf 74} (2014) no.10,  3109
   Addendum: [Eur.\ Phys.\ J.\ C {\bf 76} (2016) no.11,  642]
    [arXiv:1406.5375 [hep-ex]].
  
\bibitem{Khachatryan:2016mqs}
  V.~Khachatryan {\it et al.} [CMS Collaboration],
  {\it Measurement of the t-tbar production cross section in the e-mu
    channel in proton-proton collisions at $\sqrt{s}$ = 7 and 8 TeV},
  JHEP {\bf 1608} (2016) 029
   [arXiv:1603.02303 [hep-ex]].

\bibitem{Sirunyan:2017uhy}
  A.~M.~Sirunyan {\it et al.} [CMS Collaboration],
  {\it Measurement of the $t \bar t$ production cross section using
    events with one lepton and at least one jet in pp collisions at
    $\sqrt{s}$  = 13 TeV},
  JHEP {\bf 1709} (2017) 051
  [arXiv:1701.06228 [hep-ex]].

\bibitem{Beneke:2016cbu}
  M.~Beneke, P.~Marquard, P.~Nason and M.~Steinhauser,
  {\it On the ultimate uncertainty of the top quark pole mass},
  Phys.\ Lett.\ B {\bf 775} (2017) 63
    [arXiv:1605.03609 [hep-ph]].

\bibitem{Hoang:2017btd}
  A.~H.~Hoang, C.~Lepenik and M.~Preisser,
  {\it On the Light Massive Flavor Dependence of the Large Order
    Asymptotic Behavior and the Ambiguity of the Pole Mass},
  JHEP {\bf 1709} (2017) 099
  [arXiv:1706.08526 [hep-ph]].

\bibitem{Biswas:2010sa}
  S.~Biswas, K.~Melnikov and M.~Schulze,
  {\it Next-to-leading order QCD effects and the top quark mass
    measurements at the LHC},
  JHEP {\bf 1008} (2010) 048
    [arXiv:1006.0910 [hep-ph]].

\bibitem{Heinrich:2013qaa}
  G.~Heinrich, A.~Maier, R.~Nisius, J.~Schlenk and J.~Winter,
  {\it NLO QCD corrections to $W^{+} W^{-}b\bar{b}$ production with
    leptonic decays in the light of top quark mass and asymmetry
    measurements},
  JHEP {\bf 1406} (2014) 158
    [arXiv:1312.6659 [hep-ph]].

\bibitem{Frixione:2014ala}
  S.~Frixione and A.~Mitov,
  {\it Determination of the top quark mass from leptonic observables},
  JHEP {\bf 1409} (2014) 012
   [arXiv:1407.2763 [hep-ph]].

\bibitem{Agashe:2016bok}
  K.~Agashe, R.~Franceschini, D.~Kim and M.~Schulze,
  {\it Top quark mass determination from the energy peaks of b-jets
    and B-hadrons at NLO QCD},
  Eur.\ Phys.\ J.\ C {\bf 76} (2016) no.11,  636
    [arXiv:1603.03445 [hep-ph]].

\bibitem{Heinrich:2017bqp}
  G.~Heinrich, A.~Maier, R.~Nisius, J.~Schlenk, M.~Schulze, L.~Scyboz and J.~Winter,
  {\it NLO and off-shell effects in top quark mass determinations},  
  arXiv:1709.08615 [hep-ph].

\bibitem{Corcella:2017rpt}
  G.~Corcella, R.~Franceschini and D.~Kim,
  {\it Fragmentation Uncertainties in Hadronic Observables for
    Top-quark Mass Measurements},
  arXiv:1712.05801 [hep-ph].

\bibitem{Ravasio:2018lzi}
  S.~Ferrario Ravasio, T.~Jezo, P.~Nason and C.~Oleari,
  {\it A Theoretical Study of Top-Mass Measurements at the LHC Using
    NLO+PS Generators of Increasing Accuracy},
  arXiv:1801.03944 [hep-ph].

\bibitem{Alioli:2013mxa}
  S.~Alioli, P.~Fernandez, J.~Fuster, A.~Irles, S.~O.~Moch, P.~Uwer
  and M.~Vos,
  {\it A new observable to measure the top-quark mass at hadron
    colliders}, 
  Eur.\ Phys.\ J.\ C {\bf 73} (2013) 2438
    [arXiv:1303.6415 [hep-ph]].

\bibitem{Fuster:2017rev}
  J.~Fuster, A.~Irles, D.~Melini, P.~Uwer and M.~Vos,
  {\it Extracting the top-quark running mass using $t\bar{t} + \hbox
    {1-jet}$ events produced at the Large Hadron Collider},
  Eur.\ Phys.\ J.\ C {\bf 77} (2017) no.11,  794
    [arXiv:1704.00540 [hep-ph]].

\bibitem{Aad:2015waa}
  G.~Aad {\it et al.} [ATLAS Collaboration],
  {\it Determination of the top-quark pole mass using $ t\bar{t}+
    $ 1-jet events collected with the ATLAS experiment in 7 TeV pp
    collisions},
  JHEP {\bf 1510} (2015) 121
    [arXiv:1507.01769 [hep-ex]].

\bibitem{CMS:2016khu}
  CMS Collaboration [CMS Collaboration],
  {\it Determination of the normalised invariant mass distribution of
    $t\bar{t}+$jet and extraction of the top quark mass},
  CMS-PAS-TOP-13-006.

\bibitem{Bevilacqua:2015qha}
  G.~Bevilacqua, H.~B.~Hartanto, M.~Kraus and M.~Worek,
  {\it Top Quark Pair Production in Association with a Jet with
    Next-to-Leading-Order QCD Off-Shell Effects at the Large Hadron
    Collider},
  Phys.\ Rev.\ Lett.\  {\bf 116} (2016) no.5,  052003
    [arXiv:1509.09242 [hep-ph]].

\bibitem{Bevilacqua:2016jfk}
  G.~Bevilacqua, H.~B.~Hartanto, M.~Kraus and M.~Worek,
  {\it Off-shell Top Quarks with One Jet at the LHC: A comprehensive
    analysis at NLO QCD},
  JHEP {\bf 1611} (2016) 098
    [arXiv:1609.01659 [hep-ph]].

\bibitem{Melnikov:2011qx}
  K.~Melnikov, A.~Scharf and M.~Schulze,
  {\it Top quark pair production in association with a jet: QCD
    corrections and jet radiation in top quark decays},
  Phys.\ Rev.\ D {\bf 85} (2012) 054002
    [arXiv:1111.4991 [hep-ph]].

\bibitem{Melnikov:2010iu}
  K.~Melnikov and M.~Schulze,
  {\it NLO QCD corrections to top quark pair production in association
    with one hard jet at hadron colliders},
  Nucl.\ Phys.\ B {\bf 840} (2010) 129
  [arXiv:1004.3284 [hep-ph]].

\bibitem{Butterworth:2015oua}
  J.~Butterworth {\it et al.},
  {\it PDF4LHC recommendations for LHC Run II},
  J.\ Phys.\ G {\bf 43} (2016) 023001
   [arXiv:1510.03865 [hep-ph]].

\bibitem{Bevilacqua:2011xh}
  G.~Bevilacqua, M.~Czakon, M.~V.~Garzelli, A.~van Hameren, A.~Kardos,
  C.~G.~Papadopoulos, R.~Pittau and M.~Worek,
  {\it Helac-NLO},
  Comput.\ Phys.\ Commun.\  {\bf 184} (2013) 986
    [arXiv:1110.1499 [hep-ph]].

\bibitem{Czakon:2009ss}
  M.~Czakon, C.~G.~Papadopoulos and M.~Worek,
  {\it Polarizing the Dipoles},
  JHEP {\bf 0908} (2009) 085
    [arXiv:0905.0883 [hep-ph]].

\bibitem{Bevilacqua:2013iha}
  G.~Bevilacqua, M.~Czakon, M.~Kubocz and M.~Worek,
  {\it Complete Nagy-Soper subtraction for next-to-leading order
    calculations in QCD},
  JHEP {\bf 1310} (2013) 204
    [arXiv:1308.5605 [hep-ph]].

\bibitem{vanHameren:2009dr}
  A.~van Hameren, C.~G.~Papadopoulos and R.~Pittau,
  {\it Automated one-loop calculations: A Proof of concept},
  JHEP {\bf 0909} (2009) 106
    [arXiv:0903.4665 [hep-ph]].

\bibitem{Bevilacqua:2010qb}
  G.~Bevilacqua, M.~Czakon, A.~van Hameren, C.~G.~Papadopoulos and M.~Worek,
  {\it Complete off-shell effects in top quark pair hadroproduction
    with leptonic decay at next-to-leading order},
  JHEP {\bf 1102} (2011) 083
    [arXiv:1012.4230 [hep-ph]].

\bibitem{Melnikov:2009dn}
  K.~Melnikov and M.~Schulze,
  {\it NLO QCD corrections to top quark pair production and decay at
    hadron colliders},
  JHEP {\bf 0908} (2009) 049
  [arXiv:0907.3090 [hep-ph]].

\bibitem{Jezabek:1988iv}
  M.~Je\.zabek and J.~H.~K\"uhn,
  {\it QCD Corrections to Semileptonic Decays of Heavy Quarks},
  Nucl.\ Phys.\ B {\bf 314} (1989) 1.

\bibitem{Denner:2012yc}
  A.~Denner, S.~Dittmaier, S.~Kallweit and S.~Pozzorini,
  {\it NLO QCD corrections to off-shell top-antitop production with
    leptonic decays at hadron colliders},
  JHEP {\bf 1210} (2012) 110
    [arXiv:1207.5018 [hep-ph]].

\bibitem{Buckley:2014ana}
  A.~Buckley, J.~Ferrando, S.~Lloyd, K.~Nordstr\"om, B.~Page,
  M.~R\"ufenacht, M.~Sch\"onherr and G.~Watt,
  {\it LHAPDF6: parton density access in the LHC precision era},
  Eur.\ Phys.\ J.\ C {\bf 75} (2015) 132
    [arXiv:1412.7420 [hep-ph]].

\bibitem{Dulat:2015mca}
  S.~Dulat {\it et al.},
  {\it New parton distribution functions from a global analysis of
    quantum chromodynamics},
  Phys.\ Rev.\ D {\bf 93} (2016) no.3,  033006
    [arXiv:1506.07443 [hep-ph]].

\bibitem{Ball:2014uwa}
  R.~D.~Ball {\it et al.} [NNPDF Collaboration],
 {\it  Parton distributions for the LHC Run II},
  JHEP {\bf 1504} (2015) 040
    [arXiv:1410.8849 [hep-ph]].

\bibitem{Harland-Lang:2014zoa}
  L.~A.~Harland-Lang, A.~D.~Martin, P.~Motylinski and R.~S.~Thorne,
  {\it Parton distributions in the LHC era: MMHT 2014 PDFs},
  Eur.\ Phys.\ J.\ C {\bf 75} (2015) 5,  204
  [arXiv:1412.3989 [hep-ph]].

\bibitem{Cacciari:2008gp}
  M.~Cacciari, G.~P.~Salam and G.~Soyez,
  {\it The Anti-k(t) jet clustering algorithm},
  JHEP {\bf 0804} (2008) 063
    [arXiv:0802.1189 [hep-ph]].

\bibitem{Kiyo:2008bv}
  Y.~Kiyo, J.~H.~K\"uhn, S.~Moch, M.~Steinhauser and P.~Uwer,
  {\it Top-quark pair production near threshold at LHC},
  Eur.\ Phys.\ J.\ C {\bf 60} (2009) 375
    [arXiv:0812.0919 [hep-ph]].

\bibitem{Sirunyan:2017idq}
  A.~M.~Sirunyan {\it et al.} [CMS Collaboration],
  {\it Measurement of the top quark mass in the dileptonic $t\bar{t}$
    decay channel using the mass observables $M_{b\ell}$, $M_{T2}$,
    and $M_{b\ell\nu}$ in pp collisions at $\sqrt{s}=8$ TeV},
  Phys.\ Rev.\ D {\bf 96} (2017) no.3,  032002
    [arXiv:1704.06142 [hep-ex]].

\bibitem{Beneke:2000hk}
  M.~Beneke {\it et al.},
  {\it Top quark physics}, 
  Published in Geneva 1999, Standard model physics (and more) at
    the LHC,  419-529  [hep-ph/0003033].



\end{thebibliography}
\end{document}